\documentclass[fleqn,aps,pra,superscriptaddress,notitlepage,floatfix,twocolumn,longbibliography]{revtex4-1}
 \usepackage{amsmath}
\usepackage{graphicx}
\usepackage{setspace}
\usepackage{amssymb}
\usepackage{txfonts}
\usepackage{array}
\usepackage{gensymb}
\usepackage{bm}
\usepackage[usenames,dvipsnames]{xcolor}

\usepackage[colorlinks=true]{hyperref}
\hypersetup{breaklinks=true,allcolors=blue}

\begin{document}
\title{Transport of Spin and Mass at Normal-Superfluid Interfaces\\ in the Unitary Fermi Gas}
\author{Ding Zhang}
\affiliation{Department of Physics, Lehigh University, Bethlehem, Pennsylvania 18015, USA}
\affiliation{Department of Physics and Astronomy, Rice University, Houston, Texas 77005, USA}
\author{Ariel T. Sommer}
\affiliation{Department of Physics, Lehigh University, Bethlehem, Pennsylvania 18015, USA}
\begin{abstract}
Transport in strongly interacting Fermi gases provides a window into the non-equilibrium behavior of strongly correlated fermions. In particular, the interface between a strongly polarized normal gas and a weakly polarized superfluid at finite temperature presents a model for understanding transport at normal-superfluid and normal-superconductor interfaces. An excess of polarization in the normal phase or a deficit of polarization in the superfluid brings the system out of equilibrium, leading to transport currents across the interface.
We implement a phenomenological mean-field model of the unitary Fermi gas, and investigate the transport of spin and mass under non-equilibrium conditions. We consider independently prepared normal and superfluid regions brought into contact, and calculate the instantaneous spin and mass currents across the normal-superfluid (NS) interface. For an unpolarized superfluid, we find that spin current is suppressed below a threshold value in the driving chemical potential differences, while the threshold nearly vanishes for a critically polarized superfluid. 
The mass current can exhibit a threshold in cases where Andreev reflection vanishes, while in general Andreev reflection prevents the occurrence of a threshold in the mass current. Our results provide guidance to future experiments aiming to characterize spin and mass transport across NS interfaces.
\end{abstract}
\maketitle
\section{Introduction}
Experiments on quantum gases of atoms enable strong tests of many-body theories. Studies of ultracold Fermi gases have provided insight into the thermodynamics, excitation spectra, and bulk transport properties of strongly interacting fermions, 
 e.g.~\cite{
%thermo
 shin2008phase,navon2010equation,nascimb`ene2010exploring,nascimb`ene2011fermi-liquid,ku2012revealing,van_houcke2012feynman,
%spectra
 schirotzek2008determination,gaebler2010observation,hoinka2017goldstone,
%transport
 sommer2011universal,sommer2011spin,cao2011universal,enss2012quantum,valtolina2017exploring,enss2019universal,tajima2020spin-dipole}.  
Measurements of fermion transport through quantum point contacts~\cite{brantut2012conduction,husmann2015connecting,krinner2016mapping,kanasz-nagy2016anomalous,hausler2017scanning,corman2019quantized,ono2021observation} and Josephson junctions~\cite{valtolina2015josephson,burchianti2018connecting,zaccanti2019critical,luick2020ideal,del_pace2021tunneling} have extended atomic Fermi gas experiments into the domain of structured devices. Meanwhile, strongly correlated electron materials such as high-temperature superconductors have gained growing interest for application in devices, such as Josephson junctions~\cite{berggren2016computational,perconte2018tunable}, spin valves~\cite{visani2012equal-spin,komori2018magnetic}, and semiconductor-superconductor junction devices~\cite{bouscher2020high-tc}, that feature normal-superconductor interfaces. Experiments on cold atom-based systems that emulate normal-superconductor junctions can therefore provide valuable insight into the effects of strong correlations on transport in such devices. More fundamentally, atomic gas experiments provide a platform for controlled studies of strongly interacting systems out of equilibrium, and can therefore aid in the development of theoretical techniques for understanding the dynamics of many-body systems.

Spin-imbalanced unitary Fermi gases provide a natural model system in which to study strongly correlated normal-superfluid junctions. At low temperatures, when the difference in chemical potential between the two spin components exceeds the Chandrasekhar-Clogston limit, the system phase separates into a weakly polarized superfluid and a strongly polarized normal fluid that coexist at equilibrium~\cite{bulgac2007zero-temperature,shin2008phase,shin2008determination,baur2009theory,nascimb`ene2010exploring,pilati2008phase,liu2008finite-temperature,olsen2015phase}. Spin-imbalanced Fermi gases therefore naturally form a normal-superfluid (NS) interface akin to the ferromagnet-superconductor interfaces employed in superconducting spin valves~\cite{de_jong1995andreev,halterman2004layered,kashimura2010superfluid-ferromagnet-superfluid,alidoust2018half-metallic}. 
Transport across NS interfaces results from non-equilibrium conditions, making strongly interacting Fermi gases an interesting model of non-equilibrium behavior in strongly correlated systems. 

Several previous works have considered aspects of transport across NS interfaces in strongly interacting Fermi gases. 
Calculations of thermal conductivity across the NS interface predicted a suppression of thermal conduction across the interface 
in chemical equilibrium~\cite{van_schaeybroeck2007normal-superfluid,van_schaeybroeck2009normal-superfluid}. Analysis of evaporation dynamics in trapped spin-imbalanced gases predicted a modification of the apparent critical polarization due to non-equilibrium spin distribution 
~\cite{parish2009evaporative}. Experiments on spin-imbalanced Fermi gases observed metastability of non-equilibrium NS interfaces, which the authors attributed partly to inhibition of spin transport at the interface~\cite{liao2011metastability}. Measurements of spin transport coefficients found strong damping of the spin dipole mode in spin-balanced~\cite{sommer2011universal} and spin-imbalanced gases with and without a superfluid core~\cite{sommer2011spin}, and experiments on fermionic quantum point contacts observed suppressed spin conductance with decreasing temperature ~\cite{krinner2016mapping}. Numerical simulations have recently predicted metastable spin-polarized droplets in superfluid Fermi gases~\cite{magierski2019spin-polarized}.

In this paper, we investigate theoretically the transport of spin and mass across the NS interface in the spin-imbalanced unitary Fermi gas. 
We address three main questions: how much spin and mass current flows across the interface under a given set of conditions? Under what conditions does the superfluid excitation gap significantly inhibit spin or mass transport? And to what extent does Andreev reflection cause the mass current to behave differently from the spin current?
To address these questions, we consider the interface between normal and superfluid regions out of chemical equilibrium and calculate the instantaneous spin and mass currents by employing a phenomenological mean-field model. We consider two situations: first, the case of normal and superfluid regions separated by a tunneling barrier potential; and second, the case without a barrier, where the normal and superfluid regions are in mechanical equilibrium. Our calculations provide guidance to future experiments on non-equilibrium NS interfaces by establishing the expected magnitude and behavior of the transport currents. 

Our results show that the spin current flowing into a superfluid is suppressed below a threshold in the driving chemical potential difference. The predicted threshold is analogous to the threshold in the current-voltage (I-V) curve of normal-superconductor junctions at large barrier strength~\cite{blonder1982transition}, employed in scanning tunneling spectroscopy to measure superconducting gaps~\cite{fischer2007scanning}.
In analogy with scanning tunneling spectroscopy, the threshold is related to the minimum in the superfluid excitation spectrum. We find that, for an unpolarized superfluid in contact with a highly polarized normal region, the threshold in chemical potential differences between normal and superfluid regions matches the superfluid gap parameter (Section IV), up to a temperature-dependent correction that we identify in Section IV B. For a critically polarized superfluid (at finite temperature), the minimum in the superfluid excitation spectrum is significantly reduced, leading to a significant reduction in the threshold for transport current. The existence of a threshold supports the notion that non-equilibrium NS interfaces can persist in a metastable configuration. The transport threshold applies specifically to the non-Andreev portion of the current and therefore always affects the spin current. On the other hand, the net (mass) current can have a significant Andreev component, and therefore does not always exhibit a threshold.

The remainder of the paper is structured as follows. In Section II we introduce our phenomenological mean-field model, and in Section III we outline the calculation of the transport currents. In Section IV we present and discuss our results, and we conclude in Section V.

\begin{figure}[h]
   %\centering
    \includegraphics[width=3.5in]{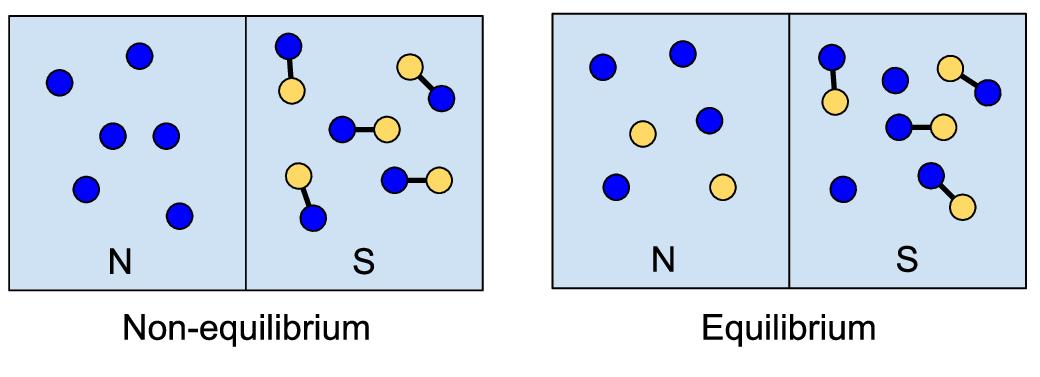}
    \caption{Schematic of non-equilibrium and equilibrium states of a phase-separated spin-imbalanced Fermi gas at finite temperature. Blue: majority (spin up), yellow: minority (spin down); lines represent Cooper pairing.}
    \label{fig:cartoon}
\end{figure}
\section{Theoretical Model}\label{sec:model}
\subsection{Description of the Problem}

We consider a unitary Fermi gas confined in a three-dimensional box potential~\cite{mukherjee2017homogeneous} at low temperature, divided into left and right regions. The confining potential has a uniform cross section perpendicular to the $z$ axis. 
In the left region ($z<0$), the gas has a large spin polarization and is in the normal phase. In the right region ($z>0$), the gas has a smaller spin polarization and is in the superfluid phase. The densities of spin-up and spin-down fermions are uniform within a given region. The temperatures of the two regions can in general differ, but we will consider the case of equal temperatures for the two regions. Due to phase separation below the tricritical point~\cite{shin2008phase,gubbels2008renormalization}, the system can be in equilibrium or out of equilibrium, depending on the degree of polarization in each region. Figure \ref{fig:cartoon} illustrates qualitatively
the equilibrium and non-equilibrium configurations of the system under consideration.
We will focus on calculating the instantaneous currents of spin-up and spin-down fermions across the interface between the two regions.

\begin{figure}[h]
    \includegraphics{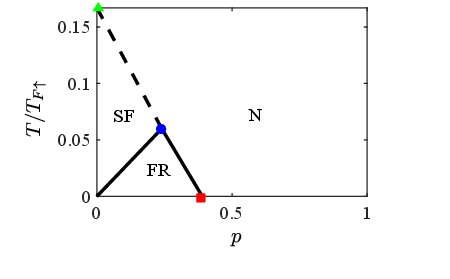}
    \caption{Phase diagram of the two-component Fermi mixture in the unitarity limit, consisting of the superfluid phase (SF), the forbidden region (FR), and the normal phase (N) in the coordinates of $p$ and $T/T_{F\uparrow}$. The dashed line denotes the approximate phase boundary beyond the tricritical temperature $T_{c3}$. Square (red): quantum critical point from Ref.~\cite{gubbels2008renormalization} at polarization $p_{c0} = 0.39$. Circle (blue): the tricritical point from Ref.~\cite{gubbels2008renormalization}, at polarization $p_{c3} = 0.24$ and temperature $T_{c3} = 0.06T_{F\uparrow}$. Triangle (green): critical temperature $T_c = 0.167 T_F$ at zero polarization from Ref.~\cite{ku2012revealing}.
    }
    \label{fig:pphasediagram}
\end{figure}

Figure \ref{fig:pphasediagram} shows the approximate phase diagram of the spin-imbalanced homogeneous unitary Fermi gas. The polarization $p=(n_\uparrow-n_\downarrow)/(n_\uparrow+n_\downarrow)$ characterizes the degree of spin imbalance, where $n_\sigma$ gives the number density of fermions with spin projection $\sigma$. The phase diagram of Fig. \ref{fig:pphasediagram} focuses on the case of a spin-up majority ($p>0$) and normalizes the temperature by the majority Fermi temperature $T_{F\uparrow}=E_{F\uparrow}/k_B$. Here $k_B$ is the Boltzmann constant and $E_{F\uparrow}=\hbar^2(6\pi^2 n_\uparrow)^{2/3}/(2m)$, where $m$ is the mass of the fermions. As in Ref.~\cite{shin2008phase}, we approximate the phase boundaries as straight lines in the $p-T$ plane.
In the non-equilibrium two-region configuration that we consider, each region is internally described by a point $(p_i,T)$ on the equilibrium phase diagram, with the same absolute temperature $T$. The two regions have differing polarization, with the left (normal) side having polarization $p_N$ and the right (superfluid) side having polarization $p_S$. 

 For a given $T/T_{F\uparrow}$, the superfluid phase has a maximum (critical) polarization of $p_{Sc}$, while the normal phase has a minimum (critical) polarization of $p_{Nc}$. Below the tricritical temperature $T_{c3}$, the normal-to-superfluid phase transition is first-order, and the polarization is discontinuous, with $p_{Sc} < p_{Nc}$. At equilibrium, a system with global polarization $(N_\uparrow-N_\downarrow)/(N_\uparrow+N_\downarrow)$ between $p_{Sc}$ and $p_{Nc}$ will exhibit phase separation into a superfluid region and a normal region (here $N_\uparrow$ and $N_\downarrow$ are the total number of spin up and spin down fermions in a homogeneous box potential). At equilibrium, the phase-separated regions attain their critical polarizations, $p_S=p_{Sc}$ and $p_N=p_{Nc}$, respectively. 
 
 Our analysis will focus on the temperature regime below the tricritical point. For context, we will briefly review a few other features of the phase diagram. Above the tricritical temperature, the phase transition is second order and the polarization is continuous ($p_{Sc} = p_{Nc}$). The superfluid region of the phase diagram above the tricritical point is predicted to feature further subdivision into a gapped superfluid and a gapless Sarma superfluid~\cite{gubbels2008renormalization,gubbels2013imbalanced}. For our analysis, we focus on temperatures below the tricritical temperature, and therefore do not consider the Sarma phase. We do not consider the  Fulde-Ferrell-Larkin-Ovchinnikov (FFLO) phase~\cite{sheehy2006bec-bcs,son2006phase,yoshida2007larkin-ovchinnikov,parish2007finite-temperature,jensen2007non-bcs,kinnunen2018fuldeferrelllarkinovchinnikov}, which is predicted to occur away from unitarity in the regime of negative scattering length~\cite{sheehy2006bec-bcs,yoshida2007larkin-ovchinnikov,parish2007finite-temperature}, because we focus on the case of unitary (resonant) interactions. An interesting p-wave superfluid phase has been predicted in highly spin-imbalanced Fermi gases~\cite{bulgac2006induced,bulgac2009induced}. Theoretical calculations predict that the p-wave phase should occur at low temperatures over a range of polarizations above a polarization of about 0.8 in the unitary Fermi gas \cite{patton2012induced}. For simplicity, we do not include the p-wave phase in our analysis, as it covers a relatively small portion of the phase diagram. However, it would be interesting to consider transport in the p-wave phase in future work.

In addition to the confining potential, we allow for a thin barrier potential in the $z=0$ plane separating the two regions. We model the barrier as a Dirac delta function, $V(z)=\Lambda\delta(z)$. For convenience, we parameterize the barrier strength as
\begin{align}
\label{eqn:kLambda}
k_{\Lambda} = 2m\Lambda/\hbar^2
\end{align}
Experimentally, such a barrier would assist in the preparation of the non-equilibrium condition that we consider here, by allowing the two regions to equilibrate separately before initiating transport, similar to Refs. \cite{valtolina2017exploring,krinner2016mapping}. The barrier strength can then be reduced, or turned to zero, to allow currents to flow as the system begins to evolve toward global equilibrium. The temperatures of two independently prepared regions will not in general be equal, but nearly equal temperatures can be achieved through fine tuning of the cooling process applied to each region during preparation. During transport measurements, maintaining a non-zero barrier may be helpful in controlling the magnitudes of the currents. We will consider particular cases of both zero and non-zero barrier strengths.

Our analysis will focus on the instantaneous currents under a given set of conditions.  Over a finite time, one would need to consider additional dynamics. For example, the flow of particles across the interface will generate entropy and heat the system~\cite{chaikin1995principles}. The final temperature could exceed the tricritical temperature, in which case phase separation would not be present in the final state. The final equilibrium state will depend on the volumes of the two initial regions, whereas the instantaneous currents that we calculate here depend only on the local properties of the two regions. Furthermore, as particles flow across the interface, the interface itself can move and will therefore not always be located at $z=0$. Under conditions in which the system heats above the tricritical temperature, the interface would not be thermodynamically stable, and could evolve away from a planar geometry, in analogy with the snake instability of solitons~\cite{cetoli2013snake,wen2013dark-soliton,scherpelz2014phase,ku2016cascade,reichl2017core,wlazlowski2018suppressed}. While these finite-time effects will be important in understanding the full time-evolution of the system, we focus here on the instantaneous response of the system and do not consider its finite-time evolution. However, our results give insight into the initial time-evolution of the system at short times.

\subsection{Phenomenological mean-field model}\label{subsec:BdG}
To carry out the calculations, we employ the Blonder-Tinkham-Klapwijk (BTK) framework  originally introduced to describe normal-superconductor interfaces~\cite{blonder1982transition}. The BTK framework describes the superconducting state using a mean-field theory, and calculates the transport of quasiparticles across a step function in the superconducting or superfluid gap, with a delta-function potential at the interface. Despite being based on mean-field theory, the BTK framework has been successfully used to model interfaces with high-$T_c$ superconductors~\cite{tanaka1995theory,fischer2007scanning,bouscher2020high-tc}, and has been extended to spin-imbalanced unitary Fermi gases~\cite{van_schaeybroeck2007normal-superfluid,van_schaeybroeck2009normal-superfluid,parish2009evaporative}. Similar to Refs.~\cite{van_schaeybroeck2007normal-superfluid,van_schaeybroeck2009normal-superfluid,parish2009evaporative}, we employ a phenomenological mean-field model to describe excitations of the strongly interacting fermion system, and obtain transport properties by studying the scattering of quasiparticles by the NS interface. To provide the most accurate predictions possible within a phenomenological model, we choose the model parameters to fit
state-of-the-art experimental~\cite{shin2008phase,schirotzek2008determination,nascimb`ene2010exploring,nascimb`ene2011fermi-liquid,ku2012revealing,zurn2013precise,hoinka2017goldstone,yan2019boiling} and theoretical~\cite{gubbels2008renormalization,combescot2008normal,prokofev2008fermi-polaron,mora2010normal} determinations of thermodynamic and spectroscopic quantities in the unitary Fermi gas.
A variety of other approaches have recently been pursued to study non-equilibrium dynamics of strongly interacting fermions, including the time-dependent superfluid local density approximation (SLDA)~\cite{bulgac2011time-dependent,forbes2011resonantly,bulgac2012unitary,bulgac2013time-dependent-1,wlazlowski2018suppressed,magierski2019spin-polarized,Kopycinski2021vortex,hossain2022rotating}, Keldysh Green's function methods~\cite{kawamura2020nonequilibrium,wang2014nonequilibrium,liu2017anomalous}, time-dependent Ginzburg-Landau theory~\cite{scherpelz2014phase}, and linear response theory~\cite{enss2012quantum,wlazlowski2013cooper,sekino2020mesoscopic,frank2020quantum}. 

We apply a model Hamiltonian of the form~\cite{van_schaeybroeck2009normal-superfluid}:
\begin{align}
\label{eqn:modelH}
H = \sum_{\sigma}&\int\mathrm{d}^3r \,\hat{\psi}_{\sigma}^{\dag}\,
 H^{(0)}_{\sigma}
\,\hat{\psi}_{\sigma}\\\nonumber 
+&\int\mathrm{d}^3r \left[\,
\Delta(z)\,\hat{\psi}_{\uparrow}^{\dag}\,\hat{\psi}_{\downarrow}^{\dag}
+\Delta^*(z)\,\hat{\psi}_{\downarrow}\,\hat{\psi}_{\uparrow}\right]
\end{align}
Here $H^{(0)}_{\sigma}$ is the single-particle grand canonical Hamiltonian for spin $\sigma$:
\begin{equation}
    H^{(0)}_{\sigma}(z) = -\frac{\hbar^2\nabla^2}{2m_{\sigma}(z)}-\mu_{\sigma}(z)+U_{\sigma}(z)+\Lambda\delta(z)
\end{equation}
The chemical potentials $\mu_{\sigma}$, effective masses $m_\sigma$, gap $\Delta$, and Hartree energies $U_{\sigma}$ are modeled as step functions that are discontinuous across the normal-superfluid interface:
\begin{align}
& \mu_\sigma(z) = \begin{cases}
\mu_{N\sigma}, & \text{for } z<0\\ 
\mu_{S\sigma}, & \text{for } z>0
\end{cases}
\end{align}
and
\begin{align}
U_\sigma(z) = \begin{cases}
U_{N\sigma}, & \text{for } z<0\\ 
U_{S\sigma}, & \text{for } z>0
\end{cases}
\end{align}
Here $\sigma=\uparrow,\downarrow$ denotes the spin.
In a given region (N or S), we also express the 
chemical potentials of spin up and down in terms of their mean value $\mu$ and deviation $h$ (also called the Zeeman field):
\begin{align}
& \mu_N = (\mu_{N\uparrow}+\mu_{N\downarrow})/2 
& h_N = (\mu_{N\uparrow}-\mu_{N\downarrow})/2\\
& \mu_S= (\mu_{S\uparrow}+\mu_{S\downarrow})/2 
& h_S = (\mu_{S\uparrow}-\mu_{S\downarrow})/2
\label{eqn:muandh}
\end{align}
In the superfluid region, a similar parametrization proves useful for the Hartree energies:
\begin{align}
& U_S = (U_{S\uparrow}+U_{S\downarrow})/2    
& U_h = (U_{S\uparrow}-U_{S\downarrow})/2   
\end{align}

Theoretical~\cite{carlson2005asymmetric,magierski2009finite-temperature,haussmann2009spectral} and experimental~\cite{schirotzek2008determination,stewart2008using} studies show that the peak of the spectral function in the unitary Fermi gas is well-described by
an effective mass, Hartree energy, and gap parameter. We therefore choose the masses, Hartree energies, and gap to reproduce known properties of the unitary Fermi gas. Without loss of generality, we consider the case where the majority is spin up.
Minority-spin quasiparticles in the spin-imbalanced normal region acquire an effective mass $m_\downarrow(z<0)=m^*$, where $m^*$ is the polaron mass~\cite{combescot2008normal}. We set the effective mass $m_\uparrow(z<0)$ of the majority spin equal to the bare mass $m$ in the spin-imbalanced normal region~\cite{nascimb`ene2010exploring,mora2010normal}. Likewise, we set the effective masses of both spin states equal to the bare mass in the superfluid phase, in accordance with quantum Monte Carlo calculations at low temperature~\cite{magierski2009finite-temperature}. For simplicity, we do not account for the modified effective mass of quasiholes in the superfluid~\cite{stewart2008using,haussmann2009spectral}. While a general mean-field Hamiltonian contains Hartree energy terms~\cite{de_gennes1989superconductivity}, the Hartree terms vanish at the mean-field level for the unitary Fermi gas, and generally for contact interactions in the continuum limit~\cite{haussmann2007thermodynamics}. In that sense, the Hartree energies in our model Hamiltonian represent effects beyond the mean-field level.

As mentioned above, we treat $\Delta$ and the $U_\sigma$ as parameters in the Hamiltonian, and choose their values to match existing experimental data and first-principles calculations, similar to the treatment of the unitary Fermi gas in Refs. ~\cite{van_schaeybroeck2009normal-superfluid,parish2009evaporative}. Our procedure therefore differs from weak-coupling self-consistent mean-field theory, where $\Delta$ and $U_\sigma$ would be defined in terms of expectation values of the field operators, and determined using gap and number equations. The gap $\Delta$ in our calculation is therefore the spectral gap parameter rather than the superfluid order parameter~\cite{mueller2017review}. We let $\Delta=0$ in the spin-imbalanced normal phase~\cite{mueller2011evolution}.
For the superfluid phase, we set $\Delta/\mu_S=1.25$ based on experimentally measured values for the unitary Fermi gas~\cite{schirotzek2008determination,ku2012revealing,zurn2013precise,hoinka2017goldstone}. The latter quantity has an experimental uncertainty on the order of 5-10\% due to uncertainty on the gap. For simplicity, we apply the same value in the presence of spin imbalance in the superfluid.

To diagonalize the model Hamiltonian (\ref{eqn:modelH}), we apply a Bogoliubov transformation to the field operators:
\begin{align}
&\hat{\psi}_{\uparrow}(\mathbf{r}) = \sum_{n} u_{n\uparrow}(\mathbf{r})\,\hat{\gamma}_{n\alpha}-v_{n\uparrow}^*(\mathbf{r})\,\hat{\gamma}_{n\beta}^\dag \label{eq.1}\\
&\hat{\psi}_{\downarrow}(\mathbf{r}) = \sum_{n} u_{n\downarrow}(\mathbf{r})\,\hat{\gamma}_{n\beta}+ v_{n\downarrow}^*(\mathbf{r})\,\hat{\gamma}_{n\alpha}^{\dag} \label{fieldoperators}%\\
%&\{\hat{\gamma}_{n\sigma},\hat{\gamma}_{n'\sigma'}^{\dag}\}=\delta_{n n'}\delta_{\sigma \sigma'}
\end{align}
The Bogoliubov operators satisfy fermionic anti-commutation relations,
\begin{equation}
\{\hat{\gamma}_{n\sigma},\hat{\gamma}_{n'\sigma'}^{\dag}\}=\delta_{n n'}\delta_{\sigma \sigma'}
\end{equation}
The Hamiltonian (\ref{eqn:modelH}) is diagonalized when the Bogoliubov modes satisfy the Bogoliubov-de Gennes (BdG)
equations~\cite{van_schaeybroeck2007normal-superfluid,van_schaeybroeck2009normal-superfluid}:
\begin{align}
&\label{eqn:alpha}\begin{pmatrix}
H^{(0)}_{\uparrow} & \Delta(z) \\
\Delta^*(z) & -H^{(0)}_{\downarrow} 
\end{pmatrix}\,
\begin{pmatrix}
u_{n\uparrow} \\
v_{n\downarrow} 
\end{pmatrix}
= E_{n\alpha}
\begin{pmatrix}
u_{n\uparrow} \\
v_{n\downarrow} 
\end{pmatrix}\\
&\begin{pmatrix}
H_{\downarrow}^{(0)} & \Delta(z) \\
\Delta^*(z) & -H_{\uparrow}^{(0)} 
\end{pmatrix}\,
\begin{pmatrix}
u_{n\downarrow} \\
v_{n\uparrow} 
\end{pmatrix}
= E_{n\beta}
\begin{pmatrix}
u_{n\downarrow} \\
v_{n\uparrow} 
\end{pmatrix}\label{eqn:beta}
\end{align}
In terms of the Bogoliubov operators, the Hamiltonian becomes: 
\begin{align}
&H = E_{gs}+\sum_{n}\left(E_{n\alpha}\hat{\gamma}_{n\alpha}^{\dag}\hat{\gamma}_{n\alpha}+E_{n\beta}\hat{\gamma}_{n\beta}^{\dag}\hat{\gamma}_{n\beta}\right)
\end{align}
Here $E_{gs}$ is the ground-state energy and $E_{n\alpha}$ and $E_{n\beta}$ are the single-particle excitation energies 

For clarity, and to introduce our notation, below we review the solutions to the BdG equations in the presence of spin imbalance~\cite{van_schaeybroeck2009normal-superfluid,van_schaeybroeck2007normal-superfluid}.
We will refer to the solutions of (\ref{eqn:alpha}) and (\ref{eqn:beta}) as the $\alpha$ and $\beta$ branch, respectively. We denote momentum in the normal-phase by $k$ and in the superfluid by $q$.

In the normal phase ($\Delta=0$), the volume-normalized eigenfunctions on both branches have the form:
\begin{align}
\begin{pmatrix}
u_{\mathbf{k}}(\mathbf{r}) \\v_{\mathbf{k}}(\mathbf{r})
\end{pmatrix}
= \frac{1}{\sqrt{\Omega}}
\begin{pmatrix}
1 \\0
\end{pmatrix}\,e^{i\mathbf{k}\cdot \mathbf{r}}\, ,
\frac{1}{\sqrt{\Omega}}
\begin{pmatrix}
0 \\1 
\end{pmatrix}
\,e^{i\mathbf{k}\cdot \mathbf{r}}
\end{align}
where $\Omega$ is the quantization volume.
The first solution requires $\hbar^2 k^2/(2m_\sigma) > \mu_{N\sigma} - U_{N\sigma}$ to give a positive excitation energy, and corresponds to a particle excitation. Likewise, the second solution requires $\hbar^2 k^2/(2m_\sigma) < \mu_{N\sigma} - U_{N\sigma}$ to give a positive excitation energy, and corresponds to a hole excitation. In the $\alpha$ branch (\ref{eqn:alpha}), the particle solution excites purely $\psi_\uparrow$ (i.e. a spin-up atom in an atomic system), and the hole solution excites $\psi_\downarrow$,  while the reverse holds in the $\beta$ branch (\ref{eqn:beta}). 

The BdG equations for a translationally invariant superfluid admit plane wave solutions of the form:
\begin{equation}
\begin{pmatrix}
u_{\mathbf{q}}(\mathbf{r}) \\v_{\mathbf{q}}(\mathbf{r})
\end{pmatrix}
= \frac{1}{\sqrt{\Omega}}\begin{pmatrix}
u(\mathbf{q})\\
v(\mathbf{q})
\end{pmatrix}\,e^{i\mathbf{q}\cdot \mathbf{r}}  
\end{equation}
The positive eigenvalues of (\ref{eqn:alpha}) and (\ref{eqn:beta}) give the energies:
\begin{align}
& E_{\alpha} = E_s-h_S+U_h > 0\label{eq.Ealpha}\\
& E_{\beta} = E_s+h_S-U_h > 0\label{eq.Ebeta}
\end{align}
where
\begin{align}
& E_s = \sqrt{\xi_s^2+|\Delta|^2} \quad\text{and}\quad
\xi_s = \left|\frac{\hbar^2 q^2}{2m}-\mu_S+U_S\right|
\label{eq.Es}
\end{align} 
At a given energy $E_{\alpha(\beta)}$, equations (\ref{eq.Ealpha}) and (\ref{eq.Ebeta}) admit up to two solutions for the magnitude $q$ of the wavevector. The smaller value $q_h$ corresponds to quasihole excitations while the larger value $q_p$ corresponds to quasiparticle excitations. We give explicit expressions for the wavevectors as functions of energy in Appendix~\ref{app:dispersion}.
The eigenmodes in both the $\alpha$ and $\beta$ branches can then be written:
\begin{align}
&\begin{pmatrix}
u(\mathbf{r}) \\v(\mathbf{r})
\end{pmatrix}
= \frac{1}{\sqrt{\Omega}}\begin{pmatrix}
u_0\\
v_0
\end{pmatrix}\,e^{i\mathbf{q}_p\cdot \mathbf{r}},\,  
\frac{1}{\sqrt{\Omega}}\begin{pmatrix}
v_0\\
u_0
\end{pmatrix}\,e^{i\mathbf{q}_h\cdot \mathbf{r}} 
\end{align}
corresponding to quasiparticles and quasiholes, respectively.
Here the quantities $u_0$ and $v_0$ are:
\begin{align}
u_0 =\sqrt{\frac{1}{2}\,\left(1+\frac{\xi_s}{E_s}\right)} \quad\text{and}\quad
v_0 = \sqrt{\frac{1}{2}\,\left(1-\frac{\xi_s}{E_s}\right)} \label{eqn:u0v0}
\end{align}
which are functions of energy $E_{\alpha(\beta)}$ on branch $\alpha(\beta)$. Because $u_0 \geq v_0$, particle-like excitations on the $\alpha$ branch involve mostly $\psi_\uparrow$ (i.e. spin-up atoms) while hole-like excitations involve mostly $\psi_\downarrow$, while the reverse holds on the $\beta$ branch.

To find the Hartree energies $U_\sigma$, we equate the expression for the densities $n_\sigma = \langle \hat{\psi}^\dag_\sigma \hat{\psi}_\sigma\rangle$ from our phenomenological mean-field model to the expected densities based on studies of the equation of state of the unitary Fermi gas. In particular, we consider the normal phase~\cite{nascimb`ene2010exploring,mora2010normal}, balanced superfluid~\cite{ku2012revealing,nascimb`ene2010exploring} , and critically polarized superfluid~\cite{gubbels2008renormalization}. Details of our procedure for determining the Hartree energies are given in Appendix \ref{app:hartree}.

\subsection{Degrees of freedom}
At a given temperature $T$, the two-region system in local equilibrium has four degrees of freedom, namely the four chemical potentials: $\mu_{N\uparrow}$, $\mu_{N\downarrow}$, $\mu_{S\uparrow}$, and $\mu_{S\downarrow}$. We non-dimensionalize all energies by dividing by $\mu_S$. The three resulting dimensionless parameters are $\mu_{N}/\mu_S$, $h_{N}/\mu_S$, and $h_S/\mu_S$. In principle, the instantaneous transport currents can be calculated for arbitrary values of those parameters. We consider a few specific cases.

We consider two cases for $\mu_N/\mu_S$.
In the first case, we consider $\mu_N = \mu_S$. The pressure in the normal and superfluid regions will be different in this case. Experimentally, the pressure differential can be supported by maintaining a non-zero barrier height between the regions. Therefore, in this case we carry out the calculation in the presence of a non-zero tunneling barrier. Experimentally, arbitrary ratios of $\mu_N/\mu_S$ can be achieved by tuning the densities in the two regions, for example by moving one of the outer walls of the trap.  

In the second case, we choose $\mu_N/\mu_S$ for a given $T/\mu_S$ to achieve mechanical equilibrium. Experimentally, this would describe a situation where the barrier between the regions has been removed and the system has had sufficient time to reach mechanical equilibrium, while still being out of chemical equilibrium~\cite{sommer2011universal}.

After fixing $\mu_N/\mu_S$, we choose the two remaining degrees of freedom, $h_N/\mu_S$ and $h_S/\mu_S$. We consider two specific cases for $h_S/\mu_S$: a spin-balanced superfluid ($h_S=0$) or a critically polarized superfluid ($h_s=h_c$). In each case, we consider the full range of $h_N$, and calculate the transport currents as functions of $h_N$.

Chemical potential differences drive particle transport. We therefore define $\delta\mu_{\sigma}$ as the chemical potential differences across the interface:
\begin{align}
\delta\mu_{\uparrow} = \mu_{N\uparrow}-\mu_{S\uparrow} \quad\text{and}\quad
\delta\mu_{\downarrow} = \mu_{S\downarrow}-\mu_{N\downarrow}
\label{eqn:deltamus}
\end{align}
The choice of signs in (\ref{eqn:deltamus}) ensures that $\delta\mu_\sigma\geq 0$.
In the special case of $\mu_N = \mu_S$, we have $\delta\mu_{\uparrow}=\delta\mu_{\downarrow}$. The relation between the $\delta\mu_\sigma$ and $h_N$ and $h_S$ depends on the equation of state and is plotted for our model in Appendix~\ref{app:Dmu_hn}. 
Experimentally, one typically measures density rather than chemical potential, so we also plot the polarization $p_N$ of the normal region versus $h_N$ in Appendix~\ref{app:pn_hn}.

\section{Scattering formulation and Current densities}
\subsection{Scattering states and coefficients}
Transport across the normal-superfluid interface can be described in terms of quasiparticle reflection and transmission coefficients~\cite{blonder1982transition}. Scattering of quasiparticles at the normal-superfluid interface of a spin-imbalanced Fermi gas has been discussed previously in Refs.~\cite{van_schaeybroeck2007normal-superfluid,van_schaeybroeck2009normal-superfluid,parish2009evaporative}. We extend previous results by including the Hartree energies and polaron effective mass in the scattering problem, and by using the resulting scattering coefficients to calculate the currents of spin up and spin down fermions across the interface.

To describe scattering at the normal-superfluid interface, we employ energy normalization with respect to the $z$-component of the momentum, rather than the volume normalization of Section \ref{subsec:BdG}. 
Energy normalization is helpful when dealing with multiple scattering channels having potentially different group velocities. 
Moving from the single-region solutions of Section \ref{subsec:BdG} to an interface problem also changes the Bogoliubov modes into scattering solutions that obey boundary conditions at the interface. We  parameterize the scattering states in terms of their total energy and transverse momentum, which are both conserved, as well as the incident (\textit{in}) channel of the scattering process. The $\alpha$ and $\beta$ branches each have four channels, corresponding to a particle or hole incident on the interface from the left or right. Note that the $\alpha$ and $\beta$ branches have no cross-coupling due to conservation of spin~\cite{de_jong1995andreev}.

We express the total current densities of spin up and spin down in terms of the contributions 
of each Bogoliubov mode:
\begin{align}
J_{\sigma} = &\frac{1}{A}\sum_{n,\, \mathbf{k}_{\perp}}\int
\mathrm{d}E\, \left(j_{\sigma n\alpha}+j_{\sigma n\beta}\right)
\end{align}
Here $n\in\{Lp, Lh, Rp, Rh\}$ runs over the four scattering channels (particle incident from the left, hole incident from the left, particle incident from the right, and hole incident from the right, respectively), $\mathbf{k}_{\perp}$ is the transverse momentum, and $E$ is the energy.
The cross-sectional area $A$ cancels upon converting the sum on $\mathbf{k}_\perp$ to an integral.
In terms of the energy-normalized mode functions, the spin-up current per unit energy from each mode is given by:
\begin{align}
j_{\uparrow n\alpha} =&\frac{\hbar}{2im}\left(\frac{\partial u_{n\uparrow}}{\partial z}u_{n\uparrow}^*-\frac{\partial u_{n\uparrow}^*}{\partial z}u_{n\uparrow}\right)\,f_{n\alpha}
\label{eqn:jupna}\\
j_{\uparrow n\beta} =&-\frac{\hbar}{2im}\left(\frac{\partial v_{n\uparrow}}{\partial z}v_{n\uparrow}^*-\frac{\partial v_{n\uparrow}^*}{\partial z}v_{n\uparrow}\right)\,(1-f_{n\beta})\label{eqn:jupnb}
\end{align}
Similarly, the contributions to the spin-down current  are:
\begin{align}
j_{\downarrow n\beta} =&\frac{\hbar}{2im^*} \left(\frac{\partial u_{n\downarrow}}{\partial z}u_{n\downarrow}^*-\frac{\partial u_{n\downarrow}^*}{\partial z}u_{n\downarrow}\right)\,f_{n\beta}\label{eqn:jdownnb}\\
j_{\downarrow n\alpha} =&-\frac{\hbar}{2im^*}\left(\frac{\partial v_{n\downarrow}}{\partial z}v_{n\downarrow}^*-\frac{\partial v_{n\downarrow}^*}{\partial z}v_{n\downarrow}\right)\,(1-f_{n\alpha})\label{eqn:jdownna}
\end{align}
Here $f_{n\alpha}$ and $f_{n\beta}$ are the occupation probabilities of the Bogoliubov modes in the $\alpha$ and $\beta$ branches, respectively. Note that the occupation probabilities depend on $E$, and the mode functions depend on $E$ and $\mathbf{k}_\perp$.

Under non-equilibrium conditions, the left and right regions will have different chemical potentials for a given spin. When solving the scattering problem, we employ the technique introduced in Ref.~\cite{blonder1982transition} of referencing all energies to the superfluid-side chemical potentials, and accounting for the non-equilibrium conditions through the quasiparticle distribution functions $f_{n}$. We give explicit expressions for the distribution functions in Appendix \ref{app:fermifn}.

We now express the Bogoliubov modes in terms of reflection and transmission coefficients. 
We write the mode functions for the $\alpha$ and $\beta$ branches as
\begin{equation}
\psi_{n\alpha}=\begin{pmatrix}u_{n\uparrow}\\ v_{n\downarrow}\end{pmatrix} \quad\text{and}\quad \psi_{n\beta}=\begin{pmatrix} u_{n\downarrow} \\ v_{n\uparrow}\end{pmatrix}
\end{equation}for the four channels $n\in\{Lp, Lh, Rp, Rh\}$. For each branch, we construct scattering states in terms of \textit{in} and \textit{out} states,  which we formally assemble into vectors (dropping the $\alpha$ and $\beta$ subscripts):
\begin{equation}
\bm{\psi}^{in(out)} = \begin{pmatrix}
\psi_{Lp} \\ \psi_{Lh} \\ \psi_{Rp}\\\psi_{Rh}
\end{pmatrix}^{in(out)}
\end{equation}
The scattering states in each of the four channels are expressed in terms of the \textit{in} and \textit{out} states and the $S$ matrix:
\begin{equation}
    \psi_n = \bm{\psi}^{in}\cdot \bm{e}_n + \bm{\psi}^{out}\cdot S \bm{e}_n 
    \label{eqn.psin}
\end{equation}
where $n$ is the channel index, $\bm{e}_n$ is the $n$-th unit vector in $\mathbb{R}^4$, and the $S$ matrix for either branch consists of 16 scattering coefficients:
\begin{equation}
S =
\begin{pmatrix}
r_{pp}^{A} & r_{ph}^B & t_{pp}^C & t_{ph}^D \\
r_{hp}^{A} & r_{hh}^B & t_{hp}^C & t_{hh}^D \\
t_{pp}^{A} & t_{ph}^B & r_{pp}^C & r_{ph}^D \\
t_{hp}^{A} & t_{hh}^B & r_{hp}^C & r_{hh}^D
\end{pmatrix}
\end{equation}
The labels $A$, $B$, $C$, and $D$ refer to the four incident scattering channels $Lp$, $Lh$, $Rp$, and $Rh$, respectively.

For the $\alpha$ branch, the \textit{in} and \textit{out} states of a particle in the left (normal) region are: 
\begin{align}
&\psi_{Lp\alpha}^{in(out)} = \sqrt{\frac{m}{2\pi\hbar^2 k_{p\uparrow}}}
\begin{pmatrix}
1 \\ 0
\end{pmatrix}
e^{\pm ik_{p\uparrow} z}e^{i\mathbf{k_{\perp}}\cdot\mathbf{r}} \theta(-z)\label{eqn:psilpa}
\end{align}
For a hole in the left region, they are:
\begin{align}
&\psi_{Lh\alpha}^{in(out)} = \sqrt{\frac{m^*}{2\pi\hbar^2 k_{h\downarrow}}}\begin{pmatrix} 0 \\1 \end{pmatrix} e^{\mp ik_{h\downarrow} z}e^{i\mathbf{k_{\perp}}\cdot\mathbf{r}} \theta(-z) \label{eqn:psilha}
\end{align}
And for the right region:
\begin{align}
&\psi_{Rp\alpha}^{in(out)} = \sqrt{\frac{m\, E_s/\xi_s}{2\pi\hbar^2 q_{p\alpha}}}
\begin{pmatrix}
u_{0} \\ v_{0}
\end{pmatrix}
e^{\mp iq_{p\alpha}z}e^{i\mathbf{k_{\perp}}\cdot\mathbf{r}} \theta(z)\label{eqn:psirpa}
\\
&\psi_{Rh\alpha}^{in(out)} =\sqrt{\frac{m\, E_s/\xi_s}{2\pi\hbar^2 q_{h\alpha}}}\begin{pmatrix}
v_{0} \\ u_{0}
\end{pmatrix}
e^{\pm iq_{h\alpha} z}e^{i\mathbf{k_{\perp}}\cdot\mathbf{r}} \theta(z)\label{eqn:psirha}
\end{align}
Here the upper and lower signs in the exponentials correspond to the \textit{in} and \textit{out} states, respectively, and $\theta(z)$ is the Heaviside step function. The wavevectors $k_{p\uparrow}$, $k_{h\downarrow}$, $q_{p\alpha}$, and $q_{h\alpha}$ are the magnitudes of the $z$ components of the wavevectors of particle and hole excitations on the $\alpha$ branch in the normal and superfluid phases; their dependence on the energy and transverse momentum is given in Appendix \ref{app:dispersion}. Expressions for the $\beta$ branch \textit{in} and \textit{out} states can be obtained by replacing $\alpha\rightarrow\beta$, $\uparrow\leftrightarrow\downarrow$ in (\ref{eqn:psilpa})-(\ref{eqn:psirha}) and $m\leftrightarrow m^*$ in (\ref{eqn:psilpa}) and (\ref{eqn:psilha}).

The scattering coefficients are obtained by imposing boundary conditions on the scattering states (\ref{eqn.psin}) at the interface. 
The mode functions must be continuous across the interface: $ \psi_{n}(z\rightarrow 0^-)=\psi_{n}(z\rightarrow 0^+)$.
For the $\alpha$ branch, the derivatives satisfy:
\begin{align}
\left.\frac{\partial\psi_{n\alpha}}{\partial z}\right|_{z\rightarrow 0^+}-
\left.\begin{pmatrix}
1 & 0 \\
0 & m/m^*
\end{pmatrix}
\frac{\partial\psi_{n\alpha}}{\partial z}\right|_{z\rightarrow 0^-} = k_\Lambda \psi_{n\alpha}(0)\label{eqn:bcalpha}
\end{align}
where $k_\Lambda$ is defined in Eqn (\ref{eqn:kLambda}).
For the $\beta$ branch, the derivatives satisfy:
\begin{align}
\left.\frac{\partial\psi_{n\beta}}{\partial z}\right|_{z\rightarrow 0^+}-
\left.\begin{pmatrix}
m/m^* & 0 \\
0 & 1
\end{pmatrix}
\frac{\partial\psi_{n\beta}}{\partial z}\right|_{z\rightarrow 0^-} = k_\Lambda \psi_{n\beta}(0)\label{eqn:bcbeta}
\end{align}
Note that we obtain the boundary conditions (\ref{eqn:bcalpha}) and (\ref{eqn:bcbeta}) using the Hermitian kinetic energy operator ordering $\frac{1}{2}\hat{p}\,m_\sigma^{-1}(z)\hat{p}$ from effective mass theory~\cite{bendaniel1966space-charge,einevoll1990operator-1,cavalcante1997form}.

Full expressions for the resulting scattering coefficients are given in Appendix \ref{app:coef}.
We find that the $S$ matrix is unitary, $S^\dag S=1$, as required by conservation of probability. We also find that the transpose satisfies $S(\Delta)^T=S(\Delta^*)$, as required by time-reversal symmetry. As $S$ has the property $S(\Delta)^* = S(\Delta^*)$, it follows that $S$ is Hermitian: $S^\dag = S$. The unitarity and Hermiticity of $S$ will assist in simplifying the expressions for the currents. In particular, the coefficients for channels $C$ and $D$ (excitation incident from the right) can be written in terms of the coefficients for channels $A$ and $B$ (excitation incident from the left), allowing us to express the currents in terms of the coefficients for channels $A$ and $B$.

The coefficient $r^A_{hp}$ for the $\alpha$ branch (which we will denote as  $r^A_{hp_\alpha}$) represents an Andreev reflection process, where a spin-up particle from the normal region is reflected as a spin-down hole. Likewise, $r^B_{ph_\alpha}$ describes the reversed process, or reverse Andreev reflection, where a spin-down hole is reflected as a spin-up particle. Meanwhile, the  coefficients $r^C_{hp}$ and $r^D_{ph}$ describe Andreev-type reflection of excitations incident from the superfluid region. Physically, a Cooper pair is created or annihilated in the superfluid during Andreev reflection to conserve particle number. Andreev reflections therefore transport mass across the interface. In the forward Andreev reflection $r^A_{hp}$, a spin-up particle and a spin-down particle leave the normal region and a Cooper pair appears in the superfluid. In reverse Andreev reflection $r^B_{ph}$, a Cooper pair disappears and the normal region gains a spin up and a spin down particle. Andreev reflection does not transport spin, however, as the total spin in each region remains unchanged. Moreover, unlike a quasiparticle transmission process, Andreev reflection does not create or annihilate a single-particle excitation in the superfluid, and therefore can occur at energies within the superfluid excitation gap.

\subsection{Current densities}
Employing the scattering states in the expressions for the current contributions (\ref{eqn:jupna})-(\ref{eqn:jdownna}) gives general expressions for the currents in terms of the $S$ matrix elements. In particular, we are interested in the net (mass) current and the spin current:
\begin{align}
J^{\text{net}} = J_{\uparrow}+J_{\downarrow} \quad\text{and}\quad
J^{\text{spin}} = J_{\uparrow}-J_{\downarrow}
\end{align}
Depending on the values of $E$ and $k_\perp$, some scattering channels can become closed, leading to different scattering regimes as described in Refs.~\cite{van_schaeybroeck2007normal-superfluid, van_schaeybroeck2009normal-superfluid}. Within intervals of $E$ and $k_\perp$ where all the channels are open (denoted Regime I in Appendix \ref{app:regimes}), the contributions to the net and spin currents from the $\alpha$ branch are given by:
\begin{widetext}
\begin{align}
&j_{\alpha}^{\text{net}} = \frac{1}{h}\left\{(1-|r_{pp_{\alpha}}^A|^2+|r_{hp_{\alpha}}^A|^2)[f(E_{\alpha}-\delta\mu_{\uparrow})-f(E_{\alpha})]-(1-|r_{hh_{\alpha}}^B|^2+|r_{ph_{\alpha}}^B|^2)[f(E_{\alpha}-\delta\mu_{\downarrow})-f(E_{\alpha})]\right\} \label{eqn:jneta}\\
&j_{\alpha}^{\text{spin}} = \frac{1}{h}\left\{(1-|r_{pp_{\alpha}}^A|^2-|r_{hp_{\alpha}}^A|^2)[f(E_{\alpha}-\delta\mu_{\uparrow})-f(E_{\alpha})]+(1-|r_{hh_{\alpha}}^B|^2-|r_{ph_{\alpha}}^B|^2)[f(E_{\alpha}-\delta\mu_{\downarrow})-f(E_{\alpha})]\right\} \label{eqn:jspina}
\end{align}
The $\beta$ branch contributions are:
\begin{align}
&j_{\beta}^{\text{net}} = \frac{1}{h}\left\{(1-|r_{pp_{\beta}}^A|^2+|r_{hp_{\beta}}^A|^2)[f(E_{\beta}+\delta\mu_{\downarrow})-f(E_{\beta})]-(1-|r_{hh_{\beta}}^B|^2+|r_{ph_{\beta}}^B|^2)[f(E_{\beta}+\delta\mu_{\uparrow})-f(E_{\beta})]\right\}\label{eqn:jnetb} \\
&j_{\beta}^{\text{spin}} = \frac{1}{h}\left\{(1-|r_{pp_{\beta}}^A|^2-|r_{hp_{\beta}}^A|^2)[f(E_{\beta})-f(E_{\beta}+\delta\mu_{\downarrow})]-(1-|r_{hh_{\beta}}^B|^2-|r_{ph_{\beta}}^B|^2)[f(E_{\beta}+\delta\mu_{\uparrow})-f(E_{\beta})]\right\}\label{eqn:jspinb}
\end{align}
\end{widetext}
In regimes where a scattering channel is closed, the corresponding scattering coefficients drop out of the expressions for the currents. Appendix \ref{app:regimes} describes the regimes in more detail. 

The current density integrands (\ref{eqn:jneta})-(\ref{eqn:jspinb}) show that the contributions from the $\beta$ branch are small compared to the $\alpha$ branch. Since $E_{\alpha}$, $E_{\beta}$, $\delta\mu_{\uparrow}$ and $\delta\mu_{\downarrow}$ are positive, all the Fermi functions in the $\beta$ currents have positive arguments, while some in the $\alpha$ currents can have negative arguments. With positive arguments, the Fermi function quickly drops to zero, leading to vanishing results for the $\beta$ currents. The $\beta$ branch was also found to have a small contribution to heat current at the interface in Ref.~\cite{van_schaeybroeck2007normal-superfluid,van_schaeybroeck2009normal-superfluid}

The dominance of the $\alpha$ branch results from the polarization of the normal phase.
Creating a large normal (non-Andreev) current of spin $\sigma$ in the $\alpha$ branch requires $\delta \mu_\sigma \geq E_{\alpha\mathrm{min}}$, where $E_{\alpha\mathrm{min}}$ is the minimum of $E_\alpha$. As discussed in the next section, this can be achieved sufficiently far from equilibrium. On the other hand, because the $\beta$ branch consists of spin up holes and spin down particles, a large normal current in the $\beta$ branch requires $\delta \mu_\sigma \leq -E_{\alpha\mathrm{min}}$, which is impossible since $\delta\mu_\sigma \geq 0$.
In addition, as mentioned earlier, we apply the superfluid chemical potentials $\mu_{S\sigma}$ to the normal side when solving the scattering problem, and implement non-equilibrium through the quasiparticle distribution functions. 
Consequently, on the normal side, the density of spin-up particles formally exceeds the density of spin-up holes, and vice versa for spin down, so that the $\alpha$ branch accounts for the majority of excitations on the normal side. 
In our final calculations, we confirm that for temperatures below $0.3\mu_S$, the $\alpha$ branch accounts for at least 99\% of the current. 

\section{Results and Discussion}
\label{sec:results}
\subsection{Interface away from mechanical equilibrium}

\begin{figure}[h]
    \includegraphics{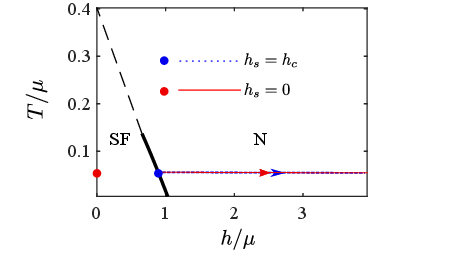}
    \caption{Thermodynamic states considered, for the case $\mu_N=\mu_S$. 
    For the sub-case $h_s=0$ (unpolarized superfluid), the red dot indicates the state of the superfluid region while the red line indicates the allowed states of the normal region, given the temperature $T= 0.05\mu_S$ and the condition $\mu_N=\mu_S$. For the other sub-case, $h_s=h_c$ (critically polarized superfluid), the blue dot indicates the state of the superfluid region while the blue dashed line indicates the allowed states of the normal region.
    The solid black and black dashed curves show the normal-superfluid phase boundary above and below the tricritical point, respectively. The maximum value of $h_N/\mu_N$ on the red solid and blue dotted lines correspond to $p_N = 0.99$, while the minimum $h_N/\mu_N$ corresponds to the critical polarization in the normal phase of $p_{Nc} = 0.34$ at $T/\mu_N=T/\mu_s = 0.05$.
    }
    \label{fig: phase diagram_barrier}
\end{figure}
\begin{figure}[h]
    \includegraphics{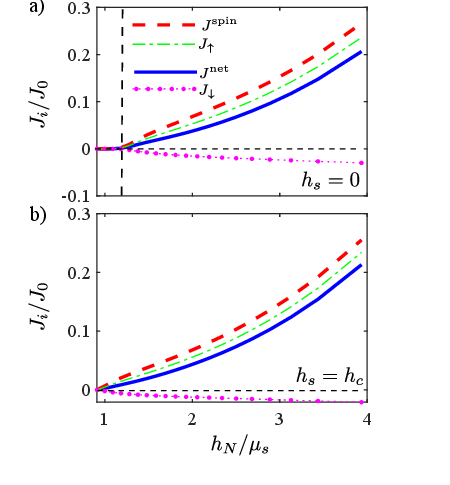}
    \caption{Net, spin, spin-up and spin-down current densities for the case $\mu_N=\mu_S$ as functions of $h_N/\mu_S$, under the conditions: (a) $h_S = 0$, and (b) $h_S = h_c$. For both plots, the temperature is $T = 0.05\mu_S$. The vertical dashed line in (a) denotes the $h_N$ value at which $\delta\mu_\uparrow=\delta\mu_\downarrow =E_{min}$, corresponding to normal-region polarization $p_N = 0.44$. The horizontal dashed line in both  plots is the $J_i = 0$ line. The horizontal axes start at $h_N=h_c(T=0.05\mu_S)=0.91\mu_s$.}
    \label{fig: J_ns12_Ba}
\end{figure}

In this section, we consider the case where the system is out of mechanical equilibrium and a Dirac delta potential barrier is applied. We analyze the particular case of $\mu_N = \mu_S$, and barrier strength $k_{\Lambda} = 20 k_s$,  where $k_s=2m\mu_S/\hbar^2$. We consider two different conditions for the superfluid, (1) $h_S = 0$ for a spin-balanced superfluid, and (2) $h_S = h_c$ for a maximally polarized superfluid. In both cases, we consider a normal region with chemical potential imbalance $h_N$ (equivalently, polarization $p_N$) greater than the equilibrium value, so that the system is out of global equilibrium.
Figure \ref{fig: phase diagram_barrier} plots the thermodynamic states under consideration on the phase diagram in terms of $T/\mu$ and $h/\mu$. In both cases, $h_s$ is held at a fixed value, while $h_N$ is varied from the critical value up to a large value corresponding to a normal-region polarization of $p_N=0.99$.

\begin{figure}[t]
    \includegraphics[width=2.5in]{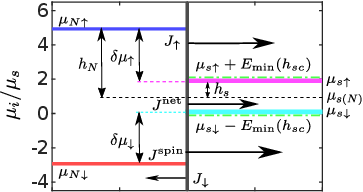}
    \caption{Chemical potentials and current densities across the interface, away from mechanical equilibrium, for parameters $p_N = 99\%$, and $h_S= h_{c}$, $\mu_N = \mu_S$, $T = 0.05\mu_S$. The left (right) side of the figure corresponds to the normal (superfluid) region. The horizontal axis is qualitative, showing the directions and relative magnitudes of the currents. The vertical axis shows the chemical potentials quantitatively. }
    \label{fig: mu_at_interface_free.eps}
\end{figure}

In Fig.~\ref{fig: J_ns12_Ba}, we show the instantaneous net and spin currents versus $h_N$. 
The currents are normalized by a factor $J_0$, given by: 
\begin{equation}
J_0 = \frac{m\,\mu_S^2}{4\pi^2\,\hbar^3} 
\end{equation}
The currents for the unpolarized superfluid exhibit a threshold at a critical value of $h_N$, as shown in Fig.~\ref{fig: J_ns12_Ba}(a). To interpret the threshold, we first note that, in the present case where $\mu_N=\mu_S$, the chemical potential differences (\ref{eqn:deltamus}) satisfy $\delta\mu_\uparrow=\delta\mu_\downarrow$, as illustrated in Fig.~\ref{fig: mu_at_interface_free.eps}. The threshold occurs at the value of $h_N$ where $\delta\mu_\uparrow$ and $\delta\mu_\downarrow$ equal the superfluid minimum excitation energy $E_{min}$. With $E_{min} = \Delta_0-h_S+U_h$, we have $E_{min}(h_S=0) = \Delta_0=1.25 \mu_s$ for the unpolarized superfluid, while $E_{min}(h_S=h_c)=0.02\mu_S$ for the critically polarized superfluid. Accordingly, the threshold in the critically polarized superfluid, Fig.~\ref{fig: J_ns12_Ba}(b), is too small to easily discern. The presence of a threshold implies that the system is metastable when $h_s=0$: the system is out of equilibrium, but mass and spin transport are strongly suppressed. Figure \ref{fig: mu_at_interface_free.eps} shows an example of a situation where the threshold is exceeded, allowing currents to flow.

As mentioned earlier, the spin current results entirely from normal (non-Andreev) transmission processes.  Normal current involves the creation or annihilation of an excitation in the superfluid. Efficient creation of excitations in the superfluid at low temperatures requires
$\delta\mu_\uparrow > E_\mathrm{min}$
for spin up excitations, and $\delta\mu_\downarrow> E_\mathrm{min}$ for spin down (holes). The energy required to excite the superfluid therefore explains the observed threshold in the spin current.

The net current consists, in general, of both normal and Andreev processes. Andreev reflection does not excite the superfluid, and therefore should exhibit no threshold effects. The presence of a threshold in the net current in Fig.~\ref{fig: J_ns12_Ba}(a) suggests that the net Andreev current vanishes in this case. To confirm that the Andreev current vanishes, we separate the net current into Andreev and normal components. We identify the Andreev current in the $\alpha$ branch as the sum of the terms in Eq. (\ref{eqn:jneta}) that are proportional to the Andreev reflection coefficients:
\begin{align}
j^{\text{Andreev}}_{\alpha} =\frac{2}{h}|r_{hp_{\alpha}}^A|^2\left[f(E_{\alpha}-\delta\mu_{\uparrow})-f(E_{\alpha}-\delta\mu_{\downarrow})\right]\label{eqn:jAndreev}
\end{align}
We have used $|r^B_{ph_\alpha}|^2=|r^A_{hp_\alpha}|^2$ from the hermiticity of the $S$-matrix to simplify the expression. We verify Eq. (\ref{eqn:jAndreev}) by considering the net current in the scattering regime where normal transmission is energetically forbidden, denoted Regime II in Appendix~\ref{app:regimes}. We find that the net $\alpha$ branch current is given by (\ref{eqn:jAndreev}) in Regime II, confirming that it captures the current due to Andreev reflection. In the present case of $\mu_N=\mu_S$, where $\delta\mu_\uparrow=\delta\mu_\downarrow$, Eqn. (\ref{eqn:jAndreev}) shows that the Andreev contribution to the net current is indeed zero, explaining the sharp threshold observed in the net current.

We now discuss a final point of interest regarding the results in Fig.~\ref{fig: J_ns12_Ba}. Although $\delta\mu_{\uparrow} = \delta\mu_{\downarrow}$ for $\mu_N = \mu_S$, the spin-up current in Fig.~\ref{fig: J_ns12_Ba} is much larger than the spin-down current. 
We attribute this asymmetry to the asymmetry in the dispersion relations between particle-like and hole-like excitations, in both the normal and superfluid phases. While the energy of a particle-like excitation is unbounded, the energy of a hole-like excitation is bounded from above. As a result, when integrating over the total energy $E_\alpha$ and transverse kinetic energy $\xi_\perp$ to obtain the currents, there are regimes in which hole-like excitations are forbidden in the normal and/or superfluid phase (Appendix~\ref{app:regimes}). As a result, the current of hole-like excitations is smaller than the current of particle-like excitations. On the $\alpha$ branch, particle-like excitations result predominantly from excitations of $\psi_\uparrow$ (i.e. spin-up atoms), while hole-like excitations result predominantly from $\psi_\downarrow$. Consequently, the spin-up current is larger than the spin-down current in this case, despite the equality of the driving chemical potential differences.

\subsection{Interface at mechanical equilibrium}
\label{sec:mecheq}
\begin{figure}[h]
    \includegraphics{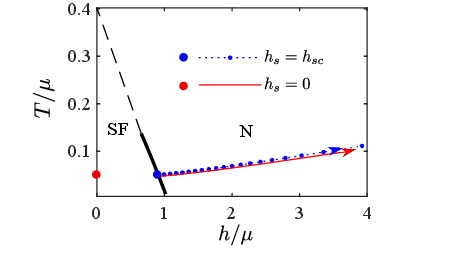}
    \caption{Thermodynamic states considered, for the case of mechanical equilibrium at temperature $T = 0.05\mu_S$. The maximum value of $h_N/\mu_N$ on the red solid (blue dotted) line corresponds to $p_N = 0.96$.
    }
    \label{fig: phase diagram_ME}
\end{figure}

\begin{figure}[h]
    \includegraphics{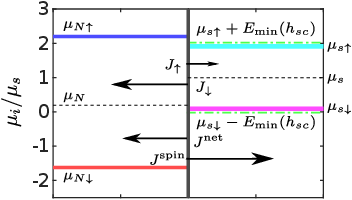}
    \caption{Chemical potentials and schematic current densities across the interface, at mechanical equilibrium, with $p_N = 0.96$, and $h_S= h_c$, $T = 0.05\mu_S$.}
    \label{fig: mu_at_interface}
\end{figure}

\begin{figure}[h]
    \includegraphics{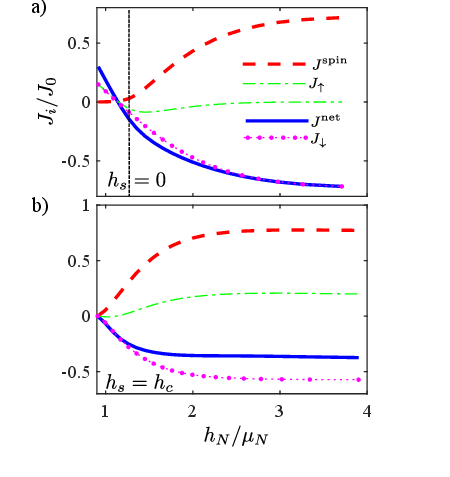}
    \caption{Net, spin, spin-up and spin-down current densities as functions of $h_N$, under the conditions: (a) $h_S = 0$, and (b) $h_S = h_c$. For both plots, temperature is $T = 0.05\mu_S$. 
    The dashed vertical line denotes the $h_N$ value at which $\delta\mu_{\downarrow} = E_{min}$.
    }
    \label{fig: J_ns12_pL_ME}
\end{figure}

\begin{figure}[h]
    \includegraphics{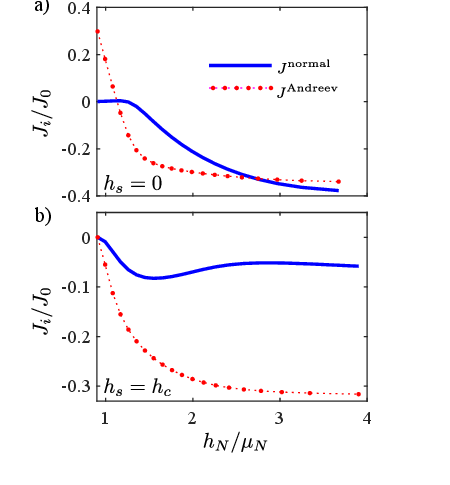}
    \caption{The normal and Andreev contributions to the net current as functions of  $h_N$, under the conditions: (a) $h_S = 0$, and (b) $h_S = h_c$. For both plots, temperature is $T = 0.05\mu_S$.}
    \label{fig: J_nNA_pL_ME}
\end{figure}

In this section, we apply our model to the case where the interface is at mechanical equilibrium and no potential barriers are applied.
We consider two conditions for the superfluid region, as in the previous section, (1) $h_S = 0$ for a spin-balanced superfluid, and (2) $h_S = h_c$ for a maximally polarized superfluid. 
We calculate the instantaneous currents as a function of normal-region chemical potential imbalance $h_N$ and point out interesting features of the results.

Figure \ref{fig: phase diagram_ME} shows the thermodynamic states considered for the normal and superfluid regions on the phase diagram in the case of mechanical equilibrium at temperature $T=0.05\mu_S$. The condition of mechanical equilibrium causes $\mu_N/\mu_S$ to depend on $h_N/\mu_S$, unlike in the previous section where $\mu_N/\mu_S$ had a fixed value. As a result, the dimensionless temperature coordinate $T/\mu_N$ of the normal region varies with $h_N/\mu_S$. The values of $T/\mu_N$ in Fig.~\ref{fig: phase diagram_ME} for an unpolarized superfluid ($h_S=0$; solid red curve) differ slightly from the case of a critically polarized superfluid ($h_S=h_c$; dotted blue curve), due to the dependence of the superfluid pressure on polarization. 

Figure \ref{fig: mu_at_interface} shows an example of the chemical potentials and current densities at large normal-region polarization, where $\mu_S>\mu_N$.

In Fig.~\ref{fig: J_ns12_pL_ME}, we show the instantaneous spin and net currents versus $h_N$ at $T = 0.05\mu_S$. As in the previous section, we observe a threshold behavior in the spin current for the unpolarized superfluid ($h_S=0$) and no significant threshold for the critically polarized superfluid ($h_S=h_c$). As before, the threshold occurs when the chemical potential difference for one of the spin states exceeds the minimum excitation energy in the superfluid, which is nearly zero for the critically polarized superfluid.
The vertical line in Fig.~\ref{fig: J_ns12_pL_ME}(a) shows the threshold for the spin current in the case of an unpolarized superfluid. The threshold is given by the point at which $\delta\mu_\downarrow=E_{min}$, which occurs at a lower polarization than $\delta\mu_\uparrow=E_{min}$.

Unlike in Fig.~\ref{fig: J_ns12_Ba}, where we considered $\mu_N=\mu_S$, here the net current does not exhibit a threshold. The absence of a threshold results from a non-zero Andreev current when $\mu_N\neq\mu_S$. Interestingly, for $h_s=0$, the sign of $\mu_N-\mu_S$ changes as $h_N$ is increased, crossing zero before the threshold, where the sign of the net current also changes.

In Fig.~\ref{fig: J_ns12_pL_ME}, the spin-up current is small compared to spin-down current at large normal-region polarization, contrary to what we found in the $\mu_S=\mu_N$ case. This is because, at large $h_N$, where $\mu_S > \mu_N$, the Andreev current flows from the superfluid into the normal region through reverse Andreev reflection. On the other hand, the normal component of the spin-up current flows in the opposite direction, because $\mu_{N\uparrow}>\mu_{S\uparrow}$. As a result, the normal and Andreev components of the spin-up current nearly cancel. Meanwhile, the normal spin-down current flows in the same direction as the Andreev current, resulting in a larger spin-down current.

\begin{figure}[h]
    \includegraphics{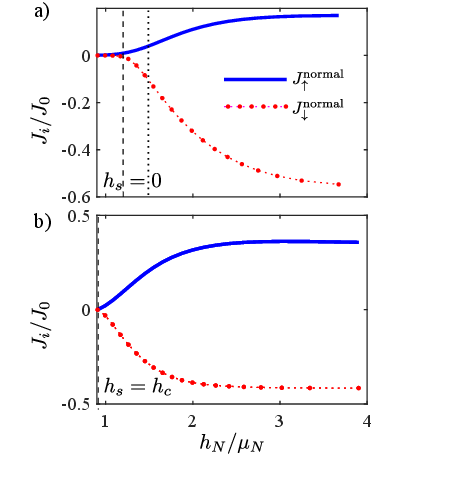}
    \caption{Normal (non-Andreev) contributions to the spin up and spin down currents on the $\alpha$ branch versus $h_N$. (a) $h_S = 0$ (b) $h_S = h_c$. For both plots, temperature is $T = 0.05\mu_S$. The dashed vertical line denotes the $h_N$ value at which $\delta\mu_{\downarrow} = E_{min}$, while the dotted vertical line indicates $\delta\mu_{\uparrow}=E_{min}$.}
    \label{fig: J_ph_ME}
\end{figure}

To confirm the interpretations described above we decompose the net current into Andreev and normal components. Figure \ref{fig: J_nNA_pL_ME} shows that the normal component exhibits a threshold in the $h_S=0$ case, while the Andreev component does not. The net current in Fig.~\ref{fig: J_ns12_pL_ME}(a), which is the sum of the normal and Andreev currents in Fig.~\ref{fig: J_nNA_pL_ME}(a), therefore does not have a threshold effect. We further decompose the $\alpha$-branch normal current into spin-up  and spin-down components in Fig.~\ref{fig: J_ph_ME}. The total contributions to these currents entering Eqns.~(\ref{eqn:jupna}) and (\ref{eqn:jdownna}) can be expressed as:
\begin{align}
&j^{\text{normal}}_{\uparrow\alpha} =\frac{1}{h}(|t_{pp_{\alpha}}^A|^2+|t_{hp_{\alpha}}^A|^2)[f(E_{\alpha}-\delta\mu_{\uparrow})-f(E_{\alpha})]\label{eqn:jnormalp}\\
&j^{\text{normal}}_{\downarrow\alpha}
=-\frac{1}{h}(|t_{hh_{\alpha}}^B|^2+|t_{ph_{\alpha}}^B|^2)[f(E_{\alpha}-\delta\mu_{\downarrow})-f(E_{\alpha})]\label{eqn:jnormalh}
\end{align}
Figure \ref{fig: J_ph_ME} and equations (\ref{eqn:jnormalp}) and (\ref{eqn:jnormalh}) confirm that the normal spin-up current is positive while the normal spin-down current is negative, as mentioned in the discussion of the relative sizes of the total spin-up and spin-down currents above.
Figure \ref{fig: J_ph_ME} also shows that the normal spin-up and spin-down currents each exhibit a threshold in the $h_S=0$ case.

Interestingly, the normal spin-up current in Fig.~\ref{fig: J_ph_ME} exhibits a threshold at a lower value of $h_N$ than expected based on the condition 
$\delta\mu_\uparrow=E_{min}$. This behavior reveals the temperature dependence of the threshold. At finite temperature, the normal current for spin $\sigma$ should become significant when $\delta\mu_\sigma \gtrsim E_{min}-k_B T$. The threshold will therefore shift to lower polarization. The size of the shift in $h_N$ depends on the sensitivity of $\delta\mu_\sigma$ to $h_N$. As shown in Fig.~\ref{fig: Dmu_hn_me} (Appendix \ref{app:Dmu_hn}), $\delta\mu_\uparrow$ has a much weaker dependence on $h_N$ than does $\delta\mu_\downarrow$.
Therefore, the threshold value of $h_N$ changes by a larger amount for spin up than for spin down. At low temperatures, a first-order Taylor expansion gives the shift of the threshold for spin $\sigma$ as $\Delta h_N^\sigma\approx -k_B T/(d \delta \mu_\sigma/dh_N)$. 
Using this formula, we confirm that the shift in the spin-up threshold should be significantly larger than the shift in the spin-down threshold. The estimated shift in the spin-down threshold ($\approx-0.06\mu_N$) is too small to observe on the scale of Fig.~\ref{fig: J_ph_ME}. The shift in the spin-up threshold ($\approx-0.3\mu_N$) coincidentally brings the spin-up threshold to about the same $h_N$ value as the spin-down threshold, in agreement with the observed behavior of the normal currents in Fig.~\ref{fig: J_ph_ME}(a).

Finally, we note that in both Fig.~\ref{fig: J_ns12_Ba}(b) and  Fig.~\ref{fig: J_ns12_pL_ME}(b), the spin current is positive when $h_S = h_c$, and, therefore, increases the polarization in the already maximally polarized superfluid region. The $z>0$ region would have to accommodate the influx of spin through phase separation, implying that the NS interface should advance to $z>0$ and the volume of the critically polarized superfluid should shrink as a function of time.

\section{Conclusions}
In conclusion, we investigated the transport of spin and mass across non-equilibrium normal-superfluid interfaces in the unitary Fermi gas. We found that, when the superfluid region is unpolarized, the spin current is strongly suppressed below a threshold value of the normal-region polarization. The threshold nearly vanishes in the limit of a critically polarized superfluid. Based on these results, we expect that, for intermediate superfluid polarization, the threshold should vary smoothly between the two limiting cases, following the variation of the minimum excitation energy of the partially polarized superfluid. 
Our results imply that non-equilibrium NS interfaces below threshold can exhibit suppressed spin transport, contributing to the metastability observed experimentally~\cite{liao2011metastability}. However, we find that Andreev reflection should allow mass current to flow even below the threshold for spin transport, except when the average chemical potentials of the normal and superfluid regions are equal.
Meanwhile, the quantitative values of the transport currents calculated here provide guidance to future experiments on NS interfaces by indicating the magnitudes of the expected currents.

An interesting question for future work will be
the long-time evolution of the NS interface. In particular, dissipation will heat the system, and finite spin conductivity will limit the rate of global equilibration. An interesting direction for future work would be to include these effects to predict the finite-time evolution of the non-equilibrium normal-superfluid mixture.
Another important challenge for future work will be to incorporate additional beyond-mean-field effects in the transport dynamics. In particular, finite quasiparticle lifetime may soften the threshold for spin transport~\cite{srikanth1992modeling,fischer2007scanning}, potentially weakening the metastability of the non-equilibrium system. 
Experimentally, future work can utilize non-equilibrium NS interfaces as a source of current to study bulk spin transport more precisely, and to explore the properties of Fermi gases under non-equilibrium conditions.

\begin{acknowledgments}	
We thank David Huse, Martin Zwierlein, Yoji Ohashi, Hiroyuki Tajima, and Henck Stoof for stimulating discussions and helpful correspondence. AS acknowledges support from the National Science Foundation (PHY-2110483).
\end{acknowledgments}

\appendix

\section{Hartree energies from thermodynamics}
\label{app:hartree}
\subsection{Polarized normal phase equation of state}
We solve for the Hartree energies in the normal phase by equating the  atomic densities in the phenomenological mean-field model to the densities given by the known equation of state at the same temperature and chemical potentials. The equation of state for the polarized normal phase is well-described by the following expression for the pressure~\cite{nascimb`ene2010exploring,mora2010normal}:
\begin{align}
    P_N = P_0(\mu_{N\uparrow})+\left(\frac{m^*}{m}\right)^{3/2}P_0\left(\mu_{N\downarrow}-A\mu_{N\uparrow}\right) \label{eqn:pressure_normal}
\end{align}
Here $P_0\,(\mu)=k_B T\lambda_{th}^{-3}F_{3/2}\left(\beta \mu\right)$ is the pressure in an ideal Fermi gas at chemical potential $\mu$, with $\lambda_{th}=\sqrt{2\pi\hbar^2/(m k_B T)}$, $F_{3/2}(x)$ the complete Fermi-Dirac integral, and $\beta=1/(k_B T)$. While we use $k_B=1$ for most of the paper, we include $k_B$ here for clarity. The polaron parameters are $A=-0.615$ and $m{^*}/m = 1.20$~\cite{nascimb`ene2010exploring, schirotzek2009observation, pilati2008phase,combescot2008normal,prokofev2008fermi-polaron,yan2019boiling}. 

We obtain the majority and minority atomic densities using $n_{\sigma} = {\partial P}/{\partial \mu_{\sigma}}$, 
\begin{align}
&n_{N\uparrow} = n_0(\mu_{N\uparrow}) -A\left(\frac{m^*}{m}\right)^{3/2}n_0\,(\mu_{N\downarrow}-A\mu_{N\uparrow})
\label{eq.nuppolaron}\\
&n_{N\downarrow} = \left(\frac{m^*}{m}\right)^{3/2}n_0\,(\mu_{N\downarrow}-A\mu_{N\uparrow}) \label{eq.ndownpolaron}
\end{align}
Where $n_0\,(\mu) =\lambda_{th}^{-3}\,\mathrm{F}_{1/2}\,({\beta}\mu)$.
Meanwhile, the phenomenological mean-field model gives the densities in terms of the Hartree energies as:
\begin{align}
&n_{N\uparrow} = n_0(\mu_{N\uparrow}+U_{N\uparrow})\label{eq.nupnmf}\\
&n_{N\downarrow} = n_0(\mu_{N\downarrow}+U_{N\downarrow}) \label{eq.ndownnmf}
\end{align}
We non-dimensionalize Eqns.~(\ref{eq.nuppolaron})-(\ref{eq.ndownnmf}), through multiplication by $\lambda_{th}^3$: 
\begin{align}
\tilde{n}_{N\sigma} = \lambda_{th}^3\,n_{N\sigma}
\end{align}
We then solve for $U_{N\uparrow}/\mu_N$ and $U_{N\downarrow}/\mu_N$ at a given $T/\mu_N$ and $h_N/\mu_N$ by equating (\ref{eq.nuppolaron}) to (\ref{eq.nupnmf}) and  (\ref{eq.ndownpolaron}) to (\ref{eq.ndownnmf}).

\subsection{Spin-balanced superfluid equation of state}
The equation of state is known accurately in the balanced case $\mu_\uparrow=\mu_\downarrow=\mu_S$~\cite{ku2012revealing}.
At low temperatures ($T<0.25\mu_S$), the balanced equation of state is well-described by the zero-temperature expression for the pressure,
\begin{align}
    P_S = \frac{2}{15\pi^2}\left(\frac{2m}{\hbar^2}\right)^{3/2}\xi^{-3/2}\mu_S^{5/2} \label{eqn:pressure_superfluid}
\end{align}
where $\xi$ is the Bertsch parameter~\cite{ku2012revealing,zurn2013precise}.
The total density $n_S$ and dimensionless density $\tilde{n}_S$ are then:
\begin{align}
    &n_S = \frac{\partial P_S}{\partial \mu_S} =  \frac{1}{3\pi^2}\left(\frac{2m}{\hbar^2}\right)^{3/2}\left(\frac{\mu_S}{\xi}\right)^{3/2} \\
    &\tilde{n}_S(\beta\mu_S) = \frac{8}{3\sqrt{\pi}}\left(\frac{\beta\mu_S}{\xi}\right)^{3/2} \label{eqn:superfluid_density}
\end{align}

The total density for the balanced superfluid in the phenomenological mean-field model is:
\begin{align}
n_S^{MF} = 2\int\frac{\mathrm{d}q\,q^2}{4\pi^2}\,&\left\{\left(1+\frac{\xi_s}{E_s}\right)\,f(E_s)\right.\nonumber \\
+&\left.\left(1-\frac{\xi_s}{E_s}\right)\,\left[1-f(E_s)\right]\right\}
\end{align}
where $f(E)= 1/(1+e\,^{\beta E})$. Equating (\ref{eqn:superfluid_density}) to the non-dimensionalized mean-field density $\tilde{n}_S^{MF}=\lambda_{th}^3 n_S^{MF}$ then gives $U_S/\mu_S$ for each $T/\mu_S$.

\subsection{Critically polarized superfluid equation of state}
We obtain an equation of state for the critically polarized superfluid below the tricritical point by exploiting the fact that it is at thermodynamic equilibrium with the critically polarized normal fluid.
To model the phase diagram, we take as input the temperature $T_{c}/T_{F\uparrow}$ at the tricritical point, and the normal and superfluid critical polarizations, $p_{Sc}$ and $p_{Nc}$, at the tricritical temperature and at zero temperature from Ref.~\cite{gubbels2008renormalization,gubbels2013imbalanced}. As in Ref.~\cite{shin2008phase}, we linearly approximate $p_{Sc}$ and $p_{Nc}$ as functions of $T/T_{F\uparrow}$. The resulting model phase diagram is shown in Fig.~\ref{fig:pphasediagram}.

We proceed in two stages to obtain the Hartree energies of the critically polarized superfluid. First, we convert the boundary of the normal phase from the variables $(p_{Nc}, T/T_{F\uparrow})$ in the polarization-temperature plane to the variables $(h_c/\mu, T/\mu)$ in the chemical potential difference-temperature plane using the normal phase equation of state. Note that, along the phase boundary, $\mu_N=\mu_S\equiv\mu$ and $h_N=h_S\equiv h_c$. Second, for each value of $(h_c/\mu, T/\mu)$ along the phase boundary, we solve for the non-dimensionalized Hartree energies $U_S/\mu_S$ and $U_h/\mu_S$ that give the correct value of $(p_{Sc}, T/T_{F\uparrow})$ in the critically polarized superfluid. For this last step, we take advantage of the observation that the density of majority-spin atoms is continuous across the phase boundary~\cite{shin2008phase}.

In the first stage, we employ the system of equations:
\begin{align}
&p_{Nc} = \frac{\tilde{n}_{N\uparrow}(\beta\mu_{\uparrow},\beta\mu_{\downarrow})-\tilde{n}_{N\downarrow}(\beta\mu_{\uparrow},\beta\mu_{\downarrow})}{\tilde{n}_{N\uparrow}(\beta\mu_{\uparrow},\beta\mu_{\downarrow})+\tilde{n}_{N\downarrow}(\beta\mu_{\uparrow},\beta\mu_{\downarrow})} \\
&\frac{T}{T_{F\uparrow}} = \frac{4\pi}{\left(6\pi^2\tilde{n}_{N\uparrow}(\beta\mu_{\uparrow},\beta\mu_{\downarrow})\right)^{2/3}}
\end{align}
Here the left-hand sides are known from the model phase diagram and the right-hand sides from the normal-phase equation of state (\ref{eq.nuppolaron}) and (\ref{eq.ndownpolaron}).
We solve for $\beta\mu_{\uparrow}$ and $\beta\mu_{\downarrow}$, which gives $h_c/\mu = (\beta\mu_{\uparrow}-\beta\mu_{\downarrow})/(\beta\mu_{\uparrow}+\beta\mu_{\downarrow})$, and $T/\mu = 2/(\beta\mu_{\uparrow}+\beta\mu_{\downarrow})$. 

In the second stage, at a given value of $(h_c/\mu,T/\mu)$, we solve for $U_S/\mu$ and $U_h/\mu$ using the system of equations:
\begin{align}
&p_{Sc} = \frac{\tilde{n}^{MF}_{S\uparrow}(U_S/\mu_S,U_h/\mu_S)-\tilde{n}^{MF}_{S\downarrow}(U_S/\mu_S,U_h/\mu_S)}{\tilde{n}^{MF}_{S\uparrow}(U_S/\mu_S,U_h/\mu_S)+\tilde{n}^{MF}_{S\downarrow}(U_S/\mu_S,U_h/\mu_S)} \\
&\frac{T}{T_{F\uparrow}} = \frac{4\pi}{\left(6\pi^2\tilde{n}^{MF}_{S\uparrow}\left(U_S/\mu_S,U_h/\mu_S\right)\right)^{2/3}}
\end{align}
The left-hand sides are again known from the phase diagram. The right-hand sides contain
the densities from the phenomenological mean-field model, which depend on the Hartree energies:
\begin{align}
n_{S\uparrow(\downarrow)}^{MF} = \int\frac{\mathrm{d}q\,q^2}{4\pi^2}\,&\left\{\left(1+\frac{\xi_s}{E_s}\right)\,f(E_{\alpha(\beta)})\right.\nonumber \\
+&\left.\left(1-\frac{\xi_s}{E_s}\right)\,\left[1-f(E_{\beta(\alpha)})\right]\right\}
\end{align}
We note that (\ref{eqn:pressure_superfluid}) has been proposed to also apply at zero temperature in the presence of imbalanced chemical potentials~\cite{chevy2006universal}. However, it does not account for the non-zero polarization of the superfluid at finite temperatures. By contrast, the procedure described above does account for finite polarization.

\section{Alpha branch dispersion relationships}\label{app:dispersion}
\begin{figure}[h]
    \includegraphics{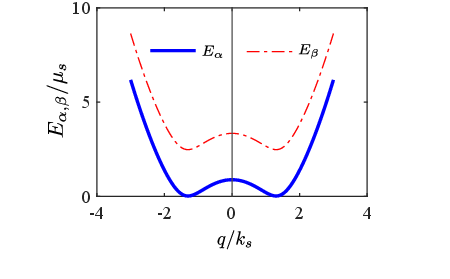}
    \caption{Dispersion curve of superfluid excitation energy for both branches at $T=0.05\mu_S$ and maximal superfluid polarization $h_S=h_c$. The solid line represents $E_{\alpha}$, and the dashed line represents $E_{\beta}$. The energy and the wavevector are normalized by $\mu_S$ and $k_s$, respectively, with $k_s=2m\mu_S/\hbar^2$ and $\mu_S$ the average chemical potential (\ref{eqn:muandh}).}
    \label{fig: Dispersion_E_k}
\end{figure}
Figure \ref{fig: Dispersion_E_k} shows the superfluid dispersion relations for $E_{\alpha(\beta)}$ versus $q_{\alpha(\beta)}$, normalized by $\mu_S$ and $k_s$ respectively.
At a given energy $E_{\alpha}$ there are up to two solutions for the magnitude of the wavevector of an $\alpha$ branch excitation, obtained from inverting Eqn. (\ref{eq.Ealpha}). Likewise, the wavevectors for the $\beta$ branch are obtained from inverting Eqn. (\ref{eq.Ebeta}). 

In the normal phase, the wavevector solutions in the $\alpha$ branch are:
\begin{align}
&k_{p\uparrow} = \sqrt{\frac{2m}{\hbar^2}\left(\mu_{S\uparrow}-U_{N\uparrow}+E_{\alpha}-\xi_{\perp}\right)}\\
&k_{h\downarrow} = \sqrt{\frac{2m^*}{\hbar^2}\left(\mu_{S\downarrow }-U_{N\downarrow}-E_{\alpha}\right)-\frac{2m}{\hbar^2}\xi_{\perp}}\label{eqn:khdown}
\end{align}
In the superfluid phase, the wavevectors are:
\begin{align}
&q_{p\alpha} = \sqrt{\frac{2m}{\hbar^2}\left(\mu_S-U_S+\sqrt{\left(E_{\alpha}-U_h+h_S\right)^2-\Delta^2}-\xi_{\perp}\right)} \\
&q_{h\alpha} = \sqrt{\frac{2m}{\hbar^2}\left(\mu_S-U_S-\sqrt{\left(E_{\alpha}-U_h+h_S\right)^2-\Delta^2}-\xi_{\perp}\right)}\label{eqn:qha}
\end{align}
For sufficiently large $E_\alpha$, the quantities inside the square roots of Eqns. (\ref{eqn:khdown}) and (\ref{eqn:qha}) become negative, causing the hole wavevectors to become imaginary and give a vanishing current.

\section{Non-equilibrium distribution functions}\label{app:fermifn}
Here we give the quasiparticle distribution functions $f_{n\alpha}$ and $f_{n\beta}$ for each channel. Using $f(E)= 1/(1+e^{\beta E})$,  $\delta\mu_{\uparrow} = \mu_{N\uparrow}-\mu_{S\uparrow}$, and $\delta\mu_{\downarrow}=\mu_{S\downarrow}-\mu_{N\downarrow}$, the $\alpha$ branch occupation numbers are:
\begin{align}
f_{Lp\alpha}(E_\alpha) &= f(E_{\alpha}-\delta\mu_{\uparrow})\label{eqn:flpa}\\
f_{Lh\alpha}(E_\alpha) &= f(E_{\alpha}-\delta\mu_{\downarrow})\label{eqn:flha}\\
f_{Rp\alpha}(E_\alpha) &= f_{Rh\alpha}(E_\alpha)=f(E_\alpha)
\end{align}
The subtraction of $\delta\mu_\sigma$ in Eqns. (\ref{eqn:flpa}) and (\ref{eqn:flha}) results from defining $E_\alpha$ relative to the superfluid chemical potentials $\mu_{S\sigma}$ for the purpose of the scattering calculation~\cite{blonder1982transition}.
For the $\beta$ branch:
\begin{align}
f_{Lp\beta}(E_\beta) &= f(E_{\beta}+\delta\mu_{\downarrow})\\
f_{Lh\beta}(E_\beta) &= f(E_{\beta}+\delta\mu_{\uparrow})\\
f_{Rp\beta}(E_\beta) &= f_{Rh\beta}(E_\beta)=f(E_\beta)
\end{align}

\section{Scattering regimes}\label{app:regimes}
\begin{center}
\begin{table*}[hbt]
\centering
\begin{tabular}{|c|c|c|}
\hline
Excitation (wavevector) & Accessible $E_{\alpha}$&Accessible $\xi_{\perp} $ (given $E_\alpha$) \\
\hline
Particle ($k_{p\uparrow}$) &$[0,\infty)$&$[0,\mu_{S\uparrow}-U_{N\uparrow}+E_{\alpha}]$\\
\hline
Hole ($k_{h\downarrow}$) &$[0,\mu_{S\downarrow}-U_{N\downarrow}]$&$[0,\frac{m^*}{m}(\mu_{S\downarrow}-U_{N\downarrow}-E_{\alpha})]$\\
\hline
Quasiparticle ($q_{p\alpha}$) &$[U_h-h_S+\Delta,\infty)$&$[0,\mu_S-U_S+\sqrt{(E_{\alpha}+h_S-U_h)^2-\Delta^2}]$\\
\hline
Quasihole ($q_{h\alpha}$) &$[U_h-h_S+\Delta, U_h-h_S+\sqrt{(U_S-\mu_S)^2+\Delta^2}]$&$[0, \mu_S-U_S-\sqrt{(E_{\alpha}+h_S-U_h)^2-\Delta^2}]$\\
\hline
\multicolumn{3}{|c|}{Equation (\ref{eq.Ealpha}) requires $E_{\alpha}\geq U_h-h_S $}\\
\hline
\end{tabular}
\caption{Conditions on $E_{\alpha}$ and $\xi_{\perp}$ that determine the scattering regimes for the $\alpha$ branch.}
\label{Tab:1}
\end{table*}
\end{center}
\begin{figure}[ht]
    \includegraphics{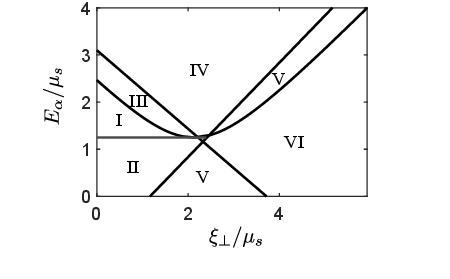}
    \caption{Alpha branch scattering regimes on the excitation energy $E_{\alpha}$ vs transverse kinetic energy $\xi_{\perp}$ plane. The case shown has $\mu_N = \mu_S$ and $p_N=0.8$. The other parameters, normalized by $\mu_S$, are $T=0.05$, $h_S = 0$, $U_S= 1.13$, $U_h = 0$, $U_{N\uparrow} = 0.16$, $U_{N\downarrow} = 2.10$.}
    \label{fig: scattering_regime}
\end{figure}
For a given excitation energy $E_\alpha$ and transverse momentum $k_\perp$, each of the four types of $\alpha$-branch excitations can be either allowed (real wavevector) or forbidden (imaginary wavevector), leading to different scattering regimes.
Table \ref{Tab:1} lists the conditions on each type of excitation. An example of the scattering regimes is shown in Fig.~\ref{fig: scattering_regime} on the plane of excitation energy vs $\xi_{\perp}\equiv \hbar^2 k_\perp^2/(2m)$. In Regime I, all four excitation types are allowed. Regime II supports only normal particle and hole modes and therefore only allows transmission by Andreev reflection. Regime III allows the particle, quasiparticle and quasihole modes, and prohibits any transmission requiring the hole mode. Regime IV allows only the particle and quasiparticle modes and supports only the transmission between a particle and a quasiparticle. Regime V allows only the particle mode and, therefore, causes total reflection. Regime VI is the energetically forbidden regime, where the transverse kinetic energy exceeds the total kinetic energy. Since the Andreev current is important for the net current contribution, we present the formula for the Regime II: 
\begin{align}
j_{II,\alpha}^{Net} = \frac{2}{h}\,|r_{hp_{\alpha}}^A|^2\,[f(E_{\alpha}-\delta\mu_{\uparrow})-f(E_{\alpha}-\delta\mu_{\downarrow})]
\end{align}
The prefactor of 2 is typical for Andreev current and indicates the transport of two fermions per scattering event.

\section{Scattering coefficients}\label{app:coef}
The scattering coefficients necessary for the determination of the currents are:
\begin{align}
r_{pp_{\alpha}}^A =& \frac{1}{c_0}\left[u_0^2\left(k_{p\uparrow}-q_{p\alpha}-ik_{\Lambda}\right)\left(\frac{m}{m^*}k_{h\downarrow}+q_{h\alpha}-ik_{\Lambda}\right)\nonumber \right.
\\&+\left. v_0^2\left(q_{p\alpha}-\frac{m}{m^*}k_{h\downarrow}+ik_{\Lambda}\right)\left(k_{p\uparrow}+q_{h\alpha}-ik_{\Lambda}\right)\right]
\end{align}

\begin{align}
r_{hp_{\alpha}}^A = &\frac{1}{c_0}2u_0v_0\sqrt{\frac{m}{m^*}k_{h\downarrow}k_{p\uparrow}}\,\left(q_{h\alpha}+q_{p\alpha}\right)e^{-iX_0}
\end{align}

\begin{align}
r_{hh_{\alpha}}^B =& \frac{1}{c_0}\left[u_0^2\left(\frac{m}{m^*}k_{h\downarrow}-q_{h\alpha}+ik_{\Lambda}\right)\left(q_{p\alpha}+k_{p\uparrow}+ik_{\Lambda}\right)\right.\nonumber\\&\left.+v_0^2\left(q_{h\alpha}-k_{p\uparrow}-ik_{\Lambda}\right)\left(q_{p\alpha}+\frac{m}{m^*}k_{h\downarrow}+ik_{\Lambda}\right)\right]
\end{align}
where
\begin{align}
c_0 =&u_0^2 \left(k_{p\uparrow}+q_{p\alpha}+ik_{\Lambda}\right)\left(\frac{m}{m^*}k_{h\downarrow}+q_{h\alpha}-ik_{\Lambda}\right)\nonumber\\&+v_0^2\left(q_e-\frac{m}{m^*}k_{h\downarrow}+ik_{\Lambda}\right)\left(k_{p\uparrow}-q_{h\alpha}+ik_{\Lambda}\right) 
\end{align}
and $X_0$ denotes the phase of the gap $\Delta=|\Delta|e^{iX_0}$. We set $X_0=0$ without loss of generality. The transmission coefficients for channel A are: 
\begin{align}
&t_{pp_{\alpha}}^A = \frac{1}{c_0}2u_0\sqrt{q_{p\alpha}k_{p\uparrow}}\,\left(q_{h\alpha}+\frac{m}{m^*}k_{h\downarrow}-ik_{\Lambda}\right)\sqrt{\frac{\xi_s}{E_s}}e^{-iX_0/2} \\
&t_{hp_{\alpha}}^A = \frac{1}{c_0}2v_0\sqrt{k_{p\uparrow}q_{h\alpha}}\,\left(q_{p\alpha}-\frac{m}{m^*}k_{h\downarrow}+ik_{\Lambda}\right)\sqrt{\frac{\xi_s}{E_s}}e^{-iX_0/2}
\end{align}
And for channel B:
\begin{align}
&t_{ph_{\alpha}}^B = \frac{1}{c_0}2v_0\sqrt{\frac{m}{m^*}k_{h\downarrow}q_{e\alpha}}\,\left(q_{h\alpha}-k_{e\uparrow}-ik_{\Lambda}\right)\sqrt{\frac{\xi_s}{E_s}}e^{iX_0/2} \\
&t_{hh_{\alpha}}^B = \frac{1}{c_0}2u_0\sqrt{\frac{m}{m^*}k_{h\downarrow}q_{h\alpha}}\,\left(q_{e\alpha}+k_{e\uparrow}+ik_{\Lambda}\right)\sqrt{\frac{\xi_s}{E_s}}e^{iX_0/2}
\end{align}
With the 7 coefficients given above, the other 9 coefficients for the $\alpha$ branch can be inferred from the symmetries of $S$-matrix. 

\section{Additional Plots}
\label{app:Tspin}\label{app:Dmu_hn}
\begin{figure}[b]
    \centering
    \includegraphics{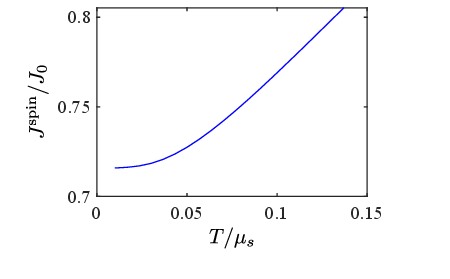}
    \caption{Spin current $J^{\text{spin}}/J_0$ plotted vs temperature $T/\mu_S\,$ under mechanical equilibrium with $p_N = 99\%$, $h_S = 0$.}
    \label{fig: J_s_T_ME}
\end{figure}

We show an example of the dependence of the spin current on temperature in Fig.~\ref{fig: J_s_T_ME}. 

\begin{figure}
    \centering
    \includegraphics{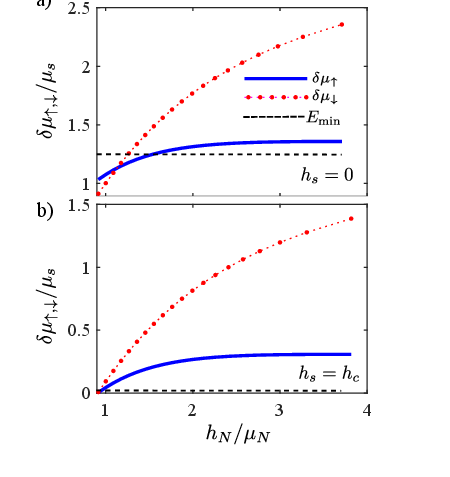}
    \caption{Normalized chemical potential differences $\delta\mu_{\uparrow(\downarrow)}$ plotted vs $h_N$ under mechanical equilibrium, at $T = 0.05\mu_S$, for (a) $h_s=0$ and (b) $h_s=h_c$. The dashed line shows the minimum of the superfluid excitation spectrum at the given value of $h_s$.}
    \label{fig: Dmu_hn_me}
\end{figure}
In Fig.~\ref{fig: Dmu_hn_me}, we show the chemical potential differences between the normal and superfluid regions versus the normal-region Zeeman field $h_N$ under mechanical equilibrium.

\label{app:pn_hn}
\begin{figure}
    \centering
    \includegraphics{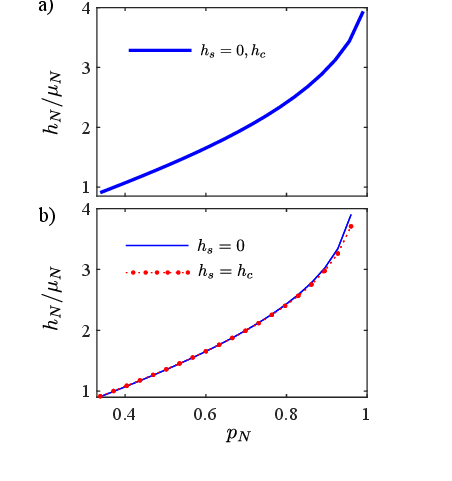}
    \caption{Zeeman field $h_N$ versus normal region polarization $p_N$ under the conditions (a) $\mu_N=\mu_S$, and (b) mechanical equilibrium. In case (a), $h_N/\mu_S$ is independent of $h_S$ because $\mu_N/\mu_S$ is constant.}
    \label{fig: p_muhN}
\end{figure}
Figure \ref{fig: p_muhN} shows the conversion between polarization and Zeeman field for the normal phase. In Fig.~\ref{fig: p_muhN}(b), the superfluid polarization leads to a slightly different conversion because the normal phase reacts to the increase in superfluid pressure at higher polarization.


%\bibliography{Ding}

\begin{thebibliography}{101}%
\makeatletter
\providecommand \@ifxundefined [1]{%
 \@ifx{#1\undefined}
}%
\providecommand \@ifnum [1]{%
 \ifnum #1\expandafter \@firstoftwo
 \else \expandafter \@secondoftwo
 \fi
}%
\providecommand \@ifx [1]{%
 \ifx #1\expandafter \@firstoftwo
 \else \expandafter \@secondoftwo
 \fi
}%
\providecommand \natexlab [1]{#1}%
\providecommand \enquote  [1]{``#1''}%
\providecommand \bibnamefont  [1]{#1}%
\providecommand \bibfnamefont [1]{#1}%
\providecommand \citenamefont [1]{#1}%
\providecommand \href@noop [0]{\@secondoftwo}%
\providecommand \href [0]{\begingroup \@sanitize@url \@href}%
\providecommand \@href[1]{\@@startlink{#1}\@@href}%
\providecommand \@@href[1]{\endgroup#1\@@endlink}%
\providecommand \@sanitize@url [0]{\catcode `\\12\catcode `\$12\catcode
  `\&12\catcode `\#12\catcode `\^12\catcode `\_12\catcode `\%12\relax}%
\providecommand \@@startlink[1]{}%
\providecommand \@@endlink[0]{}%
\providecommand \url  [0]{\begingroup\@sanitize@url \@url }%
\providecommand \@url [1]{\endgroup\@href {#1}{\urlprefix }}%
\providecommand \urlprefix  [0]{URL }%
\providecommand \Eprint [0]{\href }%
\providecommand \doibase [0]{http://dx.doi.org/}%
\providecommand \selectlanguage [0]{\@gobble}%
\providecommand \bibinfo  [0]{\@secondoftwo}%
\providecommand \bibfield  [0]{\@secondoftwo}%
\providecommand \translation [1]{[#1]}%
\providecommand \BibitemOpen [0]{}%
\providecommand \bibitemStop [0]{}%
\providecommand \bibitemNoStop [0]{.\EOS\space}%
\providecommand \EOS [0]{\spacefactor3000\relax}%
\providecommand \BibitemShut  [1]{\csname bibitem#1\endcsname}%
\let\auto@bib@innerbib\@empty
%</preamble>
\bibitem [{\citenamefont {Shin}\ \emph {et~al.}(2008)\citenamefont {Shin},
  \citenamefont {Schunck}, \citenamefont {Schirotzek},\ and\ \citenamefont
  {Ketterle}}]{shin2008phase}%
  \BibitemOpen
  \bibfield  {author} {\bibinfo {author} {\bibfnamefont {Yong-il}\ \bibnamefont
  {Shin}}, \bibinfo {author} {\bibfnamefont {Christian~H.}\ \bibnamefont
  {Schunck}}, \bibinfo {author} {\bibfnamefont {Andr{\'e}}\ \bibnamefont
  {Schirotzek}}, \ and\ \bibinfo {author} {\bibfnamefont {Wolfgang}\
  \bibnamefont {Ketterle}},\ }\bibfield  {title} {\enquote {\bibinfo {title}
  {Phase diagram of a two-component {Fermi} gas with resonant interactions},}\
  }\href {\doibase 10.1038/nature06473} {\bibfield  {journal} {\bibinfo
  {journal} {Nature}\ }\textbf {\bibinfo {volume} {451}},\ \bibinfo {pages}
  {689} (\bibinfo {year} {2008})}\BibitemShut {NoStop}%
\bibitem [{\citenamefont {Navon}\ \emph {et~al.}(2010)\citenamefont {Navon},
  \citenamefont {Nascimb{\`e}ne}, \citenamefont {Chevy},\ and\ \citenamefont
  {Salomon}}]{navon2010equation}%
  \BibitemOpen
  \bibfield  {author} {\bibinfo {author} {\bibfnamefont {N.}~\bibnamefont
  {Navon}}, \bibinfo {author} {\bibfnamefont {S.}~\bibnamefont
  {Nascimb{\`e}ne}}, \bibinfo {author} {\bibfnamefont {F.}~\bibnamefont
  {Chevy}}, \ and\ \bibinfo {author} {\bibfnamefont {C.}~\bibnamefont
  {Salomon}},\ }\bibfield  {title} {\enquote {\bibinfo {title} {The {Equation}
  of {State} of a {Low}-{Temperature} {Fermi} {Gas} with {Tunable}
  {Interactions}},}\ }\href {\doibase 10.1126/science.1187582} {\bibfield
  {journal} {\bibinfo  {journal} {Science}\ }\textbf {\bibinfo {volume}
  {328}},\ \bibinfo {pages} {729--732} (\bibinfo {year} {2010})}\BibitemShut
  {NoStop}%
\bibitem [{\citenamefont {Nascimb{\`e}ne}\ \emph {et~al.}(2010)\citenamefont
  {Nascimb{\`e}ne}, \citenamefont {Navon}, \citenamefont {Jiang}, \citenamefont
  {Chevy},\ and\ \citenamefont {Salomon}}]{nascimb`ene2010exploring}%
  \BibitemOpen
  \bibfield  {author} {\bibinfo {author} {\bibfnamefont {S.}~\bibnamefont
  {Nascimb{\`e}ne}}, \bibinfo {author} {\bibfnamefont {N.}~\bibnamefont
  {Navon}}, \bibinfo {author} {\bibfnamefont {K.~J.}\ \bibnamefont {Jiang}},
  \bibinfo {author} {\bibfnamefont {F.}~\bibnamefont {Chevy}}, \ and\ \bibinfo
  {author} {\bibfnamefont {C.}~\bibnamefont {Salomon}},\ }\bibfield  {title}
  {\enquote {\bibinfo {title} {Exploring the thermodynamics of a universal
  {Fermi} gas},}\ }\href {\doibase 10.1038/nature08814} {\bibfield  {journal}
  {\bibinfo  {journal} {Nature}\ }\textbf {\bibinfo {volume} {463}},\ \bibinfo
  {pages} {1057} (\bibinfo {year} {2010})}\BibitemShut {NoStop}%
\bibitem [{\citenamefont {Nascimb\`ene}\ \emph {et~al.}(2011)\citenamefont
  {Nascimb\`ene}, \citenamefont {Navon}, \citenamefont {Pilati}, \citenamefont
  {Chevy}, \citenamefont {Giorgini}, \citenamefont {Georges},\ and\
  \citenamefont {Salomon}}]{nascimb`ene2011fermi-liquid}%
  \BibitemOpen
  \bibfield  {author} {\bibinfo {author} {\bibfnamefont {S.}~\bibnamefont
  {Nascimb\`ene}}, \bibinfo {author} {\bibfnamefont {N.}~\bibnamefont {Navon}},
  \bibinfo {author} {\bibfnamefont {S.}~\bibnamefont {Pilati}}, \bibinfo
  {author} {\bibfnamefont {F.}~\bibnamefont {Chevy}}, \bibinfo {author}
  {\bibfnamefont {S.}~\bibnamefont {Giorgini}}, \bibinfo {author}
  {\bibfnamefont {A.}~\bibnamefont {Georges}}, \ and\ \bibinfo {author}
  {\bibfnamefont {C.}~\bibnamefont {Salomon}},\ }\bibfield  {title} {\enquote
  {\bibinfo {title} {Fermi-{Liquid} {Behavior} of the {Normal} {Phase} of a
  {Strongly} {Interacting} {Gas} of {Cold} {Atoms}},}\ }\href {\doibase
  10.1103/PhysRevLett.106.215303} {\bibfield  {journal} {\bibinfo  {journal}
  {Phys. Rev. Lett.}\ }\textbf {\bibinfo {volume} {106}},\ \bibinfo {pages}
  {215303} (\bibinfo {year} {2011})}\BibitemShut {NoStop}%
\bibitem [{\citenamefont {Ku}\ \emph {et~al.}(2012)\citenamefont {Ku},
  \citenamefont {Sommer}, \citenamefont {Cheuk},\ and\ \citenamefont
  {Zwierlein}}]{ku2012revealing}%
  \BibitemOpen
  \bibfield  {author} {\bibinfo {author} {\bibfnamefont {M.~J.~H.}\
  \bibnamefont {Ku}}, \bibinfo {author} {\bibfnamefont {A.~T.}\ \bibnamefont
  {Sommer}}, \bibinfo {author} {\bibfnamefont {L.~W.}\ \bibnamefont {Cheuk}}, \
  and\ \bibinfo {author} {\bibfnamefont {M.~W.}\ \bibnamefont {Zwierlein}},\
  }\bibfield  {title} {\enquote {\bibinfo {title} {Revealing the {Superfluid}
  {Lambda} {Transition} in the {Universal} {Thermodynamics} of a {Unitary}
  {Fermi} {Gas}},}\ }\href {\doibase 10.1126/science.1214987} {\bibfield
  {journal} {\bibinfo  {journal} {Science}\ }\textbf {\bibinfo {volume}
  {335}},\ \bibinfo {pages} {563--567} (\bibinfo {year} {2012})}\BibitemShut
  {NoStop}%
\bibitem [{\citenamefont {Van~Houcke}\ \emph {et~al.}(2012)\citenamefont
  {Van~Houcke}, \citenamefont {Werner}, \citenamefont {Kozik}, \citenamefont
  {Prokof'ev}, \citenamefont {Svistunov}, \citenamefont {Ku}, \citenamefont
  {Sommer}, \citenamefont {Cheuk}, \citenamefont {Schirotzek},\ and\
  \citenamefont {Zwierlein}}]{van_houcke2012feynman}%
  \BibitemOpen
  \bibfield  {author} {\bibinfo {author} {\bibfnamefont {K.}~\bibnamefont
  {Van~Houcke}}, \bibinfo {author} {\bibfnamefont {F.}~\bibnamefont {Werner}},
  \bibinfo {author} {\bibfnamefont {E.}~\bibnamefont {Kozik}}, \bibinfo
  {author} {\bibfnamefont {N.}~\bibnamefont {Prokof'ev}}, \bibinfo {author}
  {\bibfnamefont {B.}~\bibnamefont {Svistunov}}, \bibinfo {author}
  {\bibfnamefont {M.~J.~H.}\ \bibnamefont {Ku}}, \bibinfo {author}
  {\bibfnamefont {A.~T.}\ \bibnamefont {Sommer}}, \bibinfo {author}
  {\bibfnamefont {L.~W.}\ \bibnamefont {Cheuk}}, \bibinfo {author}
  {\bibfnamefont {A.}~\bibnamefont {Schirotzek}}, \ and\ \bibinfo {author}
  {\bibfnamefont {M.~W.}\ \bibnamefont {Zwierlein}},\ }\bibfield  {title}
  {\enquote {\bibinfo {title} {Feynman diagrams versus {Fermi}-gas {Feynman}
  emulator},}\ }\href {\doibase 10.1038/nphys2273} {\bibfield  {journal}
  {\bibinfo  {journal} {Nature Physics}\ }\textbf {\bibinfo {volume} {8}},\
  \bibinfo {pages} {366--370} (\bibinfo {year} {2012})}\BibitemShut {NoStop}%
\bibitem [{\citenamefont {Schirotzek}\ \emph {et~al.}(2008)\citenamefont
  {Schirotzek}, \citenamefont {Shin}, \citenamefont {Schunck},\ and\
  \citenamefont {Ketterle}}]{schirotzek2008determination}%
  \BibitemOpen
  \bibfield  {author} {\bibinfo {author} {\bibfnamefont {Andr{\'e}}\
  \bibnamefont {Schirotzek}}, \bibinfo {author} {\bibfnamefont {Yong-il}\
  \bibnamefont {Shin}}, \bibinfo {author} {\bibfnamefont {Christian~H.}\
  \bibnamefont {Schunck}}, \ and\ \bibinfo {author} {\bibfnamefont {Wolfgang}\
  \bibnamefont {Ketterle}},\ }\bibfield  {title} {\enquote {\bibinfo {title}
  {Determination of the {Superfluid} {Gap} in {Atomic} {Fermi} {Gases} by
  {Quasiparticle} {Spectroscopy}},}\ }\href {\doibase
  10.1103/PhysRevLett.101.140403} {\bibfield  {journal} {\bibinfo  {journal}
  {Phys. Rev. Lett.}\ }\textbf {\bibinfo {volume} {101}},\ \bibinfo {pages}
  {140403} (\bibinfo {year} {2008})}\BibitemShut {NoStop}%
\bibitem [{\citenamefont {Gaebler}\ \emph {et~al.}(2010)\citenamefont
  {Gaebler}, \citenamefont {Stewart}, \citenamefont {Drake}, \citenamefont
  {Jin}, \citenamefont {Perali}, \citenamefont {Pieri},\ and\ \citenamefont
  {Strinati}}]{gaebler2010observation}%
  \BibitemOpen
  \bibfield  {author} {\bibinfo {author} {\bibfnamefont {J.~P.}\ \bibnamefont
  {Gaebler}}, \bibinfo {author} {\bibfnamefont {J.~T.}\ \bibnamefont
  {Stewart}}, \bibinfo {author} {\bibfnamefont {T.~E.}\ \bibnamefont {Drake}},
  \bibinfo {author} {\bibfnamefont {D.~S.}\ \bibnamefont {Jin}}, \bibinfo
  {author} {\bibfnamefont {A.}~\bibnamefont {Perali}}, \bibinfo {author}
  {\bibfnamefont {P.}~\bibnamefont {Pieri}}, \ and\ \bibinfo {author}
  {\bibfnamefont {G.~C.}\ \bibnamefont {Strinati}},\ }\bibfield  {title}
  {\enquote {\bibinfo {title} {Observation of pseudogap behaviour in a strongly
  interacting {Fermi} gas},}\ }\href {\doibase 10.1038/nphys1709} {\bibfield
  {journal} {\bibinfo  {journal} {Nature Physics}\ }\textbf {\bibinfo {volume}
  {6}},\ \bibinfo {pages} {569--573} (\bibinfo {year} {2010})}\BibitemShut
  {NoStop}%
\bibitem [{\citenamefont {Hoinka}\ \emph {et~al.}(2017)\citenamefont {Hoinka},
  \citenamefont {Dyke}, \citenamefont {Lingham}, \citenamefont {Kinnunen},
  \citenamefont {Bruun},\ and\ \citenamefont {Vale}}]{hoinka2017goldstone}%
  \BibitemOpen
  \bibfield  {author} {\bibinfo {author} {\bibfnamefont {Sascha}\ \bibnamefont
  {Hoinka}}, \bibinfo {author} {\bibfnamefont {Paul}\ \bibnamefont {Dyke}},
  \bibinfo {author} {\bibfnamefont {Marcus~G.}\ \bibnamefont {Lingham}},
  \bibinfo {author} {\bibfnamefont {Jami~J.}\ \bibnamefont {Kinnunen}},
  \bibinfo {author} {\bibfnamefont {Georg~M.}\ \bibnamefont {Bruun}}, \ and\
  \bibinfo {author} {\bibfnamefont {Chris~J.}\ \bibnamefont {Vale}},\
  }\bibfield  {title} {\enquote {\bibinfo {title} {Goldstone mode and
  pair-breaking excitations in atomic {Fermi} superfluids},}\ }\href {\doibase
  10.1038/nphys4187} {\bibfield  {journal} {\bibinfo  {journal} {Nature
  Physics}\ }\textbf {\bibinfo {volume} {13}},\ \bibinfo {pages} {943--946}
  (\bibinfo {year} {2017})}\BibitemShut {NoStop}%
\bibitem [{\citenamefont {Sommer}\ \emph
  {et~al.}(2011{\natexlab{a}})\citenamefont {Sommer}, \citenamefont {Ku},
  \citenamefont {Roati},\ and\ \citenamefont
  {Zwierlein}}]{sommer2011universal}%
  \BibitemOpen
  \bibfield  {author} {\bibinfo {author} {\bibfnamefont {Ariel}\ \bibnamefont
  {Sommer}}, \bibinfo {author} {\bibfnamefont {Mark}\ \bibnamefont {Ku}},
  \bibinfo {author} {\bibfnamefont {Giacomo}\ \bibnamefont {Roati}}, \ and\
  \bibinfo {author} {\bibfnamefont {Martin~W.}\ \bibnamefont {Zwierlein}},\
  }\bibfield  {title} {\enquote {\bibinfo {title} {Universal spin transport in
  a strongly interacting {Fermi} gas},}\ }\href {\doibase 10.1038/nature09989}
  {\bibfield  {journal} {\bibinfo  {journal} {Nature}\ }\textbf {\bibinfo
  {volume} {472}},\ \bibinfo {pages} {201--204} (\bibinfo {year}
  {2011}{\natexlab{a}})}\BibitemShut {NoStop}%
\bibitem [{\citenamefont {Sommer}\ \emph
  {et~al.}(2011{\natexlab{b}})\citenamefont {Sommer}, \citenamefont {Ku},\ and\
  \citenamefont {Zwierlein}}]{sommer2011spin}%
  \BibitemOpen
  \bibfield  {author} {\bibinfo {author} {\bibfnamefont {Ariel}\ \bibnamefont
  {Sommer}}, \bibinfo {author} {\bibfnamefont {Mark}\ \bibnamefont {Ku}}, \
  and\ \bibinfo {author} {\bibfnamefont {Martin~W}\ \bibnamefont {Zwierlein}},\
  }\bibfield  {title} {\enquote {\bibinfo {title} {Spin transport in polaronic
  and superfluid {Fermi} gases},}\ }\href {\doibase
  10.1088/1367-2630/13/5/055009} {\bibfield  {journal} {\bibinfo  {journal}
  {New Journal of Physics}\ }\textbf {\bibinfo {volume} {13}},\ \bibinfo
  {pages} {055009} (\bibinfo {year} {2011}{\natexlab{b}})}\BibitemShut
  {NoStop}%
\bibitem [{\citenamefont {Cao}\ \emph {et~al.}(2011)\citenamefont {Cao},
  \citenamefont {Elliott}, \citenamefont {Joseph}, \citenamefont {Wu},
  \citenamefont {Petricka}, \citenamefont {Sch{\"a}fer},\ and\ \citenamefont
  {Thomas}}]{cao2011universal}%
  \BibitemOpen
  \bibfield  {author} {\bibinfo {author} {\bibfnamefont {C.}~\bibnamefont
  {Cao}}, \bibinfo {author} {\bibfnamefont {E.}~\bibnamefont {Elliott}},
  \bibinfo {author} {\bibfnamefont {J.}~\bibnamefont {Joseph}}, \bibinfo
  {author} {\bibfnamefont {H.}~\bibnamefont {Wu}}, \bibinfo {author}
  {\bibfnamefont {J.}~\bibnamefont {Petricka}}, \bibinfo {author}
  {\bibfnamefont {T.}~\bibnamefont {Sch{\"a}fer}}, \ and\ \bibinfo {author}
  {\bibfnamefont {J.~E.}\ \bibnamefont {Thomas}},\ }\bibfield  {title}
  {\enquote {\bibinfo {title} {Universal {Quantum} {Viscosity} in a {Unitary}
  {Fermi} {Gas}},}\ }\href {\doibase 10.1126/science.1195219} {\bibfield
  {journal} {\bibinfo  {journal} {Science}\ }\textbf {\bibinfo {volume}
  {331}},\ \bibinfo {pages} {58--61} (\bibinfo {year} {2011})}\BibitemShut
  {NoStop}%
\bibitem [{\citenamefont {Enss}\ and\ \citenamefont
  {Haussmann}(2012)}]{enss2012quantum}%
  \BibitemOpen
  \bibfield  {author} {\bibinfo {author} {\bibfnamefont {Tilman}\ \bibnamefont
  {Enss}}\ and\ \bibinfo {author} {\bibfnamefont {Rudolf}\ \bibnamefont
  {Haussmann}},\ }\bibfield  {title} {\enquote {\bibinfo {title} {Quantum
  {Mechanical} {Limitations} to {Spin} {Diffusion} in the {Unitary} {Fermi}
  {Gas}},}\ }\href {\doibase 10.1103/PhysRevLett.109.195303} {\bibfield
  {journal} {\bibinfo  {journal} {Phys. Rev. Lett.}\ }\textbf {\bibinfo
  {volume} {109}},\ \bibinfo {pages} {195303} (\bibinfo {year}
  {2012})}\BibitemShut {NoStop}%
\bibitem [{\citenamefont {Valtolina}\ \emph {et~al.}(2017)\citenamefont
  {Valtolina}, \citenamefont {Scazza}, \citenamefont {Amico}, \citenamefont
  {Burchianti}, \citenamefont {Recati}, \citenamefont {Enss}, \citenamefont
  {Inguscio}, \citenamefont {Zaccanti},\ and\ \citenamefont
  {Roati}}]{valtolina2017exploring}%
  \BibitemOpen
  \bibfield  {author} {\bibinfo {author} {\bibfnamefont {G.}~\bibnamefont
  {Valtolina}}, \bibinfo {author} {\bibfnamefont {F.}~\bibnamefont {Scazza}},
  \bibinfo {author} {\bibfnamefont {A.}~\bibnamefont {Amico}}, \bibinfo
  {author} {\bibfnamefont {A.}~\bibnamefont {Burchianti}}, \bibinfo {author}
  {\bibfnamefont {A.}~\bibnamefont {Recati}}, \bibinfo {author} {\bibfnamefont
  {T.}~\bibnamefont {Enss}}, \bibinfo {author} {\bibfnamefont {M.}~\bibnamefont
  {Inguscio}}, \bibinfo {author} {\bibfnamefont {M.}~\bibnamefont {Zaccanti}},
  \ and\ \bibinfo {author} {\bibfnamefont {G.}~\bibnamefont {Roati}},\
  }\bibfield  {title} {\enquote {\bibinfo {title} {Exploring the ferromagnetic
  behaviour of a repulsive {Fermi} gas through spin dynamics},}\ }\href
  {\doibase 10.1038/nphys4108} {\bibfield  {journal} {\bibinfo  {journal}
  {Nature Physics}\ }\textbf {\bibinfo {volume} {13}},\ \bibinfo {pages} {704}
  (\bibinfo {year} {2017})}\BibitemShut {NoStop}%
\bibitem [{\citenamefont {Enss}\ and\ \citenamefont
  {Thywissen}(2019)}]{enss2019universal}%
  \BibitemOpen
  \bibfield  {author} {\bibinfo {author} {\bibfnamefont {Tilman}\ \bibnamefont
  {Enss}}\ and\ \bibinfo {author} {\bibfnamefont {Joseph~H.}\ \bibnamefont
  {Thywissen}},\ }\bibfield  {title} {\enquote {\bibinfo {title} {Universal
  {Spin} {Transport} and {Quantum} {Bounds} for {Unitary} {Fermions}},}\ }\href
  {\doibase 10.1146/annurev-conmatphys-031218-013732} {\bibfield  {journal}
  {\bibinfo  {journal} {Annual Review of Condensed Matter Physics}\ }\textbf
  {\bibinfo {volume} {10}},\ \bibinfo {pages} {85--106} (\bibinfo {year}
  {2019})}\BibitemShut {NoStop}%
\bibitem [{\citenamefont {Tajima}\ \emph {et~al.}(2020)\citenamefont {Tajima},
  \citenamefont {Recati},\ and\ \citenamefont
  {Ohashi}}]{tajima2020spin-dipole}%
  \BibitemOpen
  \bibfield  {author} {\bibinfo {author} {\bibfnamefont {Hiroyuki}\
  \bibnamefont {Tajima}}, \bibinfo {author} {\bibfnamefont {Alessio}\
  \bibnamefont {Recati}}, \ and\ \bibinfo {author} {\bibfnamefont {Yoji}\
  \bibnamefont {Ohashi}},\ }\bibfield  {title} {\enquote {\bibinfo {title}
  {Spin-dipole mode in a trapped {Fermi} gas near unitarity},}\ }\href
  {\doibase 10.1103/PhysRevA.101.013610} {\bibfield  {journal} {\bibinfo
  {journal} {Phys. Rev. A}\ }\textbf {\bibinfo {volume} {101}},\ \bibinfo
  {pages} {013610} (\bibinfo {year} {2020})}\BibitemShut {NoStop}%
\bibitem [{\citenamefont {Brantut}\ \emph {et~al.}(2012)\citenamefont
  {Brantut}, \citenamefont {Meineke}, \citenamefont {Stadler}, \citenamefont
  {Krinner},\ and\ \citenamefont {Esslinger}}]{brantut2012conduction}%
  \BibitemOpen
  \bibfield  {author} {\bibinfo {author} {\bibfnamefont {Jean-Philippe}\
  \bibnamefont {Brantut}}, \bibinfo {author} {\bibfnamefont {Jakob}\
  \bibnamefont {Meineke}}, \bibinfo {author} {\bibfnamefont {David}\
  \bibnamefont {Stadler}}, \bibinfo {author} {\bibfnamefont {Sebastian}\
  \bibnamefont {Krinner}}, \ and\ \bibinfo {author} {\bibfnamefont {Tilman}\
  \bibnamefont {Esslinger}},\ }\bibfield  {title} {\enquote {\bibinfo {title}
  {Conduction of {Ultracold} {Fermions} {Through} a {Mesoscopic} {Channel}},}\
  }\href {\doibase 10.1126/science.1223175} {\bibfield  {journal} {\bibinfo
  {journal} {Science}\ }\textbf {\bibinfo {volume} {337}},\ \bibinfo {pages}
  {1069--1071} (\bibinfo {year} {2012})}\BibitemShut {NoStop}%
\bibitem [{\citenamefont {Husmann}\ \emph {et~al.}(2015)\citenamefont
  {Husmann}, \citenamefont {Uchino}, \citenamefont {Krinner}, \citenamefont
  {Lebrat}, \citenamefont {Giamarchi}, \citenamefont {Esslinger},\ and\
  \citenamefont {Brantut}}]{husmann2015connecting}%
  \BibitemOpen
  \bibfield  {author} {\bibinfo {author} {\bibfnamefont {Dominik}\ \bibnamefont
  {Husmann}}, \bibinfo {author} {\bibfnamefont {Shun}\ \bibnamefont {Uchino}},
  \bibinfo {author} {\bibfnamefont {Sebastian}\ \bibnamefont {Krinner}},
  \bibinfo {author} {\bibfnamefont {Martin}\ \bibnamefont {Lebrat}}, \bibinfo
  {author} {\bibfnamefont {Thierry}\ \bibnamefont {Giamarchi}}, \bibinfo
  {author} {\bibfnamefont {Tilman}\ \bibnamefont {Esslinger}}, \ and\ \bibinfo
  {author} {\bibfnamefont {Jean-Philippe}\ \bibnamefont {Brantut}},\ }\bibfield
   {title} {\enquote {\bibinfo {title} {Connecting strongly correlated
  superfluids by a quantum point contact},}\ }\href {\doibase
  10.1126/science.aac9584} {\bibfield  {journal} {\bibinfo  {journal}
  {Science}\ }\textbf {\bibinfo {volume} {350}},\ \bibinfo {pages} {1498--1501}
  (\bibinfo {year} {2015})}\BibitemShut {NoStop}%
\bibitem [{\citenamefont {Krinner}\ \emph {et~al.}(2016)\citenamefont
  {Krinner}, \citenamefont {Lebrat}, \citenamefont {Husmann}, \citenamefont
  {Grenier}, \citenamefont {Brantut},\ and\ \citenamefont
  {Esslinger}}]{krinner2016mapping}%
  \BibitemOpen
  \bibfield  {author} {\bibinfo {author} {\bibfnamefont {Sebastian}\
  \bibnamefont {Krinner}}, \bibinfo {author} {\bibfnamefont {Martin}\
  \bibnamefont {Lebrat}}, \bibinfo {author} {\bibfnamefont {Dominik}\
  \bibnamefont {Husmann}}, \bibinfo {author} {\bibfnamefont {Charles}\
  \bibnamefont {Grenier}}, \bibinfo {author} {\bibfnamefont {Jean-Philippe}\
  \bibnamefont {Brantut}}, \ and\ \bibinfo {author} {\bibfnamefont {Tilman}\
  \bibnamefont {Esslinger}},\ }\bibfield  {title} {\enquote {\bibinfo {title}
  {Mapping out spin and particle conductances in a quantum point contact},}\
  }\href {\doibase 10.1073/pnas.1601812113} {\bibfield  {journal} {\bibinfo
  {journal} {PNAS}\ }\textbf {\bibinfo {volume} {113}},\ \bibinfo {pages}
  {8144--8149} (\bibinfo {year} {2016})}\BibitemShut {NoStop}%
\bibitem [{\citenamefont {Kan{\'a}sz-Nagy}\ \emph {et~al.}(2016)\citenamefont
  {Kan{\'a}sz-Nagy}, \citenamefont {Glazman}, \citenamefont {Esslinger},\ and\
  \citenamefont {Demler}}]{kanasz-nagy2016anomalous}%
  \BibitemOpen
  \bibfield  {author} {\bibinfo {author} {\bibfnamefont {M.}~\bibnamefont
  {Kan{\'a}sz-Nagy}}, \bibinfo {author} {\bibfnamefont {L.}~\bibnamefont
  {Glazman}}, \bibinfo {author} {\bibfnamefont {T.}~\bibnamefont {Esslinger}},
  \ and\ \bibinfo {author} {\bibfnamefont {E.~A.}\ \bibnamefont {Demler}},\
  }\bibfield  {title} {\enquote {\bibinfo {title} {Anomalous {Conductances} in
  an {Ultracold} {Quantum} {Wire}},}\ }\href {\doibase
  10.1103/PhysRevLett.117.255302} {\bibfield  {journal} {\bibinfo  {journal}
  {Phys. Rev. Lett.}\ }\textbf {\bibinfo {volume} {117}},\ \bibinfo {pages}
  {255302} (\bibinfo {year} {2016})}\BibitemShut {NoStop}%
\bibitem [{\citenamefont {H{\"a}usler}\ \emph {et~al.}(2017)\citenamefont
  {H{\"a}usler}, \citenamefont {Nakajima}, \citenamefont {Lebrat},
  \citenamefont {Husmann}, \citenamefont {Krinner}, \citenamefont {Esslinger},\
  and\ \citenamefont {Brantut}}]{hausler2017scanning}%
  \BibitemOpen
  \bibfield  {author} {\bibinfo {author} {\bibfnamefont {Samuel}\ \bibnamefont
  {H{\"a}usler}}, \bibinfo {author} {\bibfnamefont {Shuta}\ \bibnamefont
  {Nakajima}}, \bibinfo {author} {\bibfnamefont {Martin}\ \bibnamefont
  {Lebrat}}, \bibinfo {author} {\bibfnamefont {Dominik}\ \bibnamefont
  {Husmann}}, \bibinfo {author} {\bibfnamefont {Sebastian}\ \bibnamefont
  {Krinner}}, \bibinfo {author} {\bibfnamefont {Tilman}\ \bibnamefont
  {Esslinger}}, \ and\ \bibinfo {author} {\bibfnamefont {Jean-Philippe}\
  \bibnamefont {Brantut}},\ }\bibfield  {title} {\enquote {\bibinfo {title}
  {Scanning {Gate} {Microscope} for {Cold} {Atomic} {Gases}},}\ }\href
  {\doibase 10.1103/PhysRevLett.119.030403} {\bibfield  {journal} {\bibinfo
  {journal} {Phys. Rev. Lett.}\ }\textbf {\bibinfo {volume} {119}},\ \bibinfo
  {pages} {030403} (\bibinfo {year} {2017})}\BibitemShut {NoStop}%
\bibitem [{\citenamefont {Corman}\ \emph {et~al.}(2019)\citenamefont {Corman},
  \citenamefont {Fabritius}, \citenamefont {H{\"a}usler}, \citenamefont
  {Mohan}, \citenamefont {Dogra}, \citenamefont {Husmann}, \citenamefont
  {Lebrat},\ and\ \citenamefont {Esslinger}}]{corman2019quantized}%
  \BibitemOpen
  \bibfield  {author} {\bibinfo {author} {\bibfnamefont {Laura}\ \bibnamefont
  {Corman}}, \bibinfo {author} {\bibfnamefont {Philipp}\ \bibnamefont
  {Fabritius}}, \bibinfo {author} {\bibfnamefont {Samuel}\ \bibnamefont
  {H{\"a}usler}}, \bibinfo {author} {\bibfnamefont {Jeffrey}\ \bibnamefont
  {Mohan}}, \bibinfo {author} {\bibfnamefont {Lena~H.}\ \bibnamefont {Dogra}},
  \bibinfo {author} {\bibfnamefont {Dominik}\ \bibnamefont {Husmann}}, \bibinfo
  {author} {\bibfnamefont {Martin}\ \bibnamefont {Lebrat}}, \ and\ \bibinfo
  {author} {\bibfnamefont {Tilman}\ \bibnamefont {Esslinger}},\ }\bibfield
  {title} {\enquote {\bibinfo {title} {Quantized conductance through a
  dissipative atomic point contact},}\ }\href {\doibase
  10.1103/PhysRevA.100.053605} {\bibfield  {journal} {\bibinfo  {journal}
  {Phys. Rev. A}\ }\textbf {\bibinfo {volume} {100}},\ \bibinfo {pages}
  {053605} (\bibinfo {year} {2019})}\BibitemShut {NoStop}%
\bibitem [{\citenamefont {Ono}\ \emph {et~al.}(2021)\citenamefont {Ono},
  \citenamefont {Higomoto}, \citenamefont {Saito}, \citenamefont {Uchino},
  \citenamefont {Nishida},\ and\ \citenamefont
  {Takahashi}}]{ono2021observation}%
  \BibitemOpen
  \bibfield  {author} {\bibinfo {author} {\bibfnamefont {Koki}\ \bibnamefont
  {Ono}}, \bibinfo {author} {\bibfnamefont {Toshiya}\ \bibnamefont {Higomoto}},
  \bibinfo {author} {\bibfnamefont {Yugo}\ \bibnamefont {Saito}}, \bibinfo
  {author} {\bibfnamefont {Shun}\ \bibnamefont {Uchino}}, \bibinfo {author}
  {\bibfnamefont {Yusuke}\ \bibnamefont {Nishida}}, \ and\ \bibinfo {author}
  {\bibfnamefont {Yoshiro}\ \bibnamefont {Takahashi}},\ }\bibfield  {title}
  {\enquote {\bibinfo {title} {Observation of spin-space quantum transport
  induced by an atomic quantum point contact},}\ }\href {\doibase
  10.1038/s41467-021-27011-2} {\bibfield  {journal} {\bibinfo  {journal} {Nat
  Commun}\ }\textbf {\bibinfo {volume} {12}},\ \bibinfo {pages} {6724}
  (\bibinfo {year} {2021})}\BibitemShut {NoStop}%
\bibitem [{\citenamefont {Valtolina}\ \emph {et~al.}(2015)\citenamefont
  {Valtolina}, \citenamefont {Burchianti}, \citenamefont {Amico}, \citenamefont
  {Neri}, \citenamefont {Xhani}, \citenamefont {Seman}, \citenamefont
  {Trombettoni}, \citenamefont {Smerzi}, \citenamefont {Zaccanti},
  \citenamefont {Inguscio},\ and\ \citenamefont
  {Roati}}]{valtolina2015josephson}%
  \BibitemOpen
  \bibfield  {author} {\bibinfo {author} {\bibfnamefont {Giacomo}\ \bibnamefont
  {Valtolina}}, \bibinfo {author} {\bibfnamefont {Alessia}\ \bibnamefont
  {Burchianti}}, \bibinfo {author} {\bibfnamefont {Andrea}\ \bibnamefont
  {Amico}}, \bibinfo {author} {\bibfnamefont {Elettra}\ \bibnamefont {Neri}},
  \bibinfo {author} {\bibfnamefont {Klejdja}\ \bibnamefont {Xhani}}, \bibinfo
  {author} {\bibfnamefont {Jorge~Amin}\ \bibnamefont {Seman}}, \bibinfo
  {author} {\bibfnamefont {Andrea}\ \bibnamefont {Trombettoni}}, \bibinfo
  {author} {\bibfnamefont {Augusto}\ \bibnamefont {Smerzi}}, \bibinfo {author}
  {\bibfnamefont {Matteo}\ \bibnamefont {Zaccanti}}, \bibinfo {author}
  {\bibfnamefont {Massimo}\ \bibnamefont {Inguscio}}, \ and\ \bibinfo {author}
  {\bibfnamefont {Giacomo}\ \bibnamefont {Roati}},\ }\bibfield  {title}
  {\enquote {\bibinfo {title} {Josephson effect in fermionic superfluids across
  the {BEC}-{BCS} crossover},}\ }\href {\doibase 10.1126/science.aac9725}
  {\bibfield  {journal} {\bibinfo  {journal} {Science}\ }\textbf {\bibinfo
  {volume} {350}},\ \bibinfo {pages} {1505--1508} (\bibinfo {year}
  {2015})}\BibitemShut {NoStop}%
\bibitem [{\citenamefont {Burchianti}\ \emph {et~al.}(2018)\citenamefont
  {Burchianti}, \citenamefont {Scazza}, \citenamefont {Amico}, \citenamefont
  {Valtolina}, \citenamefont {Seman}, \citenamefont {Fort}, \citenamefont
  {Zaccanti}, \citenamefont {Inguscio},\ and\ \citenamefont
  {Roati}}]{burchianti2018connecting}%
  \BibitemOpen
  \bibfield  {author} {\bibinfo {author} {\bibfnamefont {A.}~\bibnamefont
  {Burchianti}}, \bibinfo {author} {\bibfnamefont {F.}~\bibnamefont {Scazza}},
  \bibinfo {author} {\bibfnamefont {A.}~\bibnamefont {Amico}}, \bibinfo
  {author} {\bibfnamefont {G.}~\bibnamefont {Valtolina}}, \bibinfo {author}
  {\bibfnamefont {J.~A.}\ \bibnamefont {Seman}}, \bibinfo {author}
  {\bibfnamefont {C.}~\bibnamefont {Fort}}, \bibinfo {author} {\bibfnamefont
  {M.}~\bibnamefont {Zaccanti}}, \bibinfo {author} {\bibfnamefont
  {M.}~\bibnamefont {Inguscio}}, \ and\ \bibinfo {author} {\bibfnamefont
  {G.}~\bibnamefont {Roati}},\ }\bibfield  {title} {\enquote {\bibinfo {title}
  {Connecting {Dissipation} and {Phase} {Slips} in a {Josephson} {Junction}
  between {Fermionic} {Superfluids}},}\ }\href {\doibase
  10.1103/PhysRevLett.120.025302} {\bibfield  {journal} {\bibinfo  {journal}
  {Phys. Rev. Lett.}\ }\textbf {\bibinfo {volume} {120}},\ \bibinfo {pages}
  {025302} (\bibinfo {year} {2018})}\BibitemShut {NoStop}%
\bibitem [{\citenamefont {Zaccanti}\ and\ \citenamefont
  {Zwerger}(2019)}]{zaccanti2019critical}%
  \BibitemOpen
  \bibfield  {author} {\bibinfo {author} {\bibfnamefont {M.}~\bibnamefont
  {Zaccanti}}\ and\ \bibinfo {author} {\bibfnamefont {W.}~\bibnamefont
  {Zwerger}},\ }\bibfield  {title} {\enquote {\bibinfo {title} {Critical
  {Josephson} current in {BCS}-{BEC}--crossover superfluids},}\ }\href
  {\doibase 10.1103/PhysRevA.100.063601} {\bibfield  {journal} {\bibinfo
  {journal} {Phys. Rev. A}\ }\textbf {\bibinfo {volume} {100}},\ \bibinfo
  {pages} {063601} (\bibinfo {year} {2019})}\BibitemShut {NoStop}%
\bibitem [{\citenamefont {Luick}\ \emph {et~al.}(2020)\citenamefont {Luick},
  \citenamefont {Sobirey}, \citenamefont {Bohlen}, \citenamefont {Singh},
  \citenamefont {Mathey}, \citenamefont {Lompe},\ and\ \citenamefont
  {Moritz}}]{luick2020ideal}%
  \BibitemOpen
  \bibfield  {author} {\bibinfo {author} {\bibfnamefont {Niclas}\ \bibnamefont
  {Luick}}, \bibinfo {author} {\bibfnamefont {Lennart}\ \bibnamefont
  {Sobirey}}, \bibinfo {author} {\bibfnamefont {Markus}\ \bibnamefont
  {Bohlen}}, \bibinfo {author} {\bibfnamefont {Vijay~Pal}\ \bibnamefont
  {Singh}}, \bibinfo {author} {\bibfnamefont {Ludwig}\ \bibnamefont {Mathey}},
  \bibinfo {author} {\bibfnamefont {Thomas}\ \bibnamefont {Lompe}}, \ and\
  \bibinfo {author} {\bibfnamefont {Henning}\ \bibnamefont {Moritz}},\
  }\bibfield  {title} {\enquote {\bibinfo {title} {An ideal {Josephson}
  junction in an ultracold two-dimensional {Fermi} gas},}\ }\href {\doibase
  10.1126/science.aaz2342} {\bibfield  {journal} {\bibinfo  {journal}
  {Science}\ }\textbf {\bibinfo {volume} {369}},\ \bibinfo {pages} {89--91}
  (\bibinfo {year} {2020})}\BibitemShut {NoStop}%
\bibitem [{\citenamefont {Del~Pace}\ \emph {et~al.}(2021)\citenamefont
  {Del~Pace}, \citenamefont {Kwon}, \citenamefont {Zaccanti}, \citenamefont
  {Roati},\ and\ \citenamefont {Scazza}}]{del_pace2021tunneling}%
  \BibitemOpen
  \bibfield  {author} {\bibinfo {author} {\bibfnamefont {G.}~\bibnamefont
  {Del~Pace}}, \bibinfo {author} {\bibfnamefont {W.~J.}\ \bibnamefont {Kwon}},
  \bibinfo {author} {\bibfnamefont {M.}~\bibnamefont {Zaccanti}}, \bibinfo
  {author} {\bibfnamefont {G.}~\bibnamefont {Roati}}, \ and\ \bibinfo {author}
  {\bibfnamefont {F.}~\bibnamefont {Scazza}},\ }\bibfield  {title} {\enquote
  {\bibinfo {title} {Tunneling {Transport} of {Unitary} {Fermions} across the
  {Superfluid} {Transition}},}\ }\href {\doibase
  10.1103/PhysRevLett.126.055301} {\bibfield  {journal} {\bibinfo  {journal}
  {Phys. Rev. Lett.}\ }\textbf {\bibinfo {volume} {126}},\ \bibinfo {pages}
  {055301} (\bibinfo {year} {2021})}\BibitemShut {NoStop}%
\bibitem [{\citenamefont {Berggren}\ \emph {et~al.}(2016)\citenamefont
  {Berggren}, \citenamefont {Taylor}, \citenamefont {Mitchell}, \citenamefont
  {Hannam}, \citenamefont {Lazar},\ and\ \citenamefont {Leese
  De~Escobar}}]{berggren2016computational}%
  \BibitemOpen
  \bibfield  {author} {\bibinfo {author} {\bibfnamefont {S.}~\bibnamefont
  {Berggren}}, \bibinfo {author} {\bibfnamefont {B.~J.}\ \bibnamefont
  {Taylor}}, \bibinfo {author} {\bibfnamefont {E.~E.}\ \bibnamefont
  {Mitchell}}, \bibinfo {author} {\bibfnamefont {K.~E.}\ \bibnamefont
  {Hannam}}, \bibinfo {author} {\bibfnamefont {J.~Y.}\ \bibnamefont {Lazar}}, \
  and\ \bibinfo {author} {\bibfnamefont {A.}~\bibnamefont {Leese De~Escobar}},\
  }\bibfield  {title} {\enquote {\bibinfo {title} {Computational {Modeling} of
  bi-{Superconducting} {Quantum} {Interference} {Devices} for
  {High}-{Temperature} {Superconducting} {Prototype} {Chips}},}\ }\href
  {\doibase 10.1109/TASC.2016.2569502} {\bibfield  {journal} {\bibinfo
  {journal} {IEEE Transactions on Applied Superconductivity}\ }\textbf
  {\bibinfo {volume} {26}},\ \bibinfo {pages} {1--6} (\bibinfo {year}
  {2016})}\BibitemShut {NoStop}%
\bibitem [{\citenamefont {Perconte}\ \emph {et~al.}(2018)\citenamefont
  {Perconte}, \citenamefont {Cuellar}, \citenamefont {Moreau-Luchaire},
  \citenamefont {Piquemal-Banci}, \citenamefont {Galceran}, \citenamefont
  {Kidambi}, \citenamefont {Martin}, \citenamefont {Hofmann}, \citenamefont
  {Bernard}, \citenamefont {Dlubak}, \citenamefont {Seneor},\ and\
  \citenamefont {Villegas}}]{perconte2018tunable}%
  \BibitemOpen
  \bibfield  {author} {\bibinfo {author} {\bibfnamefont {David}\ \bibnamefont
  {Perconte}}, \bibinfo {author} {\bibfnamefont {Fabian~A.}\ \bibnamefont
  {Cuellar}}, \bibinfo {author} {\bibfnamefont {Constance}\ \bibnamefont
  {Moreau-Luchaire}}, \bibinfo {author} {\bibfnamefont {Maelis}\ \bibnamefont
  {Piquemal-Banci}}, \bibinfo {author} {\bibfnamefont {Regina}\ \bibnamefont
  {Galceran}}, \bibinfo {author} {\bibfnamefont {Piran~R.}\ \bibnamefont
  {Kidambi}}, \bibinfo {author} {\bibfnamefont {Marie-Blandine}\ \bibnamefont
  {Martin}}, \bibinfo {author} {\bibfnamefont {Stephan}\ \bibnamefont
  {Hofmann}}, \bibinfo {author} {\bibfnamefont {Rozenn}\ \bibnamefont
  {Bernard}}, \bibinfo {author} {\bibfnamefont {Bruno}\ \bibnamefont {Dlubak}},
  \bibinfo {author} {\bibfnamefont {Pierre}\ \bibnamefont {Seneor}}, \ and\
  \bibinfo {author} {\bibfnamefont {Javier~E.}\ \bibnamefont {Villegas}},\
  }\bibfield  {title} {\enquote {\bibinfo {title} {Tunable {Klein}-like
  tunnelling of high-temperature superconducting pairs into graphene},}\ }\href
  {\doibase 10.1038/nphys4278} {\bibfield  {journal} {\bibinfo  {journal}
  {Nature Physics}\ }\textbf {\bibinfo {volume} {14}},\ \bibinfo {pages}
  {25--29} (\bibinfo {year} {2018})}\BibitemShut {NoStop}%
\bibitem [{\citenamefont {Visani}\ \emph {et~al.}(2012)\citenamefont {Visani},
  \citenamefont {Sefrioui}, \citenamefont {Tornos}, \citenamefont {Leon},
  \citenamefont {Briatico}, \citenamefont {Bibes}, \citenamefont
  {Barth{\'e}l{\'e}my}, \citenamefont {Santamar{\'i}a},\ and\ \citenamefont
  {Villegas}}]{visani2012equal-spin}%
  \BibitemOpen
  \bibfield  {author} {\bibinfo {author} {\bibfnamefont {C.}~\bibnamefont
  {Visani}}, \bibinfo {author} {\bibfnamefont {Z.}~\bibnamefont {Sefrioui}},
  \bibinfo {author} {\bibfnamefont {J.}~\bibnamefont {Tornos}}, \bibinfo
  {author} {\bibfnamefont {C.}~\bibnamefont {Leon}}, \bibinfo {author}
  {\bibfnamefont {J.}~\bibnamefont {Briatico}}, \bibinfo {author}
  {\bibfnamefont {M.}~\bibnamefont {Bibes}}, \bibinfo {author} {\bibfnamefont
  {A.}~\bibnamefont {Barth{\'e}l{\'e}my}}, \bibinfo {author} {\bibfnamefont
  {J.}~\bibnamefont {Santamar{\'i}a}}, \ and\ \bibinfo {author} {\bibfnamefont
  {Javier~E.}\ \bibnamefont {Villegas}},\ }\bibfield  {title} {\enquote
  {\bibinfo {title} {Equal-spin {Andreev} reflection and long-range coherent
  transport in high-temperature superconductor/half-metallic ferromagnet
  junctions},}\ }\href {\doibase 10.1038/nphys2318} {\bibfield  {journal}
  {\bibinfo  {journal} {Nature Phys}\ }\textbf {\bibinfo {volume} {8}},\
  \bibinfo {pages} {539--543} (\bibinfo {year} {2012})}\BibitemShut {NoStop}%
\bibitem [{\citenamefont {Komori}\ \emph {et~al.}(2018)\citenamefont {Komori},
  \citenamefont {Di~Bernardo}, \citenamefont {Buzdin}, \citenamefont
  {Blamire},\ and\ \citenamefont {Robinson}}]{komori2018magnetic}%
  \BibitemOpen
  \bibfield  {author} {\bibinfo {author} {\bibfnamefont {S.}~\bibnamefont
  {Komori}}, \bibinfo {author} {\bibfnamefont {A.}~\bibnamefont {Di~Bernardo}},
  \bibinfo {author} {\bibfnamefont {A.~I.}\ \bibnamefont {Buzdin}}, \bibinfo
  {author} {\bibfnamefont {M.~G.}\ \bibnamefont {Blamire}}, \ and\ \bibinfo
  {author} {\bibfnamefont {J.~W.~A.}\ \bibnamefont {Robinson}},\ }\bibfield
  {title} {\enquote {\bibinfo {title} {Magnetic {Exchange} {Fields} and
  {Domain} {Wall} {Superconductivity} at an {All}-{Oxide}
  {Superconductor}-{Ferromagnet} {Insulator} {Interface}},}\ }\href {\doibase
  10.1103/PhysRevLett.121.077003} {\bibfield  {journal} {\bibinfo  {journal}
  {Phys. Rev. Lett.}\ }\textbf {\bibinfo {volume} {121}},\ \bibinfo {pages}
  {077003} (\bibinfo {year} {2018})}\BibitemShut {NoStop}%
\bibitem [{\citenamefont {Bouscher}\ \emph {et~al.}(2020)\citenamefont
  {Bouscher}, \citenamefont {Kang}, \citenamefont {Balasubramanian},
  \citenamefont {Panna}, \citenamefont {Yu}, \citenamefont {Chen},\ and\
  \citenamefont {Hayat}}]{bouscher2020high-tc}%
  \BibitemOpen
  \bibfield  {author} {\bibinfo {author} {\bibfnamefont {Shlomi}\ \bibnamefont
  {Bouscher}}, \bibinfo {author} {\bibfnamefont {Zhixin}\ \bibnamefont {Kang}},
  \bibinfo {author} {\bibfnamefont {Krishna}\ \bibnamefont {Balasubramanian}},
  \bibinfo {author} {\bibfnamefont {Dmitry}\ \bibnamefont {Panna}}, \bibinfo
  {author} {\bibfnamefont {Pu}~\bibnamefont {Yu}}, \bibinfo {author}
  {\bibfnamefont {Xi}~\bibnamefont {Chen}}, \ and\ \bibinfo {author}
  {\bibfnamefont {Alex}\ \bibnamefont {Hayat}},\ }\bibfield  {title} {\enquote
  {\bibinfo {title} {High-{Tc} {Cooper}-pair injection in a
  semiconductor{\textendash}superconductor structure},}\ }\href {\doibase
  10.1088/1361-648X/abae18} {\bibfield  {journal} {\bibinfo  {journal} {J.
  Phys.: Condens. Matter}\ }\textbf {\bibinfo {volume} {32}},\ \bibinfo {pages}
  {475502} (\bibinfo {year} {2020})}\BibitemShut {NoStop}%
\bibitem [{\citenamefont {Bulgac}\ and\ \citenamefont
  {Forbes}(2007)}]{bulgac2007zero-temperature}%
  \BibitemOpen
  \bibfield  {author} {\bibinfo {author} {\bibfnamefont {Aurel}\ \bibnamefont
  {Bulgac}}\ and\ \bibinfo {author} {\bibfnamefont {Michael~McNeil}\
  \bibnamefont {Forbes}},\ }\bibfield  {title} {\enquote {\bibinfo {title}
  {Zero-temperature thermodynamics of asymmetric {Fermi} gases at unitarity},}\
  }\href {\doibase 10.1103/PhysRevA.75.031605} {\bibfield  {journal} {\bibinfo
  {journal} {Phys. Rev. A}\ }\textbf {\bibinfo {volume} {75}},\ \bibinfo
  {pages} {031605(R)} (\bibinfo {year} {2007})}\BibitemShut {NoStop}%
\bibitem [{\citenamefont {Shin}(2008)}]{shin2008determination}%
  \BibitemOpen
  \bibfield  {author} {\bibinfo {author} {\bibfnamefont {Yong-il}\ \bibnamefont
  {Shin}},\ }\bibfield  {title} {\enquote {\bibinfo {title} {Determination of
  the equation of state of a polarized {Fermi} gas at unitarity},}\ }\href
  {\doibase 10.1103/PhysRevA.77.041603} {\bibfield  {journal} {\bibinfo
  {journal} {Phys. Rev. A}\ }\textbf {\bibinfo {volume} {77}},\ \bibinfo
  {pages} {041603(R)} (\bibinfo {year} {2008})}\BibitemShut {NoStop}%
\bibitem [{\citenamefont {Baur}\ \emph {et~al.}(2009)\citenamefont {Baur},
  \citenamefont {Basu}, \citenamefont {De~Silva},\ and\ \citenamefont
  {Mueller}}]{baur2009theory}%
  \BibitemOpen
  \bibfield  {author} {\bibinfo {author} {\bibfnamefont {Stefan~K.}\
  \bibnamefont {Baur}}, \bibinfo {author} {\bibfnamefont {Sourish}\
  \bibnamefont {Basu}}, \bibinfo {author} {\bibfnamefont {Theja~N.}\
  \bibnamefont {De~Silva}}, \ and\ \bibinfo {author} {\bibfnamefont {Erich~J.}\
  \bibnamefont {Mueller}},\ }\bibfield  {title} {\enquote {\bibinfo {title}
  {Theory of the normal-superfluid interface in population-imbalanced {Fermi}
  gases},}\ }\href {\doibase 10.1103/PhysRevA.79.063628} {\bibfield  {journal}
  {\bibinfo  {journal} {Phys. Rev. A}\ }\textbf {\bibinfo {volume} {79}},\
  \bibinfo {pages} {063628} (\bibinfo {year} {2009})}\BibitemShut {NoStop}%
\bibitem [{\citenamefont {Pilati}\ and\ \citenamefont
  {Giorgini}(2008)}]{pilati2008phase}%
  \BibitemOpen
  \bibfield  {author} {\bibinfo {author} {\bibfnamefont {S.}~\bibnamefont
  {Pilati}}\ and\ \bibinfo {author} {\bibfnamefont {S.}~\bibnamefont
  {Giorgini}},\ }\bibfield  {title} {\enquote {\bibinfo {title} {Phase
  {Separation} in a {Polarized} {Fermi} {Gas} at {Zero} {Temperature}},}\
  }\href {\doibase 10.1103/PhysRevLett.100.030401} {\bibfield  {journal}
  {\bibinfo  {journal} {Phys. Rev. Lett.}\ }\textbf {\bibinfo {volume} {100}},\
  \bibinfo {pages} {030401} (\bibinfo {year} {2008})}\BibitemShut {NoStop}%
\bibitem [{\citenamefont {Liu}\ \emph {et~al.}(2008)\citenamefont {Liu},
  \citenamefont {Hu},\ and\ \citenamefont
  {Drummond}}]{liu2008finite-temperature}%
  \BibitemOpen
  \bibfield  {author} {\bibinfo {author} {\bibfnamefont {Xia-Ji}\ \bibnamefont
  {Liu}}, \bibinfo {author} {\bibfnamefont {Hui}\ \bibnamefont {Hu}}, \ and\
  \bibinfo {author} {\bibfnamefont {Peter~D.}\ \bibnamefont {Drummond}},\
  }\bibfield  {title} {\enquote {\bibinfo {title} {Finite-temperature phase
  diagram of a spin-polarized ultracold {Fermi} gas in a highly elongated
  harmonic trap},}\ }\href {\doibase 10.1103/PhysRevA.78.023601} {\bibfield
  {journal} {\bibinfo  {journal} {Phys. Rev. A}\ }\textbf {\bibinfo {volume}
  {78}},\ \bibinfo {pages} {023601} (\bibinfo {year} {2008})}\BibitemShut
  {NoStop}%
\bibitem [{\citenamefont {Olsen}\ \emph {et~al.}(2015)\citenamefont {Olsen},
  \citenamefont {Revelle}, \citenamefont {Fry}, \citenamefont {Sheehy},\ and\
  \citenamefont {Hulet}}]{olsen2015phase}%
  \BibitemOpen
  \bibfield  {author} {\bibinfo {author} {\bibfnamefont {Ben~A.}\ \bibnamefont
  {Olsen}}, \bibinfo {author} {\bibfnamefont {Melissa~C.}\ \bibnamefont
  {Revelle}}, \bibinfo {author} {\bibfnamefont {Jacob~A.}\ \bibnamefont {Fry}},
  \bibinfo {author} {\bibfnamefont {Daniel~E.}\ \bibnamefont {Sheehy}}, \ and\
  \bibinfo {author} {\bibfnamefont {Randall~G.}\ \bibnamefont {Hulet}},\
  }\bibfield  {title} {\enquote {\bibinfo {title} {Phase diagram of a strongly
  interacting spin-imbalanced {Fermi} gas},}\ }\href {\doibase
  10.1103/PhysRevA.92.063616} {\bibfield  {journal} {\bibinfo  {journal} {Phys.
  Rev. A}\ }\textbf {\bibinfo {volume} {92}},\ \bibinfo {pages} {063616}
  (\bibinfo {year} {2015})}\BibitemShut {NoStop}%
\bibitem [{\citenamefont {de~Jong}\ and\ \citenamefont
  {Beenakker}(1995)}]{de_jong1995andreev}%
  \BibitemOpen
  \bibfield  {author} {\bibinfo {author} {\bibfnamefont {M.~J.~M.}\
  \bibnamefont {de~Jong}}\ and\ \bibinfo {author} {\bibfnamefont {C.~W.~J.}\
  \bibnamefont {Beenakker}},\ }\bibfield  {title} {\enquote {\bibinfo {title}
  {Andreev {Reflection} in {Ferromagnet}-{Superconductor} {Junctions}},}\
  }\href {\doibase 10.1103/PhysRevLett.74.1657} {\bibfield  {journal} {\bibinfo
   {journal} {Phys. Rev. Lett.}\ }\textbf {\bibinfo {volume} {74}},\ \bibinfo
  {pages} {1657--1660} (\bibinfo {year} {1995})}\BibitemShut {NoStop}%
\bibitem [{\citenamefont {Halterman}\ and\ \citenamefont
  {Valls}(2004)}]{halterman2004layered}%
  \BibitemOpen
  \bibfield  {author} {\bibinfo {author} {\bibfnamefont {Klaus}\ \bibnamefont
  {Halterman}}\ and\ \bibinfo {author} {\bibfnamefont {Oriol~T.}\ \bibnamefont
  {Valls}},\ }\bibfield  {title} {\enquote {\bibinfo {title} {Layered
  ferromagnet-superconductor structures: {The} $\pi$ state and proximity
  effects},}\ }\href {\doibase 10.1103/PhysRevB.69.014517} {\bibfield
  {journal} {\bibinfo  {journal} {Phys. Rev. B}\ }\textbf {\bibinfo {volume}
  {69}},\ \bibinfo {pages} {014517} (\bibinfo {year} {2004})}\BibitemShut
  {NoStop}%
\bibitem [{\citenamefont {Kashimura}\ \emph {et~al.}(2010)\citenamefont
  {Kashimura}, \citenamefont {Tsuchiya},\ and\ \citenamefont
  {Ohashi}}]{kashimura2010superfluid-ferromagnet-superfluid}%
  \BibitemOpen
  \bibfield  {author} {\bibinfo {author} {\bibfnamefont {Takashi}\ \bibnamefont
  {Kashimura}}, \bibinfo {author} {\bibfnamefont {Shunji}\ \bibnamefont
  {Tsuchiya}}, \ and\ \bibinfo {author} {\bibfnamefont {Yoji}\ \bibnamefont
  {Ohashi}},\ }\bibfield  {title} {\enquote {\bibinfo {title}
  {Superfluid-ferromagnet-superfluid junction and the $\pi$ phase in a
  superfluid {Fermi} gas},}\ }\href {\doibase 10.1103/PhysRevA.82.033617}
  {\bibfield  {journal} {\bibinfo  {journal} {Phys. Rev. A}\ }\textbf {\bibinfo
  {volume} {82}},\ \bibinfo {pages} {033617} (\bibinfo {year}
  {2010})}\BibitemShut {NoStop}%
\bibitem [{\citenamefont {Alidoust}\ and\ \citenamefont
  {Halterman}(2018)}]{alidoust2018half-metallic}%
  \BibitemOpen
  \bibfield  {author} {\bibinfo {author} {\bibfnamefont {Mohammad}\
  \bibnamefont {Alidoust}}\ and\ \bibinfo {author} {\bibfnamefont {Klaus}\
  \bibnamefont {Halterman}},\ }\bibfield  {title} {\enquote {\bibinfo {title}
  {Half-metallic superconducting triplet spin multivalves},}\ }\href {\doibase
  10.1103/PhysRevB.97.064517} {\bibfield  {journal} {\bibinfo  {journal} {Phys.
  Rev. B}\ }\textbf {\bibinfo {volume} {97}},\ \bibinfo {pages} {064517}
  (\bibinfo {year} {2018})}\BibitemShut {NoStop}%
\bibitem [{\citenamefont {Van~Schaeybroeck}\ and\ \citenamefont
  {Lazarides}(2007)}]{van_schaeybroeck2007normal-superfluid}%
  \BibitemOpen
  \bibfield  {author} {\bibinfo {author} {\bibfnamefont {Bert}\ \bibnamefont
  {Van~Schaeybroeck}}\ and\ \bibinfo {author} {\bibfnamefont {Achilleas}\
  \bibnamefont {Lazarides}},\ }\bibfield  {title} {\enquote {\bibinfo {title}
  {Normal-{Superfluid} {Interface} {Scattering} for {Polarized} {Fermion}
  {Gases}},}\ }\href {\doibase 10.1103/PhysRevLett.98.170402} {\bibfield
  {journal} {\bibinfo  {journal} {Phys. Rev. Lett.}\ }\textbf {\bibinfo
  {volume} {98}},\ \bibinfo {pages} {170402} (\bibinfo {year}
  {2007})}\BibitemShut {NoStop}%
\bibitem [{\citenamefont {Van~Schaeybroeck}\ and\ \citenamefont
  {Lazarides}(2009)}]{van_schaeybroeck2009normal-superfluid}%
  \BibitemOpen
  \bibfield  {author} {\bibinfo {author} {\bibfnamefont {Bert}\ \bibnamefont
  {Van~Schaeybroeck}}\ and\ \bibinfo {author} {\bibfnamefont {Achilleas}\
  \bibnamefont {Lazarides}},\ }\bibfield  {title} {\enquote {\bibinfo {title}
  {Normal-superfluid interface for polarized fermion gases},}\ }\href {\doibase
  10.1103/PhysRevA.79.053612} {\bibfield  {journal} {\bibinfo  {journal} {Phys.
  Rev. A}\ }\textbf {\bibinfo {volume} {79}},\ \bibinfo {pages} {053612}
  (\bibinfo {year} {2009})}\BibitemShut {NoStop}%
\bibitem [{\citenamefont {Parish}\ and\ \citenamefont
  {Huse}(2009)}]{parish2009evaporative}%
  \BibitemOpen
  \bibfield  {author} {\bibinfo {author} {\bibfnamefont {Meera~M.}\
  \bibnamefont {Parish}}\ and\ \bibinfo {author} {\bibfnamefont {David~A.}\
  \bibnamefont {Huse}},\ }\bibfield  {title} {\enquote {\bibinfo {title}
  {Evaporative depolarization and spin transport in a unitary trapped {Fermi}
  gas},}\ }\href {\doibase 10.1103/PhysRevA.80.063605} {\bibfield  {journal}
  {\bibinfo  {journal} {Phys. Rev. A}\ }\textbf {\bibinfo {volume} {80}},\
  \bibinfo {pages} {063605} (\bibinfo {year} {2009})}\BibitemShut {NoStop}%
\bibitem [{\citenamefont {Liao}\ \emph {et~al.}(2011)\citenamefont {Liao},
  \citenamefont {Revelle}, \citenamefont {Paprotta}, \citenamefont {Rittner},
  \citenamefont {Li}, \citenamefont {Partridge},\ and\ \citenamefont
  {Hulet}}]{liao2011metastability}%
  \BibitemOpen
  \bibfield  {author} {\bibinfo {author} {\bibfnamefont {Y.~A.}\ \bibnamefont
  {Liao}}, \bibinfo {author} {\bibfnamefont {M.}~\bibnamefont {Revelle}},
  \bibinfo {author} {\bibfnamefont {T.}~\bibnamefont {Paprotta}}, \bibinfo
  {author} {\bibfnamefont {A.~S.~C.}\ \bibnamefont {Rittner}}, \bibinfo
  {author} {\bibfnamefont {Wenhui}\ \bibnamefont {Li}}, \bibinfo {author}
  {\bibfnamefont {G.~B.}\ \bibnamefont {Partridge}}, \ and\ \bibinfo {author}
  {\bibfnamefont {R.~G.}\ \bibnamefont {Hulet}},\ }\bibfield  {title} {\enquote
  {\bibinfo {title} {Metastability in {Spin}-{Polarized} {Fermi} {Gases}},}\
  }\href {\doibase 10.1103/PhysRevLett.107.145305} {\bibfield  {journal}
  {\bibinfo  {journal} {Phys. Rev. Lett.}\ }\textbf {\bibinfo {volume} {107}},\
  \bibinfo {pages} {145305} (\bibinfo {year} {2011})}\BibitemShut {NoStop}%
\bibitem [{\citenamefont {Magierski}\ \emph {et~al.}(2019)\citenamefont
  {Magierski}, \citenamefont {T{\"u}zemen},\ and\ \citenamefont {Wlaz{\l
  }owski}}]{magierski2019spin-polarized}%
  \BibitemOpen
  \bibfield  {author} {\bibinfo {author} {\bibfnamefont {Piotr}\ \bibnamefont
  {Magierski}}, \bibinfo {author} {\bibfnamefont {Bu{\u g}ra}\ \bibnamefont
  {T{\"u}zemen}}, \ and\ \bibinfo {author} {\bibfnamefont {Gabriel}\
  \bibnamefont {Wlaz{\l }owski}},\ }\bibfield  {title} {\enquote {\bibinfo
  {title} {Spin-polarized droplets in the unitary {Fermi} gas},}\ }\href
  {\doibase 10.1103/PhysRevA.100.033613} {\bibfield  {journal} {\bibinfo
  {journal} {Phys. Rev. A}\ }\textbf {\bibinfo {volume} {100}},\ \bibinfo
  {pages} {033613} (\bibinfo {year} {2019})}\BibitemShut {NoStop}%
\bibitem [{\citenamefont {Blonder}\ \emph {et~al.}(1982)\citenamefont
  {Blonder}, \citenamefont {Tinkham},\ and\ \citenamefont
  {Klapwijk}}]{blonder1982transition}%
  \BibitemOpen
  \bibfield  {author} {\bibinfo {author} {\bibfnamefont {G.~E.}\ \bibnamefont
  {Blonder}}, \bibinfo {author} {\bibfnamefont {M.}~\bibnamefont {Tinkham}}, \
  and\ \bibinfo {author} {\bibfnamefont {T.~M.}\ \bibnamefont {Klapwijk}},\
  }\bibfield  {title} {\enquote {\bibinfo {title} {Transition from metallic to
  tunneling regimes in superconducting microconstrictions: {Excess} current,
  charge imbalance, and supercurrent conversion},}\ }\href {\doibase
  10.1103/PhysRevB.25.4515} {\bibfield  {journal} {\bibinfo  {journal} {Phys.
  Rev. B}\ }\textbf {\bibinfo {volume} {25}},\ \bibinfo {pages} {4515--4532}
  (\bibinfo {year} {1982})}\BibitemShut {NoStop}%
\bibitem [{\citenamefont {Fischer}\ \emph {et~al.}(2007)\citenamefont
  {Fischer}, \citenamefont {Kugler}, \citenamefont {Maggio-Aprile},
  \citenamefont {Berthod},\ and\ \citenamefont {Renner}}]{fischer2007scanning}%
  \BibitemOpen
  \bibfield  {author} {\bibinfo {author} {\bibfnamefont {{\O}ystein}\
  \bibnamefont {Fischer}}, \bibinfo {author} {\bibfnamefont {Martin}\
  \bibnamefont {Kugler}}, \bibinfo {author} {\bibfnamefont {Ivan}\ \bibnamefont
  {Maggio-Aprile}}, \bibinfo {author} {\bibfnamefont {Christophe}\ \bibnamefont
  {Berthod}}, \ and\ \bibinfo {author} {\bibfnamefont {Christoph}\ \bibnamefont
  {Renner}},\ }\bibfield  {title} {\enquote {\bibinfo {title} {Scanning
  tunneling spectroscopy of high-temperature superconductors},}\ }\href
  {\doibase 10.1103/RevModPhys.79.353} {\bibfield  {journal} {\bibinfo
  {journal} {Rev. Mod. Phys.}\ }\textbf {\bibinfo {volume} {79}},\ \bibinfo
  {pages} {353--419} (\bibinfo {year} {2007})}\BibitemShut {NoStop}%
\bibitem [{\citenamefont {Mukherjee}\ \emph {et~al.}(2017)\citenamefont
  {Mukherjee}, \citenamefont {Yan}, \citenamefont {Patel}, \citenamefont
  {Hadzibabic}, \citenamefont {Yefsah}, \citenamefont {Struck},\ and\
  \citenamefont {Zwierlein}}]{mukherjee2017homogeneous}%
  \BibitemOpen
  \bibfield  {author} {\bibinfo {author} {\bibfnamefont {Biswaroop}\
  \bibnamefont {Mukherjee}}, \bibinfo {author} {\bibfnamefont {Zhenjie}\
  \bibnamefont {Yan}}, \bibinfo {author} {\bibfnamefont {Parth~B.}\
  \bibnamefont {Patel}}, \bibinfo {author} {\bibfnamefont {Zoran}\ \bibnamefont
  {Hadzibabic}}, \bibinfo {author} {\bibfnamefont {Tarik}\ \bibnamefont
  {Yefsah}}, \bibinfo {author} {\bibfnamefont {Julian}\ \bibnamefont {Struck}},
  \ and\ \bibinfo {author} {\bibfnamefont {Martin~W.}\ \bibnamefont
  {Zwierlein}},\ }\bibfield  {title} {\enquote {\bibinfo {title} {Homogeneous
  {Atomic} {Fermi} {Gases}},}\ }\href {\doibase 10.1103/PhysRevLett.118.123401}
  {\bibfield  {journal} {\bibinfo  {journal} {Phys. Rev. Lett.}\ }\textbf
  {\bibinfo {volume} {118}},\ \bibinfo {pages} {123401} (\bibinfo {year}
  {2017})}\BibitemShut {NoStop}%
\bibitem [{\citenamefont {Gubbels}\ and\ \citenamefont
  {Stoof}(2008)}]{gubbels2008renormalization}%
  \BibitemOpen
  \bibfield  {author} {\bibinfo {author} {\bibfnamefont {K.~B.}\ \bibnamefont
  {Gubbels}}\ and\ \bibinfo {author} {\bibfnamefont {H.~T.~C.}\ \bibnamefont
  {Stoof}},\ }\bibfield  {title} {\enquote {\bibinfo {title} {Renormalization
  {Group} {Theory} for the {Imbalanced} {Fermi} {Gas}},}\ }\href {\doibase
  10.1103/PhysRevLett.100.140407} {\bibfield  {journal} {\bibinfo  {journal}
  {Phys. Rev. Lett.}\ }\textbf {\bibinfo {volume} {100}},\ \bibinfo {pages}
  {140407} (\bibinfo {year} {2008})}\BibitemShut {NoStop}%
\bibitem [{\citenamefont {Gubbels}\ and\ \citenamefont
  {Stoof}(2013)}]{gubbels2013imbalanced}%
  \BibitemOpen
  \bibfield  {author} {\bibinfo {author} {\bibfnamefont {K.~B.}\ \bibnamefont
  {Gubbels}}\ and\ \bibinfo {author} {\bibfnamefont {H.~T.~C.}\ \bibnamefont
  {Stoof}},\ }\bibfield  {title} {\enquote {\bibinfo {title} {Imbalanced
  {Fermi} gases at unitarity},}\ }\href {\doibase
  10.1016/j.physrep.2012.11.004} {\bibfield  {journal} {\bibinfo  {journal}
  {Physics Reports}\ }\textbf {\bibinfo {volume} {525}},\ \bibinfo {pages}
  {255--313} (\bibinfo {year} {2013})}\BibitemShut {NoStop}%
\bibitem [{\citenamefont {Sheehy}\ and\ \citenamefont
  {Radzihovsky}(2006)}]{sheehy2006bec-bcs}%
  \BibitemOpen
  \bibfield  {author} {\bibinfo {author} {\bibfnamefont {Daniel~E.}\
  \bibnamefont {Sheehy}}\ and\ \bibinfo {author} {\bibfnamefont {Leo}\
  \bibnamefont {Radzihovsky}},\ }\bibfield  {title} {\enquote {\bibinfo {title}
  {{BEC}-{BCS} {Crossover} in ``{Magnetized}'' {Feshbach}-{Resonantly} {Paired}
  {Superfluids}},}\ }\href {\doibase 10.1103/PhysRevLett.96.060401} {\bibfield
  {journal} {\bibinfo  {journal} {Phys. Rev. Lett.}\ }\textbf {\bibinfo
  {volume} {96}},\ \bibinfo {pages} {060401} (\bibinfo {year}
  {2006})}\BibitemShut {NoStop}%
\bibitem [{\citenamefont {Son}\ and\ \citenamefont
  {Stephanov}(2006)}]{son2006phase}%
  \BibitemOpen
  \bibfield  {author} {\bibinfo {author} {\bibfnamefont {D.~T.}\ \bibnamefont
  {Son}}\ and\ \bibinfo {author} {\bibfnamefont {M.~A.}\ \bibnamefont
  {Stephanov}},\ }\bibfield  {title} {\enquote {\bibinfo {title} {Phase diagram
  of a cold polarized {Fermi} gas},}\ }\href {\doibase
  10.1103/PhysRevA.74.013614} {\bibfield  {journal} {\bibinfo  {journal} {Phys.
  Rev. A}\ }\textbf {\bibinfo {volume} {74}},\ \bibinfo {pages} {013614}
  (\bibinfo {year} {2006})}\BibitemShut {NoStop}%
\bibitem [{\citenamefont {Yoshida}\ and\ \citenamefont
  {Yip}(2007)}]{yoshida2007larkin-ovchinnikov}%
  \BibitemOpen
  \bibfield  {author} {\bibinfo {author} {\bibfnamefont {Nobukatsu}\
  \bibnamefont {Yoshida}}\ and\ \bibinfo {author} {\bibfnamefont {S.-K.}\
  \bibnamefont {Yip}},\ }\bibfield  {title} {\enquote {\bibinfo {title}
  {Larkin-{Ovchinnikov} state in resonant {Fermi} gas},}\ }\href {\doibase
  10.1103/PhysRevA.75.063601} {\bibfield  {journal} {\bibinfo  {journal} {Phys.
  Rev. A}\ }\textbf {\bibinfo {volume} {75}},\ \bibinfo {pages} {063601}
  (\bibinfo {year} {2007})}\BibitemShut {NoStop}%
\bibitem [{\citenamefont {Parish}\ \emph {et~al.}(2007)\citenamefont {Parish},
  \citenamefont {Marchetti}, \citenamefont {Lamacraft},\ and\ \citenamefont
  {Simons}}]{parish2007finite-temperature}%
  \BibitemOpen
  \bibfield  {author} {\bibinfo {author} {\bibfnamefont {M.~M.}\ \bibnamefont
  {Parish}}, \bibinfo {author} {\bibfnamefont {F.~M.}\ \bibnamefont
  {Marchetti}}, \bibinfo {author} {\bibfnamefont {A.}~\bibnamefont
  {Lamacraft}}, \ and\ \bibinfo {author} {\bibfnamefont {B.~D.}\ \bibnamefont
  {Simons}},\ }\bibfield  {title} {\enquote {\bibinfo {title}
  {Finite-temperature phase diagram of a polarized {Fermi} condensate},}\
  }\href {\doibase 10.1038/nphys520} {\bibfield  {journal} {\bibinfo  {journal}
  {Nature Phys}\ }\textbf {\bibinfo {volume} {3}},\ \bibinfo {pages} {124--128}
  (\bibinfo {year} {2007})}\BibitemShut {NoStop}%
\bibitem [{\citenamefont {Jensen}\ \emph {et~al.}(2007)\citenamefont {Jensen},
  \citenamefont {Kinnunen},\ and\ \citenamefont
  {T{\"o}rm{\"a}}}]{jensen2007non-bcs}%
  \BibitemOpen
  \bibfield  {author} {\bibinfo {author} {\bibfnamefont {L.~M.}\ \bibnamefont
  {Jensen}}, \bibinfo {author} {\bibfnamefont {J.}~\bibnamefont {Kinnunen}}, \
  and\ \bibinfo {author} {\bibfnamefont {P.}~\bibnamefont {T{\"o}rm{\"a}}},\
  }\bibfield  {title} {\enquote {\bibinfo {title} {Non-{BCS} superfluidity in
  trapped ultracold {Fermi} gases},}\ }\href {\doibase
  10.1103/PhysRevA.76.033620} {\bibfield  {journal} {\bibinfo  {journal} {Phys.
  Rev. A}\ }\textbf {\bibinfo {volume} {76}},\ \bibinfo {pages} {033620}
  (\bibinfo {year} {2007})}\BibitemShut {NoStop}%
\bibitem [{\citenamefont {Kinnunen}\ \emph {et~al.}(2018)\citenamefont
  {Kinnunen}, \citenamefont {Baarsma}, \citenamefont {Martikainen},\ and\
  \citenamefont {T{\"o}rm{\"a}}}]{kinnunen2018fuldeferrelllarkinovchinnikov}%
  \BibitemOpen
  \bibfield  {author} {\bibinfo {author} {\bibfnamefont {Jami~J}\ \bibnamefont
  {Kinnunen}}, \bibinfo {author} {\bibfnamefont {Jildou~E}\ \bibnamefont
  {Baarsma}}, \bibinfo {author} {\bibfnamefont {Jani-Petri}\ \bibnamefont
  {Martikainen}}, \ and\ \bibinfo {author} {\bibfnamefont {P{\"a}ivi}\
  \bibnamefont {T{\"o}rm{\"a}}},\ }\bibfield  {title} {\enquote {\bibinfo
  {title} {The
  {Fulde}{\textendash}{Ferrell}{\textendash}{Larkin}{\textendash}{Ovchinnikov}
  state for ultracold fermions in lattice and harmonic potentials: a review},}\
  }\href {\doibase 10.1088/1361-6633/aaa4ad} {\bibfield  {journal} {\bibinfo
  {journal} {Rep. Prog. Phys.}\ }\textbf {\bibinfo {volume} {81}},\ \bibinfo
  {pages} {046401} (\bibinfo {year} {2018})}\BibitemShut {NoStop}%
\bibitem [{\citenamefont {Bulgac}\ \emph {et~al.}(2006)\citenamefont {Bulgac},
  \citenamefont {Forbes},\ and\ \citenamefont {Schwenk}}]{bulgac2006induced}%
  \BibitemOpen
  \bibfield  {author} {\bibinfo {author} {\bibfnamefont {Aurel}\ \bibnamefont
  {Bulgac}}, \bibinfo {author} {\bibfnamefont {Michael~McNeil}\ \bibnamefont
  {Forbes}}, \ and\ \bibinfo {author} {\bibfnamefont {Achim}\ \bibnamefont
  {Schwenk}},\ }\bibfield  {title} {\enquote {\bibinfo {title} {Induced
  ${P}$-{Wave} {Superfluidity} in {Asymmetric} {Fermi} {Gases}},}\ }\href
  {\doibase 10.1103/PhysRevLett.97.020402} {\bibfield  {journal} {\bibinfo
  {journal} {Phys. Rev. Lett.}\ }\textbf {\bibinfo {volume} {97}},\ \bibinfo
  {pages} {020402} (\bibinfo {year} {2006})}\BibitemShut {NoStop}%
\bibitem [{\citenamefont {Bulgac}\ and\ \citenamefont
  {Yoon}(2009)}]{bulgac2009induced}%
  \BibitemOpen
  \bibfield  {author} {\bibinfo {author} {\bibfnamefont {Aurel}\ \bibnamefont
  {Bulgac}}\ and\ \bibinfo {author} {\bibfnamefont {Sukjin}\ \bibnamefont
  {Yoon}},\ }\bibfield  {title} {\enquote {\bibinfo {title} {Induced ${P}$-wave
  superfluidity within the full energy- and momentum-dependent {Eliashberg}
  approximation in asymmetric dilute {Fermi} gases},}\ }\href {\doibase
  10.1103/PhysRevA.79.053625} {\bibfield  {journal} {\bibinfo  {journal} {Phys.
  Rev. A}\ }\textbf {\bibinfo {volume} {79}},\ \bibinfo {pages} {053625}
  (\bibinfo {year} {2009})}\BibitemShut {NoStop}%
\bibitem [{\citenamefont {Patton}\ and\ \citenamefont
  {Sheehy}(2012)}]{patton2012induced}%
  \BibitemOpen
  \bibfield  {author} {\bibinfo {author} {\bibfnamefont {Kelly~R.}\
  \bibnamefont {Patton}}\ and\ \bibinfo {author} {\bibfnamefont {Daniel~E.}\
  \bibnamefont {Sheehy}},\ }\bibfield  {title} {\enquote {\bibinfo {title}
  {Induced superfluidity of imbalanced {Fermi} gases near unitarity},}\ }\href
  {\doibase 10.1103/PhysRevA.85.063625} {\bibfield  {journal} {\bibinfo
  {journal} {Phys. Rev. A}\ }\textbf {\bibinfo {volume} {85}},\ \bibinfo
  {pages} {063625} (\bibinfo {year} {2012})}\BibitemShut {NoStop}%
\bibitem [{\citenamefont {Chaikin}\ and\ \citenamefont
  {Lubensky}(1995)}]{chaikin1995principles}%
  \BibitemOpen
  \bibfield  {author} {\bibinfo {author} {\bibfnamefont {P.~M.}\ \bibnamefont
  {Chaikin}}\ and\ \bibinfo {author} {\bibfnamefont {T.~C.}\ \bibnamefont
  {Lubensky}},\ }\href@noop {} {\emph {\bibinfo {title} {Principles of
  {Condensed} {Matter} {Physics}}}}\ (\bibinfo  {publisher} {Cambridge},\
  \bibinfo {year} {1995})\BibitemShut {NoStop}%
\bibitem [{\citenamefont {Cetoli}\ \emph {et~al.}(2013)\citenamefont {Cetoli},
  \citenamefont {Brand}, \citenamefont {Scott}, \citenamefont {Dalfovo},\ and\
  \citenamefont {Pitaevskii}}]{cetoli2013snake}%
  \BibitemOpen
  \bibfield  {author} {\bibinfo {author} {\bibfnamefont {A.}~\bibnamefont
  {Cetoli}}, \bibinfo {author} {\bibfnamefont {J.}~\bibnamefont {Brand}},
  \bibinfo {author} {\bibfnamefont {R.~G.}\ \bibnamefont {Scott}}, \bibinfo
  {author} {\bibfnamefont {F.}~\bibnamefont {Dalfovo}}, \ and\ \bibinfo
  {author} {\bibfnamefont {L.~P.}\ \bibnamefont {Pitaevskii}},\ }\bibfield
  {title} {\enquote {\bibinfo {title} {Snake instability of dark solitons in
  fermionic superfluids},}\ }\href {\doibase 10.1103/PhysRevA.88.043639}
  {\bibfield  {journal} {\bibinfo  {journal} {Phys. Rev. A}\ }\textbf {\bibinfo
  {volume} {88}},\ \bibinfo {pages} {043639} (\bibinfo {year}
  {2013})}\BibitemShut {NoStop}%
\bibitem [{\citenamefont {Wen}\ \emph {et~al.}(2013)\citenamefont {Wen},
  \citenamefont {Zhao},\ and\ \citenamefont {Ma}}]{wen2013dark-soliton}%
  \BibitemOpen
  \bibfield  {author} {\bibinfo {author} {\bibfnamefont {Wen}\ \bibnamefont
  {Wen}}, \bibinfo {author} {\bibfnamefont {Changqing}\ \bibnamefont {Zhao}}, \
  and\ \bibinfo {author} {\bibfnamefont {Xiaodong}\ \bibnamefont {Ma}},\
  }\bibfield  {title} {\enquote {\bibinfo {title} {Dark-soliton dynamics and
  snake instability in superfluid {Fermi} gases trapped by an anisotropic
  harmonic potential},}\ }\href {\doibase 10.1103/PhysRevA.88.063621}
  {\bibfield  {journal} {\bibinfo  {journal} {Phys. Rev. A}\ }\textbf {\bibinfo
  {volume} {88}},\ \bibinfo {pages} {063621} (\bibinfo {year}
  {2013})}\BibitemShut {NoStop}%
\bibitem [{\citenamefont {Scherpelz}\ \emph {et~al.}(2014)\citenamefont
  {Scherpelz}, \citenamefont {Padavi{\'c}}, \citenamefont {Ran{\c c}on},
  \citenamefont {Glatz}, \citenamefont {Aranson},\ and\ \citenamefont
  {Levin}}]{scherpelz2014phase}%
  \BibitemOpen
  \bibfield  {author} {\bibinfo {author} {\bibfnamefont {Peter}\ \bibnamefont
  {Scherpelz}}, \bibinfo {author} {\bibfnamefont {Karmela}\ \bibnamefont
  {Padavi{\'c}}}, \bibinfo {author} {\bibfnamefont {Adam}\ \bibnamefont {Ran{\c
  c}on}}, \bibinfo {author} {\bibfnamefont {Andreas}\ \bibnamefont {Glatz}},
  \bibinfo {author} {\bibfnamefont {Igor~S.}\ \bibnamefont {Aranson}}, \ and\
  \bibinfo {author} {\bibfnamefont {K.}~\bibnamefont {Levin}},\ }\bibfield
  {title} {\enquote {\bibinfo {title} {Phase {Imprinting} in {Equilibrating}
  {Fermi} {Gases}: {The} {Transience} of {Vortex} {Rings} and {Other}
  {Defects}},}\ }\href {\doibase 10.1103/PhysRevLett.113.125301} {\bibfield
  {journal} {\bibinfo  {journal} {Phys. Rev. Lett.}\ }\textbf {\bibinfo
  {volume} {113}},\ \bibinfo {pages} {125301} (\bibinfo {year}
  {2014})}\BibitemShut {NoStop}%
\bibitem [{\citenamefont {Ku}\ \emph {et~al.}(2016)\citenamefont {Ku},
  \citenamefont {Mukherjee}, \citenamefont {Yefsah},\ and\ \citenamefont
  {Zwierlein}}]{ku2016cascade}%
  \BibitemOpen
  \bibfield  {author} {\bibinfo {author} {\bibfnamefont {Mark J.~H.}\
  \bibnamefont {Ku}}, \bibinfo {author} {\bibfnamefont {Biswaroop}\
  \bibnamefont {Mukherjee}}, \bibinfo {author} {\bibfnamefont {Tarik}\
  \bibnamefont {Yefsah}}, \ and\ \bibinfo {author} {\bibfnamefont {Martin~W.}\
  \bibnamefont {Zwierlein}},\ }\bibfield  {title} {\enquote {\bibinfo {title}
  {Cascade of {Solitonic} {Excitations} in a {Superfluid} {Fermi} gas: {From}
  {Planar} {Solitons} to {Vortex} {Rings} and {Lines}},}\ }\href {\doibase
  10.1103/PhysRevLett.116.045304} {\bibfield  {journal} {\bibinfo  {journal}
  {Phys. Rev. Lett.}\ }\textbf {\bibinfo {volume} {116}},\ \bibinfo {pages}
  {045304} (\bibinfo {year} {2016})}\BibitemShut {NoStop}%
\bibitem [{\citenamefont {Reichl}\ and\ \citenamefont
  {Mueller}(2017)}]{reichl2017core}%
  \BibitemOpen
  \bibfield  {author} {\bibinfo {author} {\bibfnamefont {Matthew~D.}\
  \bibnamefont {Reichl}}\ and\ \bibinfo {author} {\bibfnamefont {Erich~J.}\
  \bibnamefont {Mueller}},\ }\bibfield  {title} {\enquote {\bibinfo {title}
  {Core filling and snaking instability of dark solitons in spin-imbalanced
  superfluid {Fermi} gases},}\ }\href {\doibase 10.1103/PhysRevA.95.053637}
  {\bibfield  {journal} {\bibinfo  {journal} {Phys. Rev. A}\ }\textbf {\bibinfo
  {volume} {95}},\ \bibinfo {pages} {053637} (\bibinfo {year}
  {2017})}\BibitemShut {NoStop}%
\bibitem [{\citenamefont {Wlaz{\l }owski}\ \emph {et~al.}(2018)\citenamefont
  {Wlaz{\l }owski}, \citenamefont {Sekizawa}, \citenamefont {Marchwiany},\ and\
  \citenamefont {Magierski}}]{wlazlowski2018suppressed}%
  \BibitemOpen
  \bibfield  {author} {\bibinfo {author} {\bibfnamefont {Gabriel}\ \bibnamefont
  {Wlaz{\l }owski}}, \bibinfo {author} {\bibfnamefont {Kazuyuki}\ \bibnamefont
  {Sekizawa}}, \bibinfo {author} {\bibfnamefont {Maciej}\ \bibnamefont
  {Marchwiany}}, \ and\ \bibinfo {author} {\bibfnamefont {Piotr}\ \bibnamefont
  {Magierski}},\ }\bibfield  {title} {\enquote {\bibinfo {title} {Suppressed
  {Solitonic} {Cascade} in {Spin}-{Imbalanced} {Superfluid} {Fermi} {Gas}},}\
  }\href {\doibase 10.1103/PhysRevLett.120.253002} {\bibfield  {journal}
  {\bibinfo  {journal} {Phys. Rev. Lett.}\ }\textbf {\bibinfo {volume} {120}},\
  \bibinfo {pages} {253002} (\bibinfo {year} {2018})}\BibitemShut {NoStop}%
\bibitem [{\citenamefont {Tanaka}\ and\ \citenamefont
  {Kashiwaya}(1995)}]{tanaka1995theory}%
  \BibitemOpen
  \bibfield  {author} {\bibinfo {author} {\bibfnamefont {Yukio}\ \bibnamefont
  {Tanaka}}\ and\ \bibinfo {author} {\bibfnamefont {Satoshi}\ \bibnamefont
  {Kashiwaya}},\ }\bibfield  {title} {\enquote {\bibinfo {title} {Theory of
  {Tunneling} {Spectroscopy} of $d$-{Wave} {Superconductors}},}\ }\href
  {\doibase 10.1103/PhysRevLett.74.3451} {\bibfield  {journal} {\bibinfo
  {journal} {Phys. Rev. Lett.}\ }\textbf {\bibinfo {volume} {74}},\ \bibinfo
  {pages} {3451--3454} (\bibinfo {year} {1995})}\BibitemShut {NoStop}%
\bibitem [{\citenamefont {Z{\"u}rn}\ \emph {et~al.}(2013)\citenamefont
  {Z{\"u}rn}, \citenamefont {Lompe}, \citenamefont {Wenz}, \citenamefont
  {Jochim}, \citenamefont {Julienne},\ and\ \citenamefont
  {Hutson}}]{zurn2013precise}%
  \BibitemOpen
  \bibfield  {author} {\bibinfo {author} {\bibfnamefont {G.}~\bibnamefont
  {Z{\"u}rn}}, \bibinfo {author} {\bibfnamefont {T.}~\bibnamefont {Lompe}},
  \bibinfo {author} {\bibfnamefont {A.~N.}\ \bibnamefont {Wenz}}, \bibinfo
  {author} {\bibfnamefont {S.}~\bibnamefont {Jochim}}, \bibinfo {author}
  {\bibfnamefont {P.~S.}\ \bibnamefont {Julienne}}, \ and\ \bibinfo {author}
  {\bibfnamefont {J.~M.}\ \bibnamefont {Hutson}},\ }\bibfield  {title}
  {\enquote {\bibinfo {title} {Precise {Characterization} of {$^{6}$}{Li}
  {Feshbach} {Resonances} {Using} {Trap}-{Sideband}-{Resolved} {RF}
  {Spectroscopy} of {Weakly} {Bound} {Molecules}},}\ }\href {\doibase
  10.1103/PhysRevLett.110.135301} {\bibfield  {journal} {\bibinfo  {journal}
  {Phys. Rev. Lett.}\ }\textbf {\bibinfo {volume} {110}},\ \bibinfo {pages}
  {135301} (\bibinfo {year} {2013})}\BibitemShut {NoStop}%
\bibitem [{\citenamefont {Yan}\ \emph {et~al.}(2019)\citenamefont {Yan},
  \citenamefont {Patel}, \citenamefont {Mukherjee}, \citenamefont {Fletcher},
  \citenamefont {Struck},\ and\ \citenamefont {Zwierlein}}]{yan2019boiling}%
  \BibitemOpen
  \bibfield  {author} {\bibinfo {author} {\bibfnamefont {Zhenjie}\ \bibnamefont
  {Yan}}, \bibinfo {author} {\bibfnamefont {Parth~B.}\ \bibnamefont {Patel}},
  \bibinfo {author} {\bibfnamefont {Biswaroop}\ \bibnamefont {Mukherjee}},
  \bibinfo {author} {\bibfnamefont {Richard~J.}\ \bibnamefont {Fletcher}},
  \bibinfo {author} {\bibfnamefont {Julian}\ \bibnamefont {Struck}}, \ and\
  \bibinfo {author} {\bibfnamefont {Martin~W.}\ \bibnamefont {Zwierlein}},\
  }\bibfield  {title} {\enquote {\bibinfo {title} {Boiling a {Unitary} {Fermi}
  {Liquid}},}\ }\href {\doibase 10.1103/PhysRevLett.122.093401} {\bibfield
  {journal} {\bibinfo  {journal} {Phys. Rev. Lett.}\ }\textbf {\bibinfo
  {volume} {122}},\ \bibinfo {pages} {093401} (\bibinfo {year}
  {2019})}\BibitemShut {NoStop}%
\bibitem [{\citenamefont {Combescot}\ and\ \citenamefont
  {Giraud}(2008)}]{combescot2008normal}%
  \BibitemOpen
  \bibfield  {author} {\bibinfo {author} {\bibfnamefont {R.}~\bibnamefont
  {Combescot}}\ and\ \bibinfo {author} {\bibfnamefont {S.}~\bibnamefont
  {Giraud}},\ }\bibfield  {title} {\enquote {\bibinfo {title} {Normal {State}
  of {Highly} {Polarized} {Fermi} {Gases}: {Full} {Many}-{Body} {Treatment}},}\
  }\href {\doibase 10.1103/PhysRevLett.101.050404} {\bibfield  {journal}
  {\bibinfo  {journal} {Phys. Rev. Lett.}\ }\textbf {\bibinfo {volume} {101}},\
  \bibinfo {pages} {050404} (\bibinfo {year} {2008})}\BibitemShut {NoStop}%
\bibitem [{\citenamefont {Prokof{\textquoteright}ev}\ and\ \citenamefont
  {Svistunov}(2008)}]{prokofev2008fermi-polaron}%
  \BibitemOpen
  \bibfield  {author} {\bibinfo {author} {\bibfnamefont {Nikolay}\ \bibnamefont
  {Prokof{\textquoteright}ev}}\ and\ \bibinfo {author} {\bibfnamefont {Boris}\
  \bibnamefont {Svistunov}},\ }\bibfield  {title} {\enquote {\bibinfo {title}
  {Fermi-polaron problem: {Diagrammatic} {Monte} {Carlo} method for divergent
  sign-alternating series},}\ }\href {\doibase 10.1103/PhysRevB.77.020408}
  {\bibfield  {journal} {\bibinfo  {journal} {Phys. Rev. B}\ }\textbf {\bibinfo
  {volume} {77}},\ \bibinfo {pages} {020408(R)} (\bibinfo {year}
  {2008})}\BibitemShut {NoStop}%
\bibitem [{\citenamefont {Mora}\ and\ \citenamefont
  {Chevy}(2010)}]{mora2010normal}%
  \BibitemOpen
  \bibfield  {author} {\bibinfo {author} {\bibfnamefont {Christophe}\
  \bibnamefont {Mora}}\ and\ \bibinfo {author} {\bibfnamefont
  {Fr{\'e}d{\'e}ric}\ \bibnamefont {Chevy}},\ }\bibfield  {title} {\enquote
  {\bibinfo {title} {Normal {Phase} of an {Imbalanced} {Fermi} {Gas}},}\ }\href
  {\doibase 10.1103/PhysRevLett.104.230402} {\bibfield  {journal} {\bibinfo
  {journal} {Phys. Rev. Lett.}\ }\textbf {\bibinfo {volume} {104}},\ \bibinfo
  {pages} {230402} (\bibinfo {year} {2010})}\BibitemShut {NoStop}%
\bibitem [{\citenamefont {Bulgac}\ and\ \citenamefont
  {Forbes}(2011)}]{bulgac2011time-dependent}%
  \BibitemOpen
  \bibfield  {author} {\bibinfo {author} {\bibfnamefont {Aurel}\ \bibnamefont
  {Bulgac}}\ and\ \bibinfo {author} {\bibfnamefont {Michael~McNeil}\
  \bibnamefont {Forbes}},\ }\bibfield  {title} {\enquote {\bibinfo {title}
  {Time-{Dependent} {Superfluid} {Local}-{Density} {Approximation}},}\ }in\
  \href {https://www.worldscientific.com/doi/abs/10.1142/9781848168121_0026}
  {\emph {\bibinfo {booktitle} {Quantum {Gases}}}},\ \bibinfo {series} {Cold
  {Atoms}}, Vol.~\bibinfo {volume} {1}\ (\bibinfo  {publisher} {Imperial
  College},\ \bibinfo {year} {2011})\ pp.\ \bibinfo {pages}
  {397--406}\BibitemShut {NoStop}%
\bibitem [{\citenamefont {Forbes}\ \emph {et~al.}(2011)\citenamefont {Forbes},
  \citenamefont {Gandolfi},\ and\ \citenamefont
  {Gezerlis}}]{forbes2011resonantly}%
  \BibitemOpen
  \bibfield  {author} {\bibinfo {author} {\bibfnamefont {Michael~McNeil}\
  \bibnamefont {Forbes}}, \bibinfo {author} {\bibfnamefont {Stefano}\
  \bibnamefont {Gandolfi}}, \ and\ \bibinfo {author} {\bibfnamefont
  {Alexandros}\ \bibnamefont {Gezerlis}},\ }\bibfield  {title} {\enquote
  {\bibinfo {title} {Resonantly {Interacting} {Fermions} in a {Box}},}\ }\href
  {\doibase 10.1103/PhysRevLett.106.235303} {\bibfield  {journal} {\bibinfo
  {journal} {Phys. Rev. Lett.}\ }\textbf {\bibinfo {volume} {106}},\ \bibinfo
  {pages} {235303} (\bibinfo {year} {2011})}\BibitemShut {NoStop}%
\bibitem [{\citenamefont {Bulgac}\ \emph {et~al.}(2012)\citenamefont {Bulgac},
  \citenamefont {Forbes},\ and\ \citenamefont {Magierski}}]{bulgac2012unitary}%
  \BibitemOpen
  \bibfield  {author} {\bibinfo {author} {\bibfnamefont {Aurel}\ \bibnamefont
  {Bulgac}}, \bibinfo {author} {\bibfnamefont {Michael~McNeil}\ \bibnamefont
  {Forbes}}, \ and\ \bibinfo {author} {\bibfnamefont {Piotr}\ \bibnamefont
  {Magierski}},\ }\bibfield  {title} {\enquote {\bibinfo {title} {The {Unitary}
  {Fermi} {Gas}: {From} {Monte} {Carlo} to {Density} {Functionals}},}\ }in\
  \href {https://doi.org/10.1007/978-3-642-21978-8_9} {\emph {\bibinfo
  {booktitle} {The {BCS}-{BEC} {Crossover} and the {Unitary} {Fermi} {Gas}}}},\
  \bibinfo {series and number} {Lecture {Notes} in {Physics}},\ \bibinfo
  {editor} {edited by\ \bibinfo {editor} {\bibfnamefont {Wilhelm}\ \bibnamefont
  {Zwerger}}}\ (\bibinfo  {publisher} {Springer},\ \bibinfo {address} {Berlin,
  Heidelberg},\ \bibinfo {year} {2012})\ pp.\ \bibinfo {pages}
  {305--373}\BibitemShut {NoStop}%
\bibitem [{\citenamefont {Bulgac}(2013)}]{bulgac2013time-dependent-1}%
  \BibitemOpen
  \bibfield  {author} {\bibinfo {author} {\bibfnamefont {Aurel}\ \bibnamefont
  {Bulgac}},\ }\bibfield  {title} {\enquote {\bibinfo {title} {Time-{Dependent}
  {Density} {Functional} {Theory} and the {Real}-{Time} {Dynamics} of {Fermi}
  {Superfluids}},}\ }\href {\doibase 10.1146/annurev-nucl-102212-170631}
  {\bibfield  {journal} {\bibinfo  {journal} {Annual Review of Nuclear and
  Particle Science}\ }\textbf {\bibinfo {volume} {63}},\ \bibinfo {pages}
  {97--121} (\bibinfo {year} {2013})}\BibitemShut {NoStop}%
\bibitem [{\citenamefont {Kopyci{\'n}ski}\ \emph {et~al.}(2021)\citenamefont
  {Kopyci{\'n}ski}, \citenamefont {Pudelko},\ and\ \citenamefont {Wlaz{\l
  }owski}}]{Kopycinski2021vortex}%
  \BibitemOpen
  \bibfield  {author} {\bibinfo {author} {\bibfnamefont {Jakub}\ \bibnamefont
  {Kopyci{\'n}ski}}, \bibinfo {author} {\bibfnamefont {Wojciech~R.}\
  \bibnamefont {Pudelko}}, \ and\ \bibinfo {author} {\bibfnamefont {Gabriel}\
  \bibnamefont {Wlaz{\l }owski}},\ }\bibfield  {title} {\enquote {\bibinfo
  {title} {Vortex lattice in spin-imbalanced unitary {Fermi} gas},}\ }\href
  {\doibase 10.1103/PhysRevA.104.053322} {\bibfield  {journal} {\bibinfo
  {journal} {Phys. Rev. A}\ }\textbf {\bibinfo {volume} {104}},\ \bibinfo
  {pages} {053322} (\bibinfo {year} {2021})}\BibitemShut {NoStop}%
\bibitem [{\citenamefont {Hossain}\ \emph {et~al.}(2022)\citenamefont
  {Hossain}, \citenamefont {Kobuszewski}, \citenamefont {Forbes}, \citenamefont
  {Magierski}, \citenamefont {Sekizawa},\ and\ \citenamefont {Wlaz{\l
  }owski}}]{hossain2022rotating}%
  \BibitemOpen
  \bibfield  {author} {\bibinfo {author} {\bibfnamefont {Khalid}\ \bibnamefont
  {Hossain}}, \bibinfo {author} {\bibfnamefont {Konrad}\ \bibnamefont
  {Kobuszewski}}, \bibinfo {author} {\bibfnamefont {Michael~McNeil}\
  \bibnamefont {Forbes}}, \bibinfo {author} {\bibfnamefont {Piotr}\
  \bibnamefont {Magierski}}, \bibinfo {author} {\bibfnamefont {Kazuyuki}\
  \bibnamefont {Sekizawa}}, \ and\ \bibinfo {author} {\bibfnamefont {Gabriel}\
  \bibnamefont {Wlaz{\l }owski}},\ }\bibfield  {title} {\enquote {\bibinfo
  {title} {Rotating quantum turbulence in the unitary {Fermi} gas},}\ }\href
  {\doibase 10.1103/PhysRevA.105.013304} {\bibfield  {journal} {\bibinfo
  {journal} {Phys. Rev. A}\ }\textbf {\bibinfo {volume} {105}},\ \bibinfo
  {pages} {013304} (\bibinfo {year} {2022})}\BibitemShut {NoStop}%
\bibitem [{\citenamefont {Kawamura}\ \emph {et~al.}(2020)\citenamefont
  {Kawamura}, \citenamefont {Hanai}, \citenamefont {Kagamihara}, \citenamefont
  {Inotani},\ and\ \citenamefont {Ohashi}}]{kawamura2020nonequilibrium}%
  \BibitemOpen
  \bibfield  {author} {\bibinfo {author} {\bibfnamefont {Taira}\ \bibnamefont
  {Kawamura}}, \bibinfo {author} {\bibfnamefont {Ryo}\ \bibnamefont {Hanai}},
  \bibinfo {author} {\bibfnamefont {Daichi}\ \bibnamefont {Kagamihara}},
  \bibinfo {author} {\bibfnamefont {Daisuke}\ \bibnamefont {Inotani}}, \ and\
  \bibinfo {author} {\bibfnamefont {Yoji}\ \bibnamefont {Ohashi}},\ }\bibfield
  {title} {\enquote {\bibinfo {title} {Nonequilibrium strong-coupling theory
  for a driven-dissipative ultracold {Fermi} gas in the {BCS}-{BEC} crossover
  region},}\ }\href {\doibase 10.1103/PhysRevA.101.013602} {\bibfield
  {journal} {\bibinfo  {journal} {Phys. Rev. A}\ }\textbf {\bibinfo {volume}
  {101}},\ \bibinfo {pages} {013602} (\bibinfo {year} {2020})}\BibitemShut
  {NoStop}%
\bibitem [{\citenamefont {Wang}\ \emph {et~al.}(2014)\citenamefont {Wang},
  \citenamefont {Agarwalla}, \citenamefont {Li},\ and\ \citenamefont
  {Thingna}}]{wang2014nonequilibrium}%
  \BibitemOpen
  \bibfield  {author} {\bibinfo {author} {\bibfnamefont {Jian-Sheng}\
  \bibnamefont {Wang}}, \bibinfo {author} {\bibfnamefont {Bijay~Kumar}\
  \bibnamefont {Agarwalla}}, \bibinfo {author} {\bibfnamefont {Huanan}\
  \bibnamefont {Li}}, \ and\ \bibinfo {author} {\bibfnamefont {Juzar}\
  \bibnamefont {Thingna}},\ }\bibfield  {title} {\enquote {\bibinfo {title}
  {Nonequilibrium {Green}{\textquoteright}s function method for quantum thermal
  transport},}\ }\href {\doibase 10.1007/s11467-013-0340-x} {\bibfield
  {journal} {\bibinfo  {journal} {Front. Phys.}\ }\textbf {\bibinfo {volume}
  {9}},\ \bibinfo {pages} {673--697} (\bibinfo {year} {2014})}\BibitemShut
  {NoStop}%
\bibitem [{\citenamefont {Liu}\ \emph {et~al.}(2017)\citenamefont {Liu},
  \citenamefont {Zhai},\ and\ \citenamefont {Zhang}}]{liu2017anomalous}%
  \BibitemOpen
  \bibfield  {author} {\bibinfo {author} {\bibfnamefont {Boyang}\ \bibnamefont
  {Liu}}, \bibinfo {author} {\bibfnamefont {Hui}\ \bibnamefont {Zhai}}, \ and\
  \bibinfo {author} {\bibfnamefont {Shizhong}\ \bibnamefont {Zhang}},\
  }\bibfield  {title} {\enquote {\bibinfo {title} {Anomalous conductance of a
  strongly interacting {Fermi} gas through a quantum point contact},}\ }\href
  {\doibase 10.1103/PhysRevA.95.013623} {\bibfield  {journal} {\bibinfo
  {journal} {Phys. Rev. A}\ }\textbf {\bibinfo {volume} {95}},\ \bibinfo
  {pages} {013623} (\bibinfo {year} {2017})}\BibitemShut {NoStop}%
\bibitem [{\citenamefont {Wlaz{\l }owski}\ \emph {et~al.}(2013)\citenamefont
  {Wlaz{\l }owski}, \citenamefont {Magierski}, \citenamefont {Drut},
  \citenamefont {Bulgac},\ and\ \citenamefont {Roche}}]{wlazlowski2013cooper}%
  \BibitemOpen
  \bibfield  {author} {\bibinfo {author} {\bibfnamefont {Gabriel}\ \bibnamefont
  {Wlaz{\l }owski}}, \bibinfo {author} {\bibfnamefont {Piotr}\ \bibnamefont
  {Magierski}}, \bibinfo {author} {\bibfnamefont {Joaqu{\'i}n~E.}\ \bibnamefont
  {Drut}}, \bibinfo {author} {\bibfnamefont {Aurel}\ \bibnamefont {Bulgac}}, \
  and\ \bibinfo {author} {\bibfnamefont {Kenneth~J.}\ \bibnamefont {Roche}},\
  }\bibfield  {title} {\enquote {\bibinfo {title} {Cooper {Pairing} {Above} the
  {Critical} {Temperature} in a {Unitary} {Fermi} {Gas}},}\ }\href {\doibase
  10.1103/PhysRevLett.110.090401} {\bibfield  {journal} {\bibinfo  {journal}
  {Phys. Rev. Lett.}\ }\textbf {\bibinfo {volume} {110}},\ \bibinfo {pages}
  {090401} (\bibinfo {year} {2013})}\BibitemShut {NoStop}%
\bibitem [{\citenamefont {Sekino}\ \emph {et~al.}(2020)\citenamefont {Sekino},
  \citenamefont {Tajima},\ and\ \citenamefont {Uchino}}]{sekino2020mesoscopic}%
  \BibitemOpen
  \bibfield  {author} {\bibinfo {author} {\bibfnamefont {Yuta}\ \bibnamefont
  {Sekino}}, \bibinfo {author} {\bibfnamefont {Hiroyuki}\ \bibnamefont
  {Tajima}}, \ and\ \bibinfo {author} {\bibfnamefont {Shun}\ \bibnamefont
  {Uchino}},\ }\bibfield  {title} {\enquote {\bibinfo {title} {Mesoscopic spin
  transport between strongly interacting {Fermi} gases},}\ }\href {\doibase
  10.1103/PhysRevResearch.2.023152} {\bibfield  {journal} {\bibinfo  {journal}
  {Phys. Rev. Research}\ }\textbf {\bibinfo {volume} {2}},\ \bibinfo {pages}
  {023152} (\bibinfo {year} {2020})}\BibitemShut {NoStop}%
\bibitem [{\citenamefont {Frank}\ \emph {et~al.}(2020)\citenamefont {Frank},
  \citenamefont {Zwerger},\ and\ \citenamefont {Enss}}]{frank2020quantum}%
  \BibitemOpen
  \bibfield  {author} {\bibinfo {author} {\bibfnamefont {Bernhard}\
  \bibnamefont {Frank}}, \bibinfo {author} {\bibfnamefont {Wilhelm}\
  \bibnamefont {Zwerger}}, \ and\ \bibinfo {author} {\bibfnamefont {Tilman}\
  \bibnamefont {Enss}},\ }\bibfield  {title} {\enquote {\bibinfo {title}
  {Quantum critical thermal transport in the unitary {Fermi} gas},}\ }\href
  {\doibase 10.1103/PhysRevResearch.2.023301} {\bibfield  {journal} {\bibinfo
  {journal} {Phys. Rev. Research}\ }\textbf {\bibinfo {volume} {2}},\ \bibinfo
  {pages} {023301} (\bibinfo {year} {2020})}\BibitemShut {NoStop}%
\bibitem [{\citenamefont {Carlson}\ and\ \citenamefont
  {Reddy}(2005)}]{carlson2005asymmetric}%
  \BibitemOpen
  \bibfield  {author} {\bibinfo {author} {\bibfnamefont {J.}~\bibnamefont
  {Carlson}}\ and\ \bibinfo {author} {\bibfnamefont {Sanjay}\ \bibnamefont
  {Reddy}},\ }\bibfield  {title} {\enquote {\bibinfo {title} {Asymmetric
  {Two}-{Component} {Fermion} {Systems} in {Strong} {Coupling}},}\ }\href
  {\doibase 10.1103/PhysRevLett.95.060401} {\bibfield  {journal} {\bibinfo
  {journal} {Phys. Rev. Lett.}\ }\textbf {\bibinfo {volume} {95}},\ \bibinfo
  {pages} {060401} (\bibinfo {year} {2005})}\BibitemShut {NoStop}%
\bibitem [{\citenamefont {Magierski}\ \emph {et~al.}(2009)\citenamefont
  {Magierski}, \citenamefont {Wlaz{\l }owski}, \citenamefont {Bulgac},\ and\
  \citenamefont {Drut}}]{magierski2009finite-temperature}%
  \BibitemOpen
  \bibfield  {author} {\bibinfo {author} {\bibfnamefont {Piotr}\ \bibnamefont
  {Magierski}}, \bibinfo {author} {\bibfnamefont {Gabriel}\ \bibnamefont
  {Wlaz{\l }owski}}, \bibinfo {author} {\bibfnamefont {Aurel}\ \bibnamefont
  {Bulgac}}, \ and\ \bibinfo {author} {\bibfnamefont {Joaqu{\'i}n~E.}\
  \bibnamefont {Drut}},\ }\bibfield  {title} {\enquote {\bibinfo {title}
  {Finite-{Temperature} {Pairing} {Gap} of a {Unitary} {Fermi} {Gas} by
  {Quantum} {Monte} {Carlo} {Calculations}},}\ }\href {\doibase
  10.1103/PhysRevLett.103.210403} {\bibfield  {journal} {\bibinfo  {journal}
  {Phys. Rev. Lett.}\ }\textbf {\bibinfo {volume} {103}},\ \bibinfo {pages}
  {210403} (\bibinfo {year} {2009})}\BibitemShut {NoStop}%
\bibitem [{\citenamefont {Haussmann}\ \emph {et~al.}(2009)\citenamefont
  {Haussmann}, \citenamefont {Punk},\ and\ \citenamefont
  {Zwerger}}]{haussmann2009spectral}%
  \BibitemOpen
  \bibfield  {author} {\bibinfo {author} {\bibfnamefont {R.}~\bibnamefont
  {Haussmann}}, \bibinfo {author} {\bibfnamefont {M.}~\bibnamefont {Punk}}, \
  and\ \bibinfo {author} {\bibfnamefont {W.}~\bibnamefont {Zwerger}},\
  }\bibfield  {title} {\enquote {\bibinfo {title} {Spectral functions and rf
  response of ultracold fermionic atoms},}\ }\href {\doibase
  10.1103/PhysRevA.80.063612} {\bibfield  {journal} {\bibinfo  {journal} {Phys.
  Rev. A}\ }\textbf {\bibinfo {volume} {80}},\ \bibinfo {pages} {063612}
  (\bibinfo {year} {2009})}\BibitemShut {NoStop}%
\bibitem [{\citenamefont {Stewart}\ \emph {et~al.}(2008)\citenamefont
  {Stewart}, \citenamefont {Gaebler},\ and\ \citenamefont
  {Jin}}]{stewart2008using}%
  \BibitemOpen
  \bibfield  {author} {\bibinfo {author} {\bibfnamefont {J.~T.}\ \bibnamefont
  {Stewart}}, \bibinfo {author} {\bibfnamefont {J.~P.}\ \bibnamefont
  {Gaebler}}, \ and\ \bibinfo {author} {\bibfnamefont {D.~S.}\ \bibnamefont
  {Jin}},\ }\bibfield  {title} {\enquote {\bibinfo {title} {Using photoemission
  spectroscopy to probe a strongly interacting {Fermi} gas},}\ }\href {\doibase
  10.1038/nature07172} {\bibfield  {journal} {\bibinfo  {journal} {Nature}\
  }\textbf {\bibinfo {volume} {454}},\ \bibinfo {pages} {744--747} (\bibinfo
  {year} {2008})}\BibitemShut {NoStop}%
\bibitem [{\citenamefont {de~Gennes}(1989)}]{de_gennes1989superconductivity}%
  \BibitemOpen
  \bibfield  {author} {\bibinfo {author} {\bibfnamefont {P.~G.}\ \bibnamefont
  {de~Gennes}},\ }\href@noop {} {\emph {\bibinfo {title} {Superconductivity of
  {Metals} and {Alloys}}}}\ (\bibinfo  {publisher} {Addison-Wesley},\ \bibinfo
  {address} {Redwood City},\ \bibinfo {year} {1989})\BibitemShut {NoStop}%
\bibitem [{\citenamefont {Haussmann}\ \emph {et~al.}(2007)\citenamefont
  {Haussmann}, \citenamefont {Rantner}, \citenamefont {Cerrito},\ and\
  \citenamefont {Zwerger}}]{haussmann2007thermodynamics}%
  \BibitemOpen
  \bibfield  {author} {\bibinfo {author} {\bibfnamefont {R.}~\bibnamefont
  {Haussmann}}, \bibinfo {author} {\bibfnamefont {W.}~\bibnamefont {Rantner}},
  \bibinfo {author} {\bibfnamefont {S.}~\bibnamefont {Cerrito}}, \ and\
  \bibinfo {author} {\bibfnamefont {W.}~\bibnamefont {Zwerger}},\ }\bibfield
  {title} {\enquote {\bibinfo {title} {Thermodynamics of the {BCS}-{BEC}
  crossover},}\ }\href {\doibase 10.1103/PhysRevA.75.023610} {\bibfield
  {journal} {\bibinfo  {journal} {Phys. Rev. A}\ }\textbf {\bibinfo {volume}
  {75}},\ \bibinfo {pages} {023610} (\bibinfo {year} {2007})}\BibitemShut
  {NoStop}%
\bibitem [{\citenamefont {Mueller}(2017)}]{mueller2017review}%
  \BibitemOpen
  \bibfield  {author} {\bibinfo {author} {\bibfnamefont {Erich~J.}\
  \bibnamefont {Mueller}},\ }\bibfield  {title} {\enquote {\bibinfo {title}
  {Review of pseudogaps in strongly interacting {Fermi} gases},}\ }\href
  {\doibase 10.1088/1361-6633/aa7e53} {\bibfield  {journal} {\bibinfo
  {journal} {Rep. Prog. Phys.}\ }\textbf {\bibinfo {volume} {80}},\ \bibinfo
  {pages} {104401} (\bibinfo {year} {2017})}\BibitemShut {NoStop}%
\bibitem [{\citenamefont {Mueller}(2011)}]{mueller2011evolution}%
  \BibitemOpen
  \bibfield  {author} {\bibinfo {author} {\bibfnamefont {Erich~J.}\
  \bibnamefont {Mueller}},\ }\bibfield  {title} {\enquote {\bibinfo {title}
  {Evolution of the pseudogap in a polarized {Fermi} gas},}\ }\href {\doibase
  10.1103/PhysRevA.83.053623} {\bibfield  {journal} {\bibinfo  {journal} {Phys.
  Rev. A}\ }\textbf {\bibinfo {volume} {83}},\ \bibinfo {pages} {053623}
  (\bibinfo {year} {2011})}\BibitemShut {NoStop}%
\bibitem [{\citenamefont {BenDaniel}\ and\ \citenamefont
  {Duke}(1966)}]{bendaniel1966space-charge}%
  \BibitemOpen
  \bibfield  {author} {\bibinfo {author} {\bibfnamefont {D.~J.}\ \bibnamefont
  {BenDaniel}}\ and\ \bibinfo {author} {\bibfnamefont {C.~B.}\ \bibnamefont
  {Duke}},\ }\bibfield  {title} {\enquote {\bibinfo {title} {Space-{Charge}
  {Effects} on {Electron} {Tunneling}},}\ }\href {\doibase
  10.1103/PhysRev.152.683} {\bibfield  {journal} {\bibinfo  {journal} {Phys.
  Rev.}\ }\textbf {\bibinfo {volume} {152}},\ \bibinfo {pages} {683--692}
  (\bibinfo {year} {1966})}\BibitemShut {NoStop}%
\bibitem [{\citenamefont {Einevoll}(1990)}]{einevoll1990operator-1}%
  \BibitemOpen
  \bibfield  {author} {\bibinfo {author} {\bibfnamefont {G.~T.}\ \bibnamefont
  {Einevoll}},\ }\bibfield  {title} {\enquote {\bibinfo {title} {Operator
  ordering in effective-mass theory for heterostructures. {II}. {Strained}
  systems},}\ }\href {\doibase 10.1103/PhysRevB.42.3497} {\bibfield  {journal}
  {\bibinfo  {journal} {Phys. Rev. B}\ }\textbf {\bibinfo {volume} {42}},\
  \bibinfo {pages} {3497--3502} (\bibinfo {year} {1990})}\BibitemShut {NoStop}%
\bibitem [{\citenamefont {Cavalcante}\ \emph {et~al.}(1997)\citenamefont
  {Cavalcante}, \citenamefont {Costa~Filho}, \citenamefont {Filho},
  \citenamefont {de~Almeida},\ and\ \citenamefont
  {Freire}}]{cavalcante1997form}%
  \BibitemOpen
  \bibfield  {author} {\bibinfo {author} {\bibfnamefont {F.~S.~A.}\
  \bibnamefont {Cavalcante}}, \bibinfo {author} {\bibfnamefont {R.~N.}\
  \bibnamefont {Costa~Filho}}, \bibinfo {author} {\bibfnamefont {J.~Ribeiro}\
  \bibnamefont {Filho}}, \bibinfo {author} {\bibfnamefont {C.~A.~S.}\
  \bibnamefont {de~Almeida}}, \ and\ \bibinfo {author} {\bibfnamefont {V.~N.}\
  \bibnamefont {Freire}},\ }\bibfield  {title} {\enquote {\bibinfo {title}
  {Form of the quantum kinetic-energy operator with spatially varying effective
  mass},}\ }\href {\doibase 10.1103/PhysRevB.55.1326} {\bibfield  {journal}
  {\bibinfo  {journal} {Phys. Rev. B}\ }\textbf {\bibinfo {volume} {55}},\
  \bibinfo {pages} {1326--1328} (\bibinfo {year} {1997})}\BibitemShut {NoStop}%
\bibitem [{\citenamefont {Srikanth}\ and\ \citenamefont
  {Raychaudhuri}(1992)}]{srikanth1992modeling}%
  \BibitemOpen
  \bibfield  {author} {\bibinfo {author} {\bibfnamefont {H.}~\bibnamefont
  {Srikanth}}\ and\ \bibinfo {author} {\bibfnamefont {A.K.}\ \bibnamefont
  {Raychaudhuri}},\ }\bibfield  {title} {\enquote {\bibinfo {title} {Modeling
  tunneling data of normal metal-oxide superconductor point contact
  junctions},}\ }\href {\doibase 10.1016/0921-4534(92)90600-H} {\bibfield
  {journal} {\bibinfo  {journal} {Physica C: Superconductivity}\ }\textbf
  {\bibinfo {volume} {190}},\ \bibinfo {pages} {229--233} (\bibinfo {year}
  {1992})}\BibitemShut {NoStop}%
\bibitem [{\citenamefont {Schirotzek}\ \emph {et~al.}(2009)\citenamefont
  {Schirotzek}, \citenamefont {Wu}, \citenamefont {Sommer},\ and\ \citenamefont
  {Zwierlein}}]{schirotzek2009observation}%
  \BibitemOpen
  \bibfield  {author} {\bibinfo {author} {\bibfnamefont {Andr{\'e}}\
  \bibnamefont {Schirotzek}}, \bibinfo {author} {\bibfnamefont {Cheng-Hsun}\
  \bibnamefont {Wu}}, \bibinfo {author} {\bibfnamefont {Ariel}\ \bibnamefont
  {Sommer}}, \ and\ \bibinfo {author} {\bibfnamefont {Martin~W.}\ \bibnamefont
  {Zwierlein}},\ }\bibfield  {title} {\enquote {\bibinfo {title} {Observation
  of {Fermi} {Polarons} in a {Tunable} {Fermi} {Liquid} of {Ultracold}
  {Atoms}},}\ }\href {\doibase 10.1103/PhysRevLett.102.230402} {\bibfield
  {journal} {\bibinfo  {journal} {Phys. Rev. Lett.}\ }\textbf {\bibinfo
  {volume} {102}},\ \bibinfo {pages} {230402} (\bibinfo {year}
  {2009})}\BibitemShut {NoStop}%
\bibitem [{\citenamefont {Chevy}(2006)}]{chevy2006universal}%
  \BibitemOpen
  \bibfield  {author} {\bibinfo {author} {\bibfnamefont {F.}~\bibnamefont
  {Chevy}},\ }\bibfield  {title} {\enquote {\bibinfo {title} {Universal phase
  diagram of a strongly interacting {Fermi} gas with unbalanced spin
  populations},}\ }\href {\doibase 10.1103/PhysRevA.74.063628} {\bibfield
  {journal} {\bibinfo  {journal} {Phys. Rev. A}\ }\textbf {\bibinfo {volume}
  {74}},\ \bibinfo {pages} {063628} (\bibinfo {year} {2006})}\BibitemShut
  {NoStop}%
\end{thebibliography}

\end{document}